\documentclass[12pt]{article}

\usepackage[T1]{fontenc}
\usepackage[utf8]{inputenc}

\usepackage{graphicx}
\usepackage{amsmath,amssymb,amsfonts,mathtools}
\usepackage[version=4]{mhchem}
\usepackage{booktabs,multirow}
\usepackage{siunitx}
\usepackage{url}
\usepackage[hidelinks]{hyperref}
\usepackage{xcolor}
\usepackage{subcaption}
\usepackage{enumitem}
\usepackage{setspace}
\usepackage{microtype}
\usepackage[numbers,sort&compress,super]{natbib}
\usepackage{etoolbox}
\usepackage[section]{placeins}

\usepackage[
  left=1.2in,
  right=1.2in,
  top=1.2in,
  bottom=1.2in
]{geometry}

\graphicspath{{Figures/}}

\begin{document}

\begin{center}
{\LARGE Kinetically Arrested Twin-Domain State in}\\[0.5em]
{\LARGE Formamidinium Lead Iodide}\\[1.5em]

{\large
Xia Liang$^{1,2}$, Milos Dubajic$^{3}$, Zezhu Zeng$^{4}$, Yang Lu$^{3}$,\\[0.25em]
Johan Klarbring$^{5}$, Samuel D.\ Stranks$^{3}$, Aron Walsh$^{1}$
}\\[1em]

{\footnotesize
$^{1}$Department of Materials, Imperial College London, South Kensington Campus, London SW7 2AZ, UK\\[0.25em]
$^{2}$School of Energy and Chemical Engineering, Ulsan National Institute
of Science and Technology (UNIST), Ulsan 44919, Korea\\[0.25em]
$^{3}$Department of Chemical Engineering and Biotechnology, University of Cambridge, Philippa Fawcett Drive, Cambridge, CB3 0AS, UK\\[0.25em]
$^{4}$Physical and Theoretical Chemistry Laboratory, Department of Chemistry, University of Oxford, OX1 3QZ Oxford, UK\\[0.25em]
$^{5}$Department of Physics, Chemistry and Biology (IFM), Link\"{o}ping University, SE-581 83, Link\"{o}ping, Sweden
}\\[0.75em]

{\footnotesize Correspondence: xialiang@unist.ac.kr; md942@cam.ac.uk}
\end{center}

\begin{abstract}
Hybrid lead halide perovskites exhibit a delicate interplay between average crystallographic symmetry, local structural disorder and A-site orientational dynamics, giving rise to unusual vibrational and electronic behaviour. Here, we combine large-scale molecular dynamics with a density-functional-theory-accurate machine learning force field to resolve the structural dynamics of perovskites across mesoscopic length scales. In formamidinium lead iodide (\ce{FAPbI3}), we identify a high-temperature $\alpha$ phase with dynamic local order and correlated tilt nanodomains, an ordered $\gamma$ phase with long-range $a^{+}a^{+}a^{+}$ tilt coherence, and, below $\sim$100~K, a history-dependent $\gamma'$ state consisting of locally $\gamma$-like nanoscale regions separated by sharp twin-like boundaries. This low-temperature disordered state is not a distinct bulk polymorph, but a kinetically arrested metastable twin-domain network selected by the interplay between shallow tilt energetics and slowing FA reorientation. This picture provides a consistent explanation for the low-temperature diffuse scattering features observed experimentally, and accounts for the broadened low-energy vibrational response found in the simulations. Furthermore, this unique structural landscape imprints a spatially varying electronic disorder that directly impacts macroscopic optoelectronic properties, evidenced by an anomalous increase in the Urbach energy at low temperatures. Our results reconcile the debated low-temperature behaviour of \ce{FAPbI3} in terms of competition between ordered and arrested structural states, and show more broadly that in hybrid perovskites the organic cation can actively select the macroscopic structural and electronic response through its reorientation kinetics, placing thermal history on equal footing with composition as a determinant of structural and optoelectronic properties.
\end{abstract}

\section{Introduction}
Lead halide perovskites combine soft structure, strong anharmonicity and exceptional optoelectronic performance, making local structural disorder central to both their fundamental physics and functional behaviour~\cite{fapi_exp_science_yang,perov_tilting_stable_science}. In these materials, octahedral tilting, molecular motion and dynamic symmetry breaking are strongly coupled to carrier transport, band-edge broadening and phase stability, so understanding how local order forms, evolves and freezes is essential for connecting structure to function~\cite{disorder_shortening_egger_2019,liquid_glass_polaron_zhu_2017}. Recent work has further shown that this disorder is often highly structured rather than random, with dynamic nanodomains providing a concrete real-space manifestation of local order in hybrid perovskites and revealing distinct local textures for different A-site cations~\cite{yaffe2017local,MA_local_order_2023_toney,rethink_Asite_science_2022}.

At the same time, the low-temperature structure of black-phase \ce{FAPbI3} remains unresolved. Despite its near-ideal band gap and technological relevance, its structural behaviour is unusually delicate: the black perovskite phase is metastable, and its phase sequence is sensitive to thermal history and processing~\cite{rotational_dynamics_fapi_jpcl_2023,entropy_fapi_sciadv_2016,cubic_fapi_stability_cordero}. Prior studies have proposed multiple low-temperature structural descriptions, including tetragonal and orthorhombic assignments~\cite{rotational_dynamics_fapi_jpcl_2023,mixed_lead_halide_perovskite_review_Mantas}, while recent machine-learned-potential simulations have identified $a^{-}b^{-}b^{-}$ as the thermodynamic ground state but find that cooling instead leads to a metastable $a^{-}a^{-}c^{+}$ phase~\cite{lowT_fapi_dutta_jacs_2025}. No consensus has therefore emerged on whether the low-temperature black phase is best described as a unique ordered polymorph or as a history-dependent state shaped by local disorder~\cite{entropy_fapi_sciadv_2016,rotational_dynamics_fapi_jpcl_2023}.

This question is particularly acute in hybrid perovskites because the final structural state need not be determined by local octahedral tilting energetics alone. In \ce{FAPbI3}, the inorganic tilt network is strongly coupled to the orientational degrees of freedom of the FA cation, so the locally preferred tilt motif does not necessarily determine the long-range ordered state that is ultimately reached on cooling. These coupled effects must therefore be investigated explicitly. Small-cell calculations and average structural probes may therefore favour apparently well-defined ordered phases while missing metastable mesoscale textures that emerge only at larger length scales and under specific thermal histories~\cite{phonon_exciton_perov_natphys_2024,dynamic_domian_milos,lowT_fapi_dutta_jacs_2025}.

Here, we combine large-scale molecular dynamics with X-ray diffuse scattering and vibrational analysis to resolve the low-temperature structural behaviour of black-phase \ce{FAPbI3} across mesoscopic length scales. We show that \ce{FAPbI3} passes through a high-temperature $\alpha$ phase with dynamic local order, an ordered $\gamma$ phase with long-range $a^{+}a^{+}a^{+}$ tilt coherence, and, below $\sim$100~K, a lower-temperature $\gamma'$ regime that remains locally $\gamma$-like but develops additional mesoscale structural heterogeneity. This picture is consistent with the measured diffuse scattering and low-energy vibrational signatures, and further reveals a spatially heterogeneous electronic landscape associated with the arrested low-temperature state.

\section{Large-cell dynamics reveal a heterogeneous \\low-temperature state in \texorpdfstring{\ce{FAPbI3}}{FAPbI3}}

Using large-scale MD simulations analysed with \textsc{PDynA}~\cite{pdyna2023liang}, we resolve three distinct regimes in the structural dynamics of \ce{FAPbI3}. In the high-temperature $\alpha$ phase ($a^{0}a^{0}a^{0}$; Pm$\bar{3}$m), the inorganic framework exhibits dynamic local order, with correlated tilt nanodomains forming within an average cubic structure. On cooling into the $\gamma$ phase ($a^{+}a^{+}a^{+}$), which forms at $\sim$200~K, long-range in-phase tilt coherence develops along all three crystallographic directions, consistent with the in-phase cubic Im$\bar{3}$ framework, which we also experimentally verified with single crystal diffraction (Supplementary Fig.~\ref{si:sg_aaa}). Below approximately 100~K, the equilibrium low-temperature state no longer corresponds to stronger global order. Instead, the system adopts a lower-temperature $\gamma'$ regime in which the local structure remains predominantly $\gamma$-like but develops additional structural heterogeneity. 

\begin{figure}[t]
    \centering
    \includegraphics[width=0.45\textwidth]{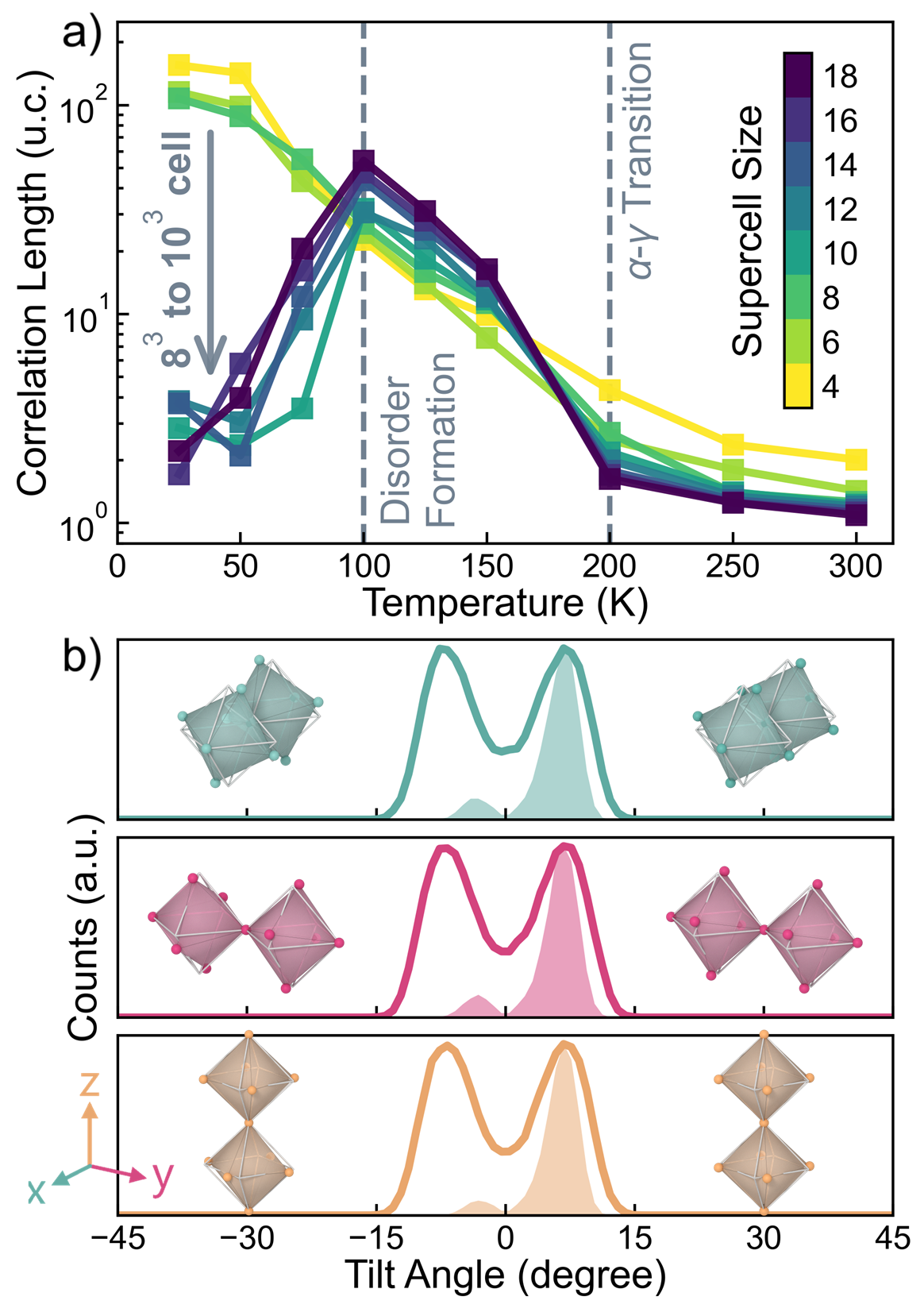}
    \caption{Large-cell simulations reveal a heterogeneous low-temperature state in \ce{FAPbI3}. (a) Temperature dependence of the maximum octahedral-tilt correlation length, showing a pronounced maximum near 100~K and strong supercell-size dependence in the lower-temperature $\gamma'$ regime. (b) Direction-resolved distributions of apparent tilt angles and nearest-neighbour tilting correlations. Solid lines show tilt angle distributions; shaded areas indicate the first-neighbour correlation product, where positive and negative values correspond to in-phase and out-of-phase correlations, respectively. }
    \label{res:fig1}
\end{figure}

This low-temperature crossover is evident in the temperature dependence of the tilt correlation length and the underlying local tilt populations. The tilting correlation length (Fig.~\ref{res:fig1}a) increases on cooling through the $\alpha$-to-$\gamma$ transition, reaches a maximum near 100~K, and then decreases again at lower temperatures. The local tilt distributions (Fig.~\ref{res:fig1}b) show that the low-temperature state remains dominated by the in-phase motif characteristic of the $\gamma$ phase, demonstrating that most octahedra still belong to locally $\gamma$-like environments. At the same time, a smaller but finite population of out-of-phase-correlated neighbours emerges, indicating that the low-temperature state is not uniformly ordered but contains competing local correlations within an otherwise predominantly $a^{+}a^{+}a^{+}$ framework.

The inferred correlation lengths correspond to characteristic domain sizes of roughly 2--20 unit cells ($\sim$1.2--12~nm), indicating a mesoscale domain structure rather than a uniform bulk-ordered crystal. A pronounced finite-size effect further supports this interpretation: the $\gamma'$ state emerges only when the simulation cell exceeds approximately $10\times10\times10$ pseudocubic units, whereas smaller cells remain trapped in a fully correlated $\gamma$-like configuration. These results therefore support a heterogeneous low-temperature state rather than a unique ordered low-temperature phase~\cite{static_dynamic_fapi_egger_2023}, a conclusion further supported below by its reciprocal space signatures.

\section{Shallow tilt energetics and FA dynamics frustrate long-range ordering}

To rationalise the origin of the low-temperature heterogeneous state, we evaluated the tilt-dependent potential energy surface predicted by the MLFF in a $2\times2\times2$ supercell with randomised molecular orientations, such that the resulting landscape probes the local energetic response to octahedral tilting without imposing long-range molecular order (noting that these are local potential energies rather than finite-temperature free energies; see Supplementary Section~\ref{sec:pes}). Here we restrict the discussion to the subspace spanned by octahedral tilt amplitudes and idealised tilt patterns~\cite{tilt_pes_johan}. 

\begin{figure}[t]
    \centering
    \includegraphics[width=0.50\textwidth]{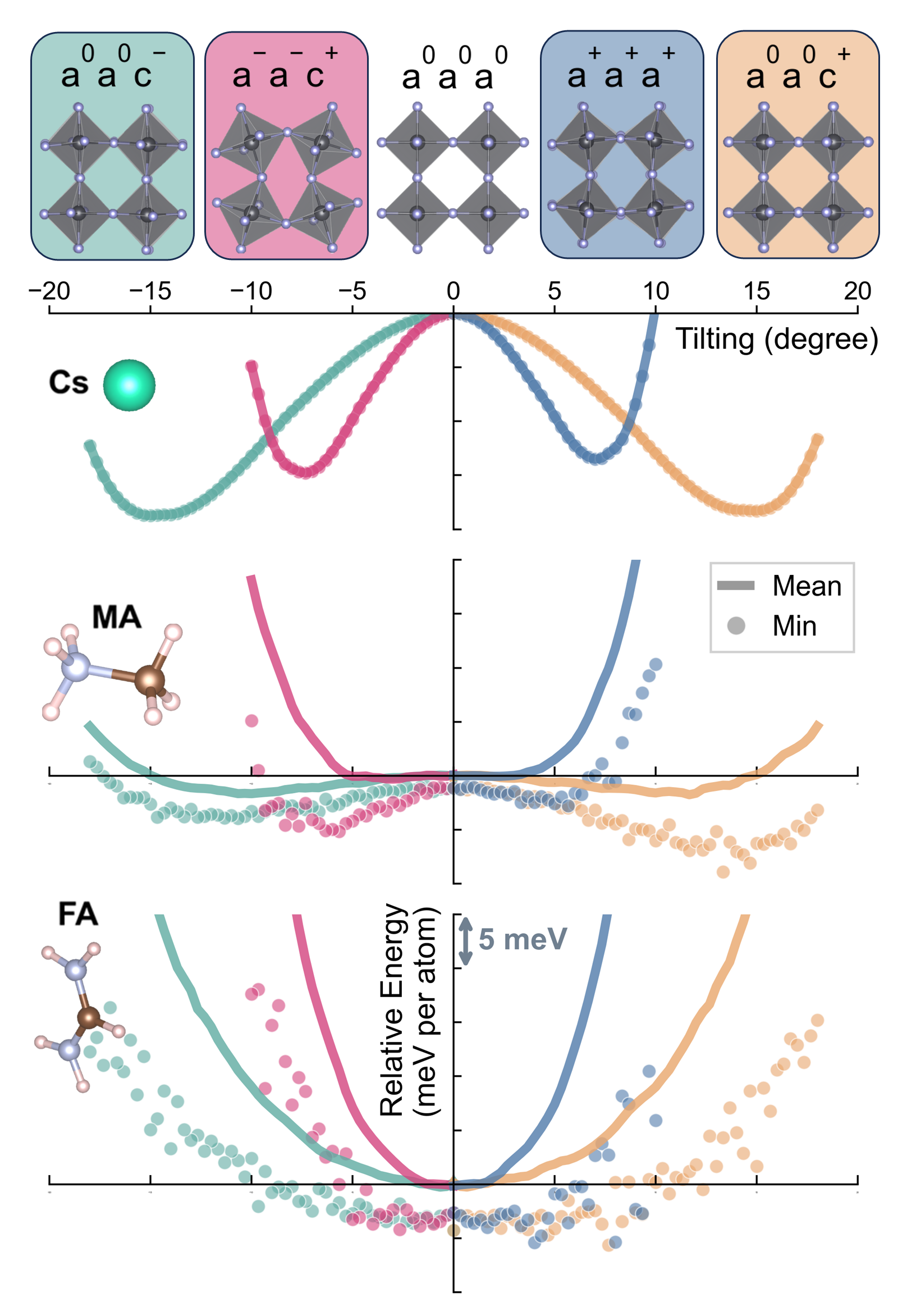}
    \caption{Tilt-dependent potential energy surfaces of iodide perovskites. Relative energies of \ce{CsPbI3}, \ce{MAPbI3}, and \ce{FAPbI3} are shown as a function of tilt angle for the idealised modes $a^{0}a^{0}c^{-}$, $a^{-}a^{-}c^{+}$, $a^{+}a^{+}a^{+}$, and $a^{0}a^{0}c^{+}$. The zero-tilt limit corresponds to the cubic $a^{0}a^{0}a^{0}$ structure. Solid lines show the mean over 50 independent runs, and points indicate the minimum-energy structures. The molecular perovskites display a shallow, frustrated tilt landscape in which the number of active tilt axes matters more than the in-phase versus out-of-phase registry alone.}
    \label{res:pes}
\end{figure}

The potential energy surface is shallow and only weakly separates symmetry-distinct tilt modes. Instead, it is organised mainly by the number of active tilt axes: $a^{0}a^{0}c^{-}$ and $a^{0}a^{0}c^{+}$ remain close in energy, as do $a^{+}a^{+}a^{+}$ and $a^{-}a^{-}c^{+}$. This shows that the local energetic cost is governed primarily by the magnitude and dimensionality of octahedral rotation, while the in-phase versus out-of-phase registry is only weakly differentiated locally and must therefore be selected cooperatively through longer-range distortions and coupling to the A-site degrees of freedom. A comparison across \ce{CsPbI3}, \ce{MAPbI3}, and \ce{FAPbI3} further highlights the role of the A-site species, as shown in Fig.~\ref{res:pes}. In \ce{CsPbI3}, the landscape is comparatively steep and well defined, closer to a conventional lattice dynamical picture. By contrast, the molecular perovskites exhibit a flatter and more frustrated manifold, indicating that the tilt energetics are strongly reshaped by coupling to molecular displacement and rotation.

This weak local energetic discrimination points to a kinetic origin of the $\gamma'$ state. In \ce{FAPbI3}, the characteristic FA reorientation time is approximately an order of magnitude longer than the corresponding octahedral-tilt decorrelation time across the temperature range considered here (Supplementary Figs.~\ref{si:structure}c and \ref{si:tilt_time}), so the inorganic framework can reorganise locally more rapidly than the molecular subsystem can reorient cooperatively. On cooling, this mismatch becomes increasingly consequential: the ordered $\gamma$ phase remains lower in energy than $\gamma'$ (Supplementary Fig.~\ref{si:epot}), indicating that the low-temperature heterogeneous state arises from kinetic arrest of the ordering process rather than from a lower static energy~\cite{entropy_fapi_sciadv_2016}.

\begin{figure}[htbp]
    \centering
    \includegraphics[width=0.9\textwidth]{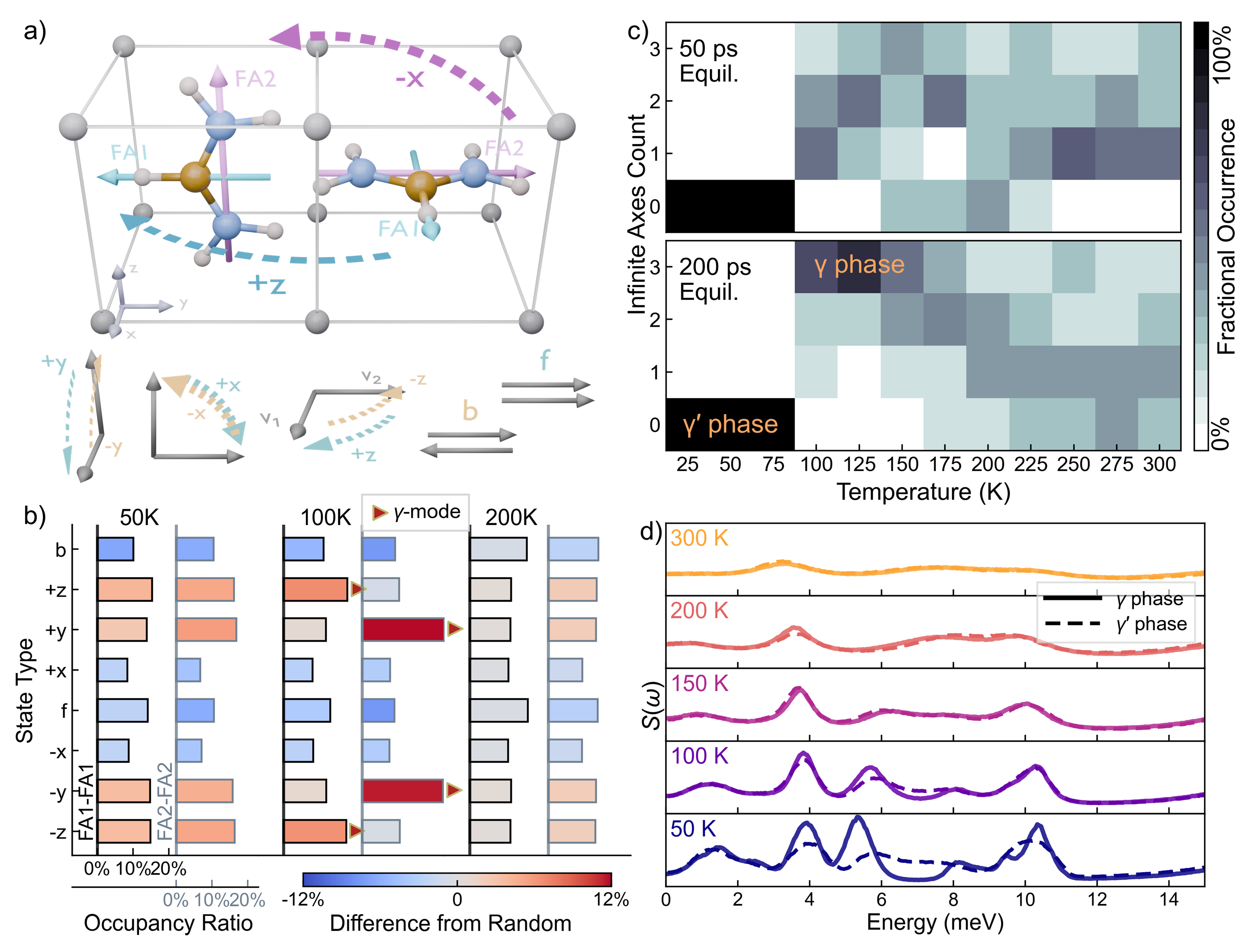}
    \caption{FA orientational correlations select the low-temperature structural outcome in \ce{FAPbI3}. (a) Definition of the FA1 and FA2 molecular vectors and the discrete pair-correlation states used to classify nearest-neighbour molecular configurations. (b) Relative populations of FA1--FA1 and FA2--FA2 correlation states at selected temperatures along the $x$ direction; colours indicate deviation from the random benchmark. (c) Fractional occurrence of low-temperature structural outcomes under controlled equilibration protocols, classified by the number of crystallographic directions with effectively infinite tilt correlation length. (d) Neutron $q$-integrated vibrational spectra $S(\omega)$ for structures initialised from ordered $\gamma$ and disordered $\gamma'$ states. The ordered $\gamma$ phase exhibits sharper low-energy features, whereas the $\gamma'$ spectra are broader and more reminiscent of the higher-temperature response, consistent with weaker long-range molecular order in the twin-domain state.}
    \label{res:mo}
\end{figure}

We therefore classify nearest-neighbour FA pairs by their relative molecular vector orientations (see Fig.~\ref{res:mo}a). This scheme is justified by the fact that FA orientations relax towards a $\langle100\rangle$ manifold, as demonstrated in Supplementary Fig.~\ref{si:molecule} and discussed in the Methods. In the $\alpha$ phase, the orientational correlation landscape remains close to a random benchmark, indicating that molecular orientations are only weakly correlated despite the presence of dynamic local tilt correlations. On entering the $\gamma$ phase, however, a symmetry-selected anisotropic molecular ordering pattern emerges, consistent with long-range $a^{+}a^{+}a^{+}$ tilt coherence. In the $\gamma'$ state, by contrast, this anisotropic molecular order no longer develops globally, and the structure instead freezes into disconnected, locally $\gamma$-like orientational variants.

To test whether the final low-temperature structure is controlled by the prepared molecular state, we performed controlled MD protocols in which the FA orientations were equilibrated for different hold times before applying a temperature ramp to 50~K, where the structural dynamics are effectively frozen. The resulting phase map (Fig.~\ref{res:mo}c) shows that the low-temperature outcome is strongly history dependent. For equilibration temperatures below approximately 100~K, no substantial molecular reorganisation occurs and the system consistently relaxes into the heterogeneous $\gamma'$ state, regardless of whether the hold time is 50 or 200~ps. In the approximate range 100--175~K, however, the outcome becomes sensitive to the hold time: equilibration for 50~ps already permits partial reorganisation towards the ordered $\gamma$ topology, while 200~ps much more effectively promotes a fully coherent long-range-ordered structure. At still higher temperatures, orientational correlations weaken and multiple low-temperature topologies become accessible. These orientation-control simulations therefore identify arrested FA reorientation as the kinetic bottleneck that selects the low-temperature heterogeneous state.

This crossover is captured in our simulations by the sharp rise in molecular reorientation times and by the broader calculated neutron spectra of the $\gamma'$ state relative to the ordered $\gamma$ phase (Fig.~\ref{res:mo}d). The corresponding Arrhenius analysis indicates an effective reorientational barrier for FA motion (Fig.~\ref{si:barrier}), such that at lower temperature the FA subsystem can no longer reorganise efficiently and the structural outcome becomes effectively fixed~\cite{rotational_dynamics_fapi_jpcl_2023}, whereas at higher temperature barrier crossing becomes possible and a wider range of low-temperature topologies can be accessed. The distinct outcomes obtained after 50 and 200~ps equilibration further show that, although both hold times are short on laboratory timescales, they are sufficient on the MD timescale to resolve these differences in kinetic accessibility. The low-temperature structural outcome is therefore selected by the interplay between a shallow local tilt manifold and the slowing FA orientational dynamics required for long-range $\gamma$-phase coherence.

\section{Reciprocal-space and electronic fingerprints of the arrested twin-domain state}

\begin{figure}[htbp]
    \centering
    \includegraphics[width=0.96\textwidth]{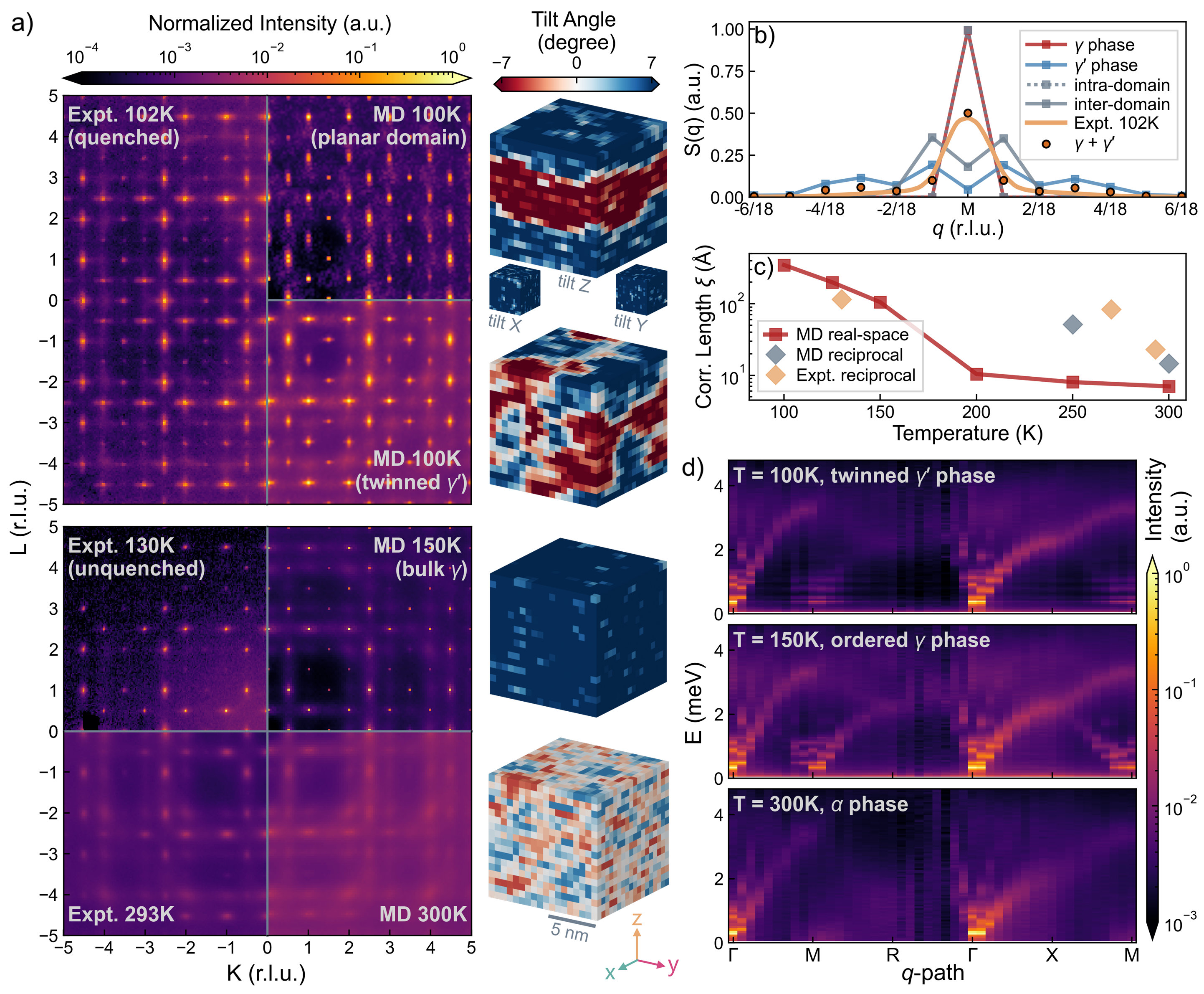}
    \caption{Reciprocal-space signatures of the $\gamma'$ twin-domain state in \ce{FAPbI3}. (a) X-ray diffuse scattering on $L=1.5$~r.l.u. half-integer reciprocal-space planes for experiment and simulation, together with one-to-one corresponding MD real-space maps of the local octahedral tilts. For the planar-domain structure, additional maps of the $x$- and $y$-tilt components are shown to highlight its anisotropic domain topology. The top panels compare the quenched low-temperature experimental pattern with two simulated low-temperature limits, namely a planar twin-domain structure and the twinned $\gamma'$ state. The bottom panels compare the experiment with the bulk ordered $\gamma$ phase at intermediate temperature and the dynamically disordered $\alpha$ phase at high temperature (intensity rescaled for display purposes). The simulated $\gamma'$ diffuse map is averaged over four independent trajectories and three symmetry-equivalent planes (12 patterns total; Supplementary Fig.~\ref{si:diffuse}). (b) Line profiles around symmetry-equivalent $M$ points for the ordered $\gamma$ phase, the twinned $\gamma'$ state, the planar-domain structure, experiment, and a 50:50 $\gamma+\gamma'$ combination. (c) Correlation lengths extracted from real-space tilting correlations and reciprocal-space peak widths for simulation and experiment. (d) Energy-resolved dynamical structure factor $S(\mathbf{q},\omega)$ along a high-symmetry path for the twinned $\gamma'$, ordered $\gamma$, and $\alpha$ phases.}
    \label{res:dsf}
\end{figure}

To connect the real-space analysis with experimentally accessible observables, we computed the dynamical structure factor $S(\mathbf{q},\omega)$ following our previous framework~\cite{dynamic_domian_milos}. The calculated diffuse scattering cleanly distinguishes the three structural regimes (Fig.~\ref{res:dsf}a): dynamic $M$-point intensity in the high-temperature $\alpha$ phase, strongly suppressed diffuse scattering in the ordered $\gamma$ phase, and broad, configuration-dependent features near the Bragg peaks and $M$ points in the low-temperature $\gamma'$ state. These low-temperature signatures are incompatible with both the dynamic nanodomain disorder of the $\alpha$ phase and the long-range coherence of the ordered $\gamma$ phase, and instead point to frozen spatial heterogeneity within an otherwise locally $\gamma$-like framework. 

This assignment is supported directly by the real-space tilt maps and their experimental counterparts. The simulated $\gamma'$ structures break up into locally $\gamma$-like regions separated by sharp orientational discontinuities, and the corresponding diffuse scattering pattern reproduces the low-temperature features observed in the quenched sample, whereas the ordered $\gamma$ and dynamically disordered $\alpha$ structures each match their respective experimental counterparts at intermediate and high temperature. This interpretation is also consistent with the thermal protocols: rapid quenching bypasses sustained equilibration in the intermediate $\gamma$ regime, leaving insufficient time for the FA subsystem to develop the anisotropic alignment required for long-range $\gamma$-phase coherence.

To isolate the role of the interfaces themselves, we constructed an artificial planar-domain structure containing a single twin boundary. In this case, the reciprocal-space anisotropy becomes explicit: the $HK$ half-integer plane remains essentially identical to the ordered $\gamma$ phase, whereas the $HL$ and $KL$ planes develop pronounced satellite intensity around the $M$ points, and the near-$M$ profile acquires a characteristic double-peak line shape (Fig.~\ref{res:dsf}b). These results show that the additional diffuse weight in the $\gamma'$ state originates from twin boundaries rather than from a different local tilt motif. The low-temperature $\gamma'$ regime is therefore best understood as a kinetically arrested twin-domain state, consisting of locally $\gamma$-like regions separated by sharp interfaces. Although the experimental line profile is not reproduced exactly by either the ideal ordered $\gamma$ or the fully twinned $\gamma'$ limit alone, a simple 50:50 combination of the two simulated signals closely tracks the measured intensity, consistent with the experimental low-temperature sample containing both nearly ordered $\gamma$-like regions and twinned metastable domains. This comparison should be viewed as a simplified physical illustration rather than as a unique decomposition or exact phase fraction of the experimental signal.

Real-space and reciprocal-space correlation lengths follow the same overall trend and lie on the same order of magnitude (Fig.~\ref{res:dsf}c), supporting a consistent picture of the structural crossover across both representations. The comparison should not be interpreted point-by-point, however, because the reciprocal-space widths become resolution-limited for the sharpest low-temperature peaks, while the real-space correlation length can approach the finite simulation-cell size in strongly ordered configurations. In addition, the simulated transition temperature is depressed relative to experiment, a known limitation of current MLFF-based phase-transition modelling~\cite{universal_softening_deng,exact_mlff_stefan_natcomm}. Even with these caveats, both analyses indicate that the low-temperature quenched state retains shorter-range coherence than the fully ordered $\gamma$ phase.

The energy-resolved dynamical structure factor highlights the same distinction (Fig.~\ref{res:dsf}d). The clearest contrast appears at the zone-boundary $M$ point, which is dominated by octahedral-tilt fluctuations. In the ordered $\gamma$ phase, the low-temperature response remains comparatively sharp, consistent with coherent long-range $a^{+}a^{+}a^{+}$ order. By contrast, the twinned $\gamma'$ state exhibits a substantially broader low-energy response at the same wave vector. Because this broadening persists even at low temperature, it cannot be attributed to thermal disorder alone, but instead reflects the static structural heterogeneity introduced by the arrested twin-domain topology~\cite{lanigan2021two,overdamped_phonon_Eg_prl_2025}. This dynamical contrast is also consistent with the molecular behaviour: the ordered $\gamma$ phase, with its well-developed anisotropic FA correlation pattern, yields sharper low-energy spectral features, whereas the partially ordered $\gamma'$ state produces broader spectra (Fig.~\ref{res:mo}d) that more closely resemble the anomalous low-temperature neutron response reported experimentally~\cite{lowT_fapi_dutta_jacs_2025}. In this sense, both the calculated dynamical structure factor and the neutron data support the same picture of the arrested low-temperature state with incomplete long-range molecular and tilt coherence.

The same arrested twin-domain texture is also electronically consequential. Structural heterogeneity in hybrid halide perovskites is known to modulate the electronic structure through coupling between lattice distortions, molecular motion and band edge states~\cite{dewolf_absorption_jpcl_2014,anhar_elecphocoup_npjcm_2023,lat_dyn_eletronics_jpcl_herz}. To probe this, we partition each large MD supercell into many $2\times2\times2$ subcells and perform DFT calculations for each local environment, thereby constructing a spatially resolved electronic landscape tied directly to the mesoscale structural texture. Across these subcells, the local band gap varies substantially and is found to correlate most strongly with the octahedron-averaged tilt norm (Supplementary Fig.~\ref{si:feature}), showing that local octahedral tilting provides the dominant structural handle on the band-edge variation.

\begin{figure}[htbp]
    \centering
    \includegraphics[width=0.9\textwidth]{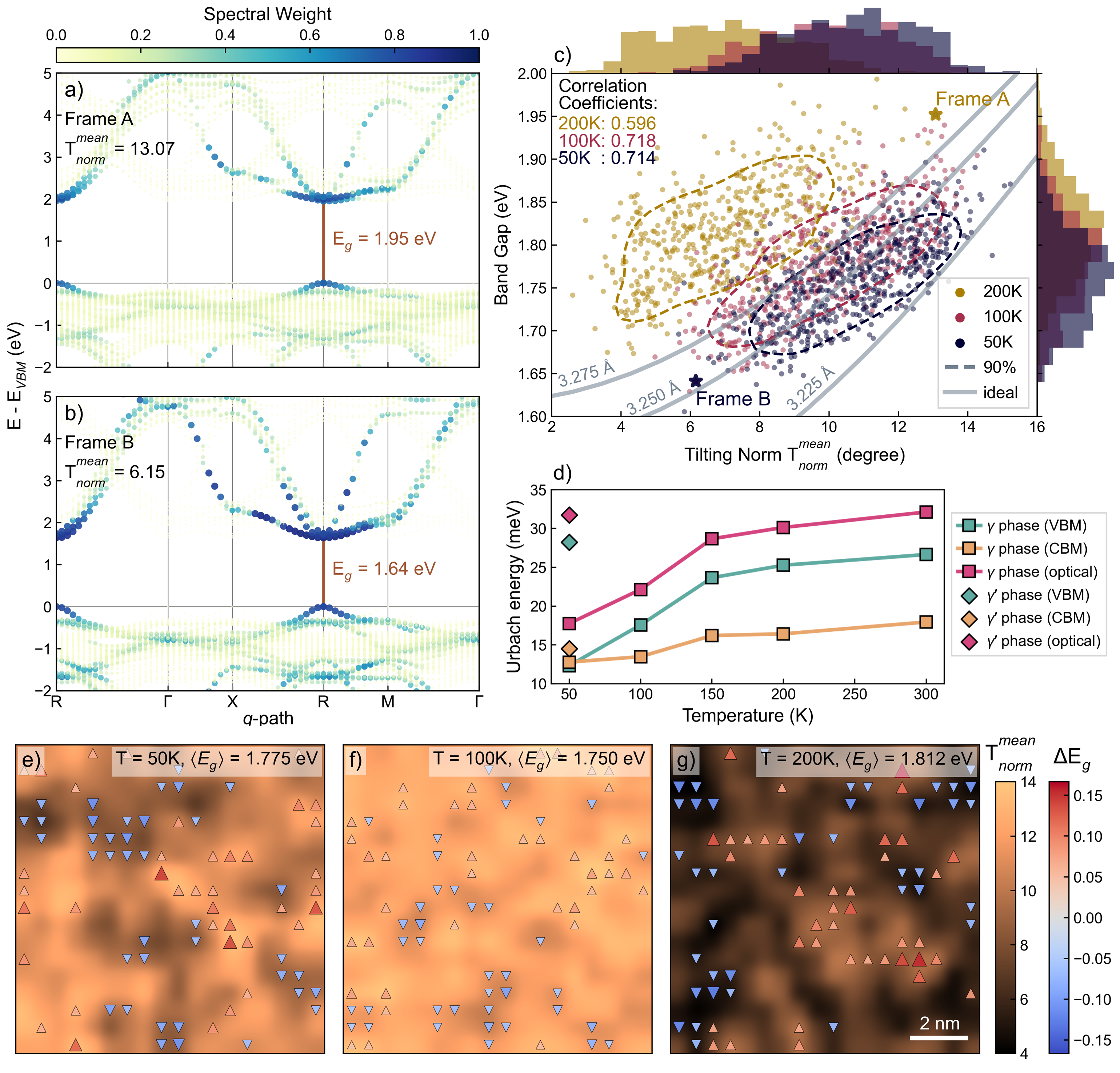}
    \caption{Electronic fingerprints of mesoscale tilt heterogeneity in \ce{FAPbI3}. (a,b) Unfolded band structures for representative subcells with large and small local tilt magnitudes. (c) Local band gap as a function of the subcell tilt metric for snapshots sampled at 50, 100, and 200~K; contours enclose 90\% of the data and stars denote the two representative subcells in (a,b). The distributions of local band gaps and tilt metrics are shown at the right and top edges of the map, respectively. Grey lines indicate idealised reference relations with varying Pb--I bond lengths. (d) Urbach energies extracted from the distributions of local valence and conduction band edge energies, together with the resulting effective optical Urbach energy. While the overall increase with temperature reflects growing thermal disorder, the low-temperature $\gamma'$ state retains substantial band-tail broadening due to static twin-domain heterogeneity. (e--g) Real-space maps of local tilt magnitude and local band gap deviation, showing that electronic extrema track the morphology of the tilt-domain network.}
    \label{res:electro}
\end{figure}

Representative unfolded band structures for weakly and strongly tilted subcells are shown in Fig.~\ref{res:electro}a,b. Although the absolute gap values should not be interpreted quantitatively because of the approximate electronic-structure treatment (see Methods and Supplementary Section~\ref{sec:elec}), the relative variation is clear: strongly tilted local environments generally exhibit smaller band gaps, whereas more weakly tilted environments retain larger gaps. Across the full dataset, this relation becomes especially clear when the local band gap is plotted against the octahedron-averaged tilt norm (Fig.~\ref{res:electro}c), establishing a direct connection between the local structural motif and the local electronic response. 

This relation is also visible directly in real space (Fig.~\ref{res:electro}e--g). Planar slices through the supercell show that local band gap extrema track the morphology of the tilt-domain network: regions with different local electronic gaps are not randomly distributed, but are spatially organised by the twin-domain texture. In practice, the electronic landscape is slaved to the mesoscale structural landscape, both in the dynamically fluctuating high-temperature regime and, more strikingly, in the low-temperature $\gamma'$ state where the domain pattern becomes quasi-static on the simulation timescale.

The temperature dependence of the extracted Urbach energies is shown in Fig.~\ref{res:electro}d. As expected, the overall band-tail broadening increases with temperature as thermal disorder grows. The extracted Urbach energies are somewhat larger than the room-temperature values typically reported for high-quality \ce{FAPbI3} films and devices, which are in the range of $\sim$14--22~meV depending on sample quality, processing route, and measurement protocol~\cite{urbach_fapi_nsr_2022,fapi_cvd_science_2020}. This offset is nevertheless consistent with the fact that the present values are derived from a structurally heterogeneous arrested state rather than from an equilibrium high-quality thin-film absorption edge. The low-temperature $\gamma'$ state is nevertheless anomalous: its Urbach broadening remains comparable to that of the 300~K structure despite the large reduction in thermal motion. This shows that if cooling drives the system into the arrested $\gamma'$ topology rather than the ordered $\gamma$ phase, static twin-domain heterogeneity can replace thermal disorder as the dominant source of low-temperature band-tail broadening. The arrested twin-domain texture therefore explains not only the anomalous low-temperature reciprocal-space signatures of black-phase \ce{FAPbI3}, but also the persistence of substantial electronic disorder deep into the low-temperature regime.

\section{Discussion}

We have shown that the low-temperature structure of black-phase \ce{FAPbI3} is best understood not as a unique long-range-ordered polymorph, but as a kinetically arrested twin-domain state composed of locally ordered $a^{+}a^{+}a^{+}$ regions separated by sharp twin-like interfaces. Large-cell molecular dynamics reveal that this $\gamma'$ state emerges only when the system is allowed to develop mesoscale structural heterogeneity, establishing twinning as an intrinsic large-length-scale feature rather than a small-cell artefact.

The origin of this behaviour lies in the coupled dynamics of octahedral tilting and FA reorientation. In \ce{FAPbI3}, the local tilt-energy landscape is shallow and only weakly constrains the final tilt registry, so the long-range structural outcome must be selected cooperatively through the coupled evolution of the inorganic tilt network and the molecular orientational degrees of freedom. As the FA subsystem slows on cooling, this ordering process can become arrested, trapping the crystal in a twin-domain topology rather than allowing it to reach a fully coherent ordered state.

This arrested structural state leaves clear reciprocal-space, vibrational, and electronic fingerprints. It reproduces the diffuse-scattering features of quenched low-temperature samples, and the artificial planar-domain control shows that the characteristic additional diffuse weight arises from the twin boundaries themselves rather than from a different local tilt motif. The same arrested topology also broadens the low-energy dynamical response and generates a spatially heterogeneous electronic landscape, with local band gaps that track the mesoscale tilt texture and with substantial low-temperature Urbach broadening that persists despite the reduction in thermal motion. In this sense, if cooling drives the system into the arrested $\gamma'$ topology rather than the ordered $\gamma$ phase, static structural heterogeneity can replace thermal disorder as a dominant source of low-temperature electronic broadening. This also has direct implications for the interpretation and reproducibility of measurements in \ce{FAPbI3}: differences in reported Urbach energies, photoluminescence linewidths, carrier lifetimes, and scattering signatures across nominally similar samples may in part reflect distinct arrested structural textures rather than only differences in material quality or defect density. Thermal history should therefore be considered alongside composition and defect chemistry as a determinant of structure and optoelectronic behaviour in hybrid halide perovskites.

More broadly, our results place static twinning alongside dynamic nanodomain formation as a fundamental mode of local order in lead halide perovskites. They show that, in hybrid organic--inorganic compounds, molecular A-site dynamics do not merely decorate an inorganic framework, but can actively determine how disorder is selected, correlated, and frozen into the structure. In this respect, black-phase \ce{FAPbI3} joins a broader class of kinetically governed functional materials, including structural glasses and relaxor ferroelectrics, in which the observed state is determined not only by thermodynamic stability but also by the pathway through which the material is cooled. This perspective may prove useful for understanding the pronounced processing sensitivity and sample-to-sample variability widely observed in hybrid perovskites.

\section*{Methods}

\subsection{Force Field Training}

Although the present study focuses primarily on \ce{FAPbI3}, with smaller comparative analyses of \ce{MAPbI3} and \ce{CsPbI3}, the MLFF was trained over the broader chemical space $(\mathrm{Cs}_{1-x-y}\mathrm{MA}_x\mathrm{FA}_y)\allowbreak \mathrm{Pb}\allowbreak (\mathrm{Br}_m\mathrm{I}_{1-m})_3$. The aim is to construct a single unified MLFF that enables direct comparison across chemically related compounds within one internally consistent potential energy framework, while also providing a reusable model for future studies of mixed-A-site and mixed-halide lead perovskites. For this seven-element chemical space, however, our previous workflow based primarily on direct on-the-fly sampling becomes prohibitively expensive~\cite{picosecond_lifetime_MA_Eduardo,pdyna2023liang,azr_perovskite_mantas}. We therefore adopt a two-stage strategy in which a preliminary surrogate MLFF is first constructed from a limited but chemically representative DFT dataset, and is then used to generate a much larger set of candidate structures for final data selection and retraining.

As the starting point for this workflow, an on-the-fly DFT training is performed for a single mixed composition, \ce{Cs2MA3FA3Pb8Br12I12}, in a $2\times2\times2$ supercell. This composition lies close to the centre of the doubly mixed composition space and therefore contains representative local A-site and halide interactions. The A-sites are randomly arranged, and the halides are distributed using the ICET package~\cite{icet_original}. The initial inorganic framework is assigned the $a^{0}a^{0}a^{0}$ tilt pattern, since the equilibrium tilt state of this mixed composition is not known a priori. Four on-the-fly MD simulations are then carried out at 100, 300, 500, and 700~K to sample a broad thermal range with a tractable number of DFT calculations.

All DFT calculations were performed in VASP using the projector-augmented-wave method~\cite{pseudopot_paw_kresse} and the r$^2$SCAN meta-GGA functional~\cite{r2scan_original1}. For the on-the-fly MD stage, we used a plane-wave cutoff of 500~eV, a $\Gamma$-centred $2\times2\times2$ $k$-point mesh, Gaussian electronic smearing with a width of 0.05~eV, and an electronic self-consistency threshold of $10^{-5}$~eV. The on-the-fly trajectories were propagated in the $NpT$ ensemble with a 0.5~fs time step and Langevin dynamics.

A total of 2{,}255 snapshot frames from these four on-the-fly trajectories are combined to train a preliminary MACE model~\cite{mace_original}. The data are split into training and validation sets using a 90:10 ratio, and the model is trained for 2{,}000 epochs. Because this model is used only as a surrogate sampler for a broad and sparsely covered configuration space, we employ a relatively expressive architecture despite its higher evaluation cost. Specifically, the preliminary model uses a 7~\AA\ radial cutoff, 128 feature channels, and message equivariance up to $L=1$. The resulting validation accuracy is sufficient for structure generation and diversity sampling, but this preliminary model is not used for the final production simulations.

The preliminary model is then used to sample the doubly mixed compositional space using a structured surrogate-MD workflow over representative mixed compositions and initial tilt states. The detailed compositional sampling scheme is summarised in Supplementary Fig.~\ref{si:doubly_config} and Supplementary Table~\ref{si:profiles}. This produces a total of 175 surrogate MD simulations in the $NpT$ ensemble, each 300~ps in duration. We extract 200 frames from each simulation by saving one structure per picosecond over the final 200~ps, resulting in 35{,}000 candidate structures.

To reduce the cost of DFT labelling, the candidate pool was down-selected using the DIRECT method~\cite{direct_original}. From the 35{,}000 surrogate-sampled frames, 3{,}000 structures were selected as the main mixed-composition DFT-labelled dataset, and 1{,}000 additional endpoint-family structures were included for each of \ce{CsPbX3}, \ce{MAPbX3}, and \ce{FAPbX3}~\cite{mixed_halide_liang_2025_chem-mater}. These endpoint families span the corresponding halide subspaces, \emph{i.e.} \ce{Pb(Br_mI_{1-m})3} with fixed A-site identity. The final training set therefore comprised 6{,}000 mixed- and endpoint-family structures. This dataset construction process is further evaluated in the Supplementary Section~\ref{sec:config}. For the final single-point labelling calculations, the same general DFT setup was retained, but with a higher plane-wave cutoff of 600~eV and a tighter electronic self-consistency threshold of $10^{-7}$~eV. The $k$-point sampling was chosen according to the supercell size, using a $\Gamma$-centred $2\times2\times2$ mesh for the $2\times2\times2$ cells and a $\Gamma$-centred $1\times1\times2$ mesh for the $4\times4\times2$ cells. These calculations were carried out as static single-point evaluations to obtain energies, forces, and stresses for the final training set.

The final production MACE model is trained on this curated dataset using a 90:10 train--test split and a training length of 1{,}000 epochs. The model uses a 5~\AA\ cutoff, 64 feature channels, and message equivariance up to $L=0$, which substantially reduces the inference cost while retaining sufficient accuracy for large-scale MD.

\subsection{Molecular Dynamics}

Large-scale MD simulations are carried out in LAMMPS~\cite{lammps_original} through the MACE-MLIAP interface, combined with the cuEquivariance-enabled workflow for efficient GPU-accelerated production simulations. Unless otherwise noted, we use an $18\times18\times18$ pseudocubic supercell (69,984 atoms) for constant-temperature simulations and a $16\times16\times16$ pseudocubic supercell (49,152 atoms) for temperature-ramp simulations, where the smaller cell was used to increase throughput. The MD time step is 1~fs throughout. Production runs are performed on a single NVIDIA A100 GPU, yielding a typical performance of approximately 0.42~ns/day.

The initial structures of \ce{FAPbI3} are constructed with an ideal $a^{0}a^{0}a^{0}$ inorganic framework. Preliminary tests show that FA orientations relax towards a $\langle100\rangle$ manifold, and all production simulations reported in the main text therefore start from initial molecular configurations chosen within this orientational basin; the precise construction of these initial states is described in the Supplementary Information.

The equilibrium structural dynamics of \ce{FAPbI3} at each temperature are determined from independent constant-temperature MD simulations starting from these initial structures, rather than by continuing a single cooling trajectory from higher temperature. Each system is first equilibrated for 200~ps in the $NpT$ ensemble at the target temperature, followed by 50~ps in the $NVT$ ensemble. Production data are then collected from a subsequent 200~ps $NVE$ trajectory, with snapshots saved every 1~ps. Temperature and pressure are controlled using a Nosé--Hoover thermostat and barostat with damping constants of 100~fs and 200~fs, respectively.

Heating and cooling protocols are initialised from the final snapshot of an equilibrated trajectory and are propagated under a temperature ramp of 0.25~K/ps in the $NpT$ ensemble. After the temperature ramp, each system is further evolved using the same $NVT+NVE$ sequence described above. To ensure statistical robustness, independent trajectories with distinct initial conditions were used where required. 

\subsection{Electronic Structure and Disorder Analysis}

Electronic structure calculations were performed on $2\times2\times2$ subcells extracted directly from representative MD snapshots of \ce{FAPbI3}. The atomic positions and lattice parameters of these subcells were taken from the MD configurations without further structural relaxation, so that the calculated band structures reflect the instantaneous local environments generated by the finite-temperature trajectories. Band structures were computed using the projector-augmented-wave method as implemented in VASP, with the PBEsol exchange-correlation functional~\cite{PBEsol}. A plane-wave cutoff of 500\,eV, Gaussian smearing of 0.05\,eV, and a $\Gamma$-centred $2\times2\times2$ $k$-point mesh were used for all $2\times2\times2$ subcells. The electronic self-consistency threshold was set to $10^{-7}$\,eV. Wavefunctions were written and used for subsequent band unfolding onto the primitive Brillouin zone~\cite{band_unfolding_prb_2012}. PBEsol was chosen for its balanced structural accuracy and computational tractability across the large number of subcell evaluations required. The electronic analysis therefore focuses on relative variations of local band-edge energies across distinct structural environments rather than on quantitative prediction of the experimental bulk gap. 

Electronic analyses were performed on a single MD snapshot at each temperature, corresponding to $\sim$500 distinct $2\times2\times2$ subcells in total. For each $2\times2\times2$ subcell, the valence band maximum (VBM), conduction band minimum (CBM), and band gap were extracted from the DFT band structure. To relate local electronic variation to local structure, the octahedron-averaged tilt norm $T_{\mathrm{norm}}^{\mathrm{mean}}$ was used as the structural descriptor.

To quantify band-tail broadening, the sampled band-edge energies were converted into normalised histogram-based density-of-states representations. The exponential tail near the band edge was fitted using the Urbach form~\cite{urbach_original_1953,dow_urbach_1972},
\begin{equation}
D(E)\propto \exp\!\left(\frac{E-E_{\min}}{E_U}\right),
\end{equation}
where $D(E)$ is the density of states in the band-edge tail region, $E_{\min}$ is the minimum sampled band-edge energy, and $E_U$ is the Urbach energy. Taking the logarithm yields
\begin{equation}
\ln D(E)=\frac{E-E_{\min}}{E_U}+C,
\end{equation}
so that
\begin{equation}
E_U=\left(\frac{d\ln D}{dE}\right)^{-1}.
\end{equation}
Only the lowest-energy tail of each histogram was used in the regression, and the fitting uncertainty was estimated from the standard error of the fitted slope. This procedure yields separate Urbach energies associated with fluctuations of the conduction-band and valence-band edges, $E_U^{\mathrm{CBM}}$ and $E_U^{\mathrm{VBM}}$. Because optical absorption depends on transitions between these states, an effective optical Urbach energy was estimated as
\begin{equation}
E_U^{\mathrm{opt}}=
\sqrt{\left(E_U^{\mathrm{CBM}}\right)^2+\left(E_U^{\mathrm{VBM}}\right)^2}.
\end{equation}
Additional details of the fitting procedure, uncertainty estimation, and comparison among alternative structural descriptors are provided in the Supplementary Section~\ref{sec:elec}. 

\subsection{Experimental Details}

X-ray diffuse scattering measurements on \ce{FAPbI3} at \SI{102}{\kelvin}  were carried out at the I19-1 beamline at Diamond Light Source using X-rays of \SI{12.9}{\kilo\eV}. We have performed a quenching procedure in which we set the cryostream to \SI{102}{\kelvin} and directly inserted the sample into the cryostream from the room-temperature environment.  Based on a lumped-capacitance estimate for a $\sim$300~$\mu$m crystal in a nitrogen cryostream (Biot number $\mathrm{Bi} \ll 0.1$), the effective cooling rate during the quench is approximately 200--500~K/s, roughly $10^{3}$--$10^{4}$ times faster than a standard cryostream ramp. X-ray diffuse scattering measurements on \ce{FAPbI3} at \SI{130}{\kelvin} were carried out on the I15 beamline at Diamond Light Source using X-rays of \SI{72}{\kilo\eV}. The sample was cooled to \SI{130}{\kelvin} with a cooling rate of 3 K/min. In both experiments, each diffraction image was collected with a Pilatus2M detector (1679 $\times$ 1475 pixels, 172 $\times$ 172 $\mu\text{m}^2$ pixel size). The single crystals, mounted on the goniometer using a cryo loop for intensity measurements, were coated with immersion oil type NVH and then transferred to the nitrogen stream generated by an Oxford Cryostream 800 series. CrysAlisPro was used for indexing, determination and refinement of the orientation matrix, and then the precession images were unwrapped using CrysAlisPro, with polarisation correction. The images were generated such that the full reciprocal volume was recovered and loaded into HDF5 (.h5) files using the rspace3d Python package~\cite{rspace3d}. Outlier rejection and symmetrisation were also applied in the software to obtain the final data cube. 

To connect the real-space analysis with experimentally accessible observables, we computed the dynamical structure factor $S(\mathbf{q},\omega)$ following our previous framework~\cite{dynamic_domian_milos}. The reciprocal-space resolution is $1/18$~r.l.u., and both X-ray and neutron signals were sampled over 0--5~r.l.u. with an energy resolution of approximately 0.05~meV.

\section*{Acknowledgements}
We thank X. Fan, W. Baldwin, Z. Li, and K. Orr for helpful discussions. We also acknowledge T. Selby, B. Gallant, and P. Holzhey for assistance with data collection and crystal synthesis. We thank the Leverhulme Trust (RPG-2021-191) for funding. Via our membership of the UK's HEC Materials Chemistry Consortium, which is funded by EPSRC (EP/X035859/1), this work used the ARCHER2 UK National Supercomputing Service (http://www.archer2.ac.uk). We are also grateful to the UK Materials and Molecular Modelling Hub for computational resources, which is partially funded by EPSRC (EP/T022213/1, EP/W032260/1 and EP/P020194/1). X.L acknowledges support from the innoCore program of the Ministry of Science and ICT (1.260005.01). Z.Z. acknowledges support from a Newton International Fellowship from The Royal Society (No.~NIF/R1/254124). Y.L. acknowledges financial support from the Engineering and Physical Sciences Research Council (EPSRC, EP/V06164X/1). S.D.S. acknowledges the Royal Society and Tata Group (grant no. UF150033, URF/R/221026). We thank Diamond Light Source for access and support in the use of beamlines I19-1 (proposal CY41632-1), I19-2 (proposal CY36628-1) and I15 (proposal CY38508-2). This work is supported by a European Research Council grant (VAPOURISE, 101169608). Views and opinions expressed are however those of the authors only and do not necessarily reflect those of the European Union or the European Research Council Executive Agency. Neither the European Union nor the granting authority can be held responsible for them.

\section*{Data Availability}
The \textsc{PDynA} package used in this work is open-source and available online at \url{https://github.com/WMD-group/PDynA} (DOI: 10.5281/zenodo.7948045). A repository with full training data and MLFF parameters will be made available upon publication. 

\FloatBarrier
\clearpage
{\footnotesize\singlespacing
\bibliography{REF}
}

\clearpage
\appendix

\setcounter{figure}{0}
\renewcommand{\thefigure}{S\arabic{figure}}
\setcounter{table}{0}
\renewcommand{\thetable}{S\arabic{table}}

\setcounter{subsection}{0}
\renewcommand{\thesubsection}{S\arabic{subsection}}

\section*{Supplementary Information}

\subsection{Force Field Validation}
\renewcommand{\arraystretch}{1.35}
\begin{table}[ht]
\centering
\caption{Root mean squared errors of energy, forces and stress tensors of the unified MLFF covering the chemical configuration space $(\mathrm{Cs}_{1-x-y}\mathrm{MA}_x\mathrm{FA}_y)\allowbreak \mathrm{Pb}\allowbreak (\mathrm{Br}_m\mathrm{I}_{1-m})_3$.}
\begin{tabular}[t]{c@{\hskip 0.4in}c@{\hskip 0.3in}c@{\hskip 0.3in}c}
\multirow{2}{*}{Material} & \multicolumn{3}{c}{RMSE on test set} \\
\cline{2-4}
 & Energy (\text{meV/atom}) & Force (meV/\AA) & Stress (meV/\AA$^{3}$) \\
\hline
\hline
\ce{CsPbBr3} & 0.46 & 9.14 & 0.343 \\
\ce{CsPbI3}  & 0.42 & 11.15 & 0.163 \\
\ce{MAPbBr3} & 1.42 & 16.53 & 0.219 \\
\ce{MAPbI3}  & 0.68 & 16.41 & 0.275 \\
\ce{FAPbBr3} & 1.14 & 18.73 & 0.225 \\
\ce{FAPbI3}  & 1.04 & 19.10 & 0.366 \\
\hline
\end{tabular}
\renewcommand{\arraystretch}{1.0}
\label{errors}
\end{table}

\begin{figure}[t]
    \centering
    
    \includegraphics[width=0.9\textwidth]{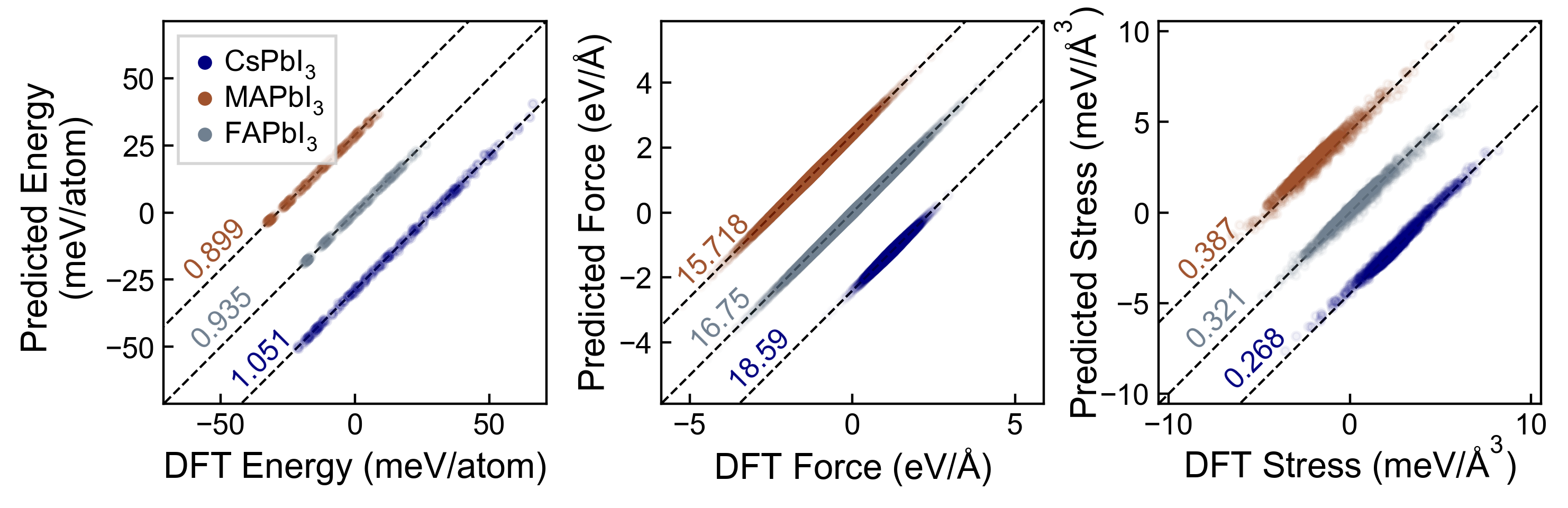}
    \caption{Parity plots for MLFF predictions on 200 MD snapshots drawn from the cubic-phase regimes of \ce{CsPbI3}, \ce{MAPbI3}, and \ce{FAPbI3}, with DFT reference values recomputed using the same settings as in the data-generation workflow. Left: energies; middle: forces; right: stresses. For visual clarity, the \ce{CsPbI3} and \ce{MAPbI3} populations are shifted relative to \ce{FAPbI3}, and separate parity lines are shown for each material. In the energy panel, the values are additionally shifted to a common scale because \ce{CsPbI3} has a larger energy per atom. The numbers printed next to each population denote the corresponding RMSE values in meV units (meV/atom for energy, meV/\AA{} for forces and meV/\AA$^{3}$ for stresses).}
    \label{si:parity_prod}
\end{figure}

The errors in Table~\ref{errors} indicate that the unified MLFF retains near-DFT fidelity across all six halide-perovskite end members. In particular, the energy RMSE remains below 1.5~meV/atom and the force RMSE below 20~meV/\AA\ for every composition, which places the model well within the accuracy regime typically regarded as high quality for MLFFs of soft ionic solids. These values are comparable to, and in several cases lower than, those reported for recent MLFFs trained on chemically related perovskite systems~\cite{vaspmlff_perov,unified_perovskite_FF_biswas_2026,mlff_perovskite_fransson_jpcc_2023}. The small spread across compositions further suggests that the unified model does not sacrifice transferability to achieve accuracy for a single compound.

Figure~\ref{si:parity_prod} shows parity plots for 200 MD snapshots drawn from the cubic-phase regimes of \ce{CsPbI3}, \ce{MAPbI3}, and \ce{FAPbI3}, with DFT energies, forces, and stresses recomputed using the same settings as in the data-generation workflow. The resulting errors remain at the same overall level as those obtained on the held-out test sets, confirming that the model remains accurate on representative structures sampled during production MD rather than only on the train/test partition.

\begin{figure}[t]
    \centering
    
    \includegraphics[width=0.6\textwidth]{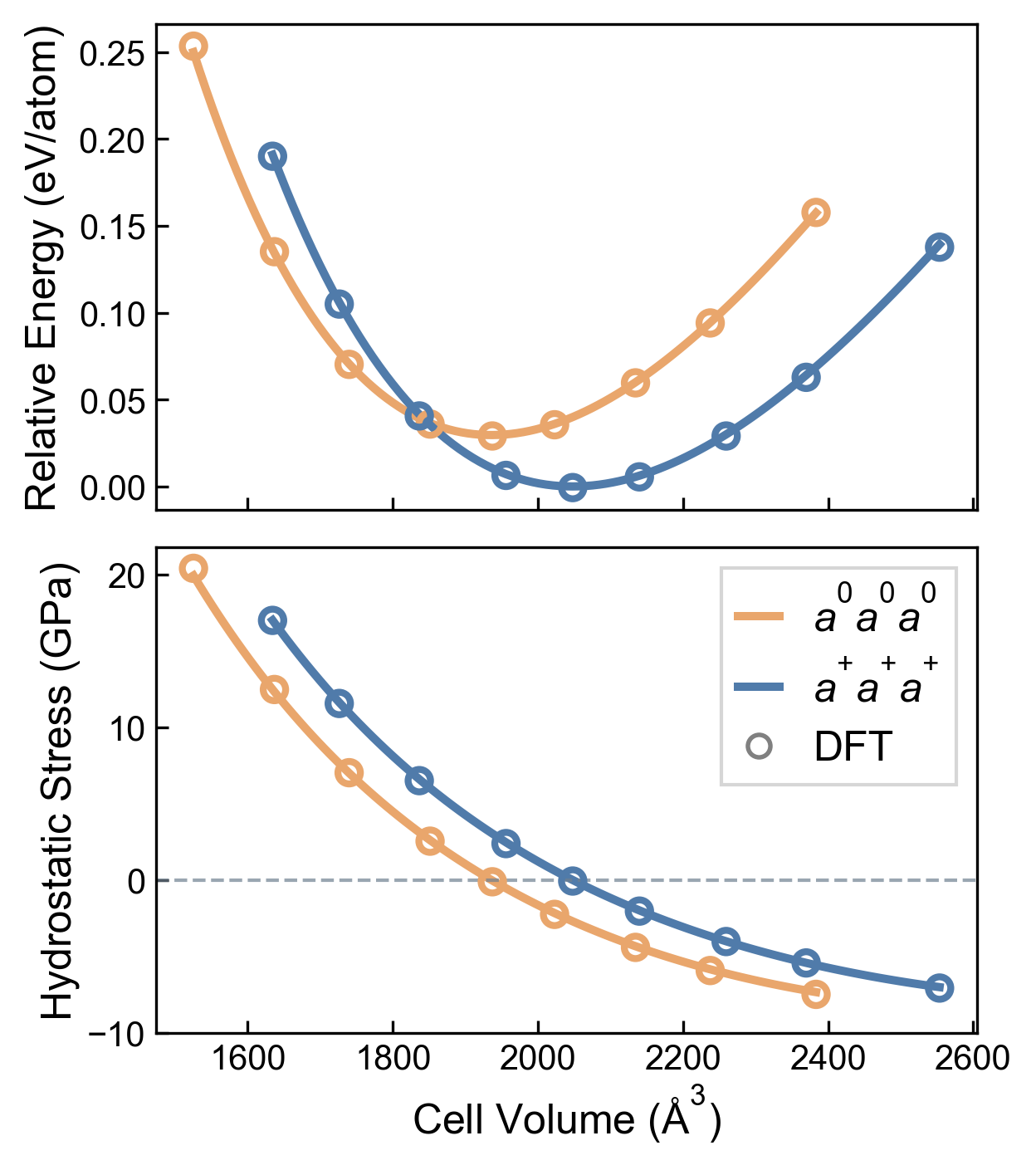}
    \caption{Energy--volume and hydrostatic-stress--volume relations for \ce{FAPbI3} along the cubic $a^{0}a^{0}a^{0}$ and ordered $a^{+}a^{+}a^{+}$ structural branches. Solid lines show MLFF predictions over an isotropic volume range from 80\% to 125\% of the reference cell volume, and open circles denote DFT reference calculations on selected structures. }
    \label{si:ev}
\end{figure}

A second validation is provided by the energy--volume and hydrostatic-stress--volume curves for \ce{FAPbI3} in the cubic $a^{0}a^{0}a^{0}$ and ordered $a^{+}a^{+}a^{+}$ structures (Fig.~\ref{si:ev}). Over a broad isotropic volume range from 80\% to 125\% of the reference cell volume, the MLFF reproduces the DFT-computed energies and hydrostatic stresses closely for both structural branches. This agreement confirms that the model captures not only local forces but also the near-equilibrium curvature and relative mechanical response of distinct structural states relevant to the present study.

\begin{figure}[htb]
    \centering
    
    \includegraphics[width=0.99\textwidth]{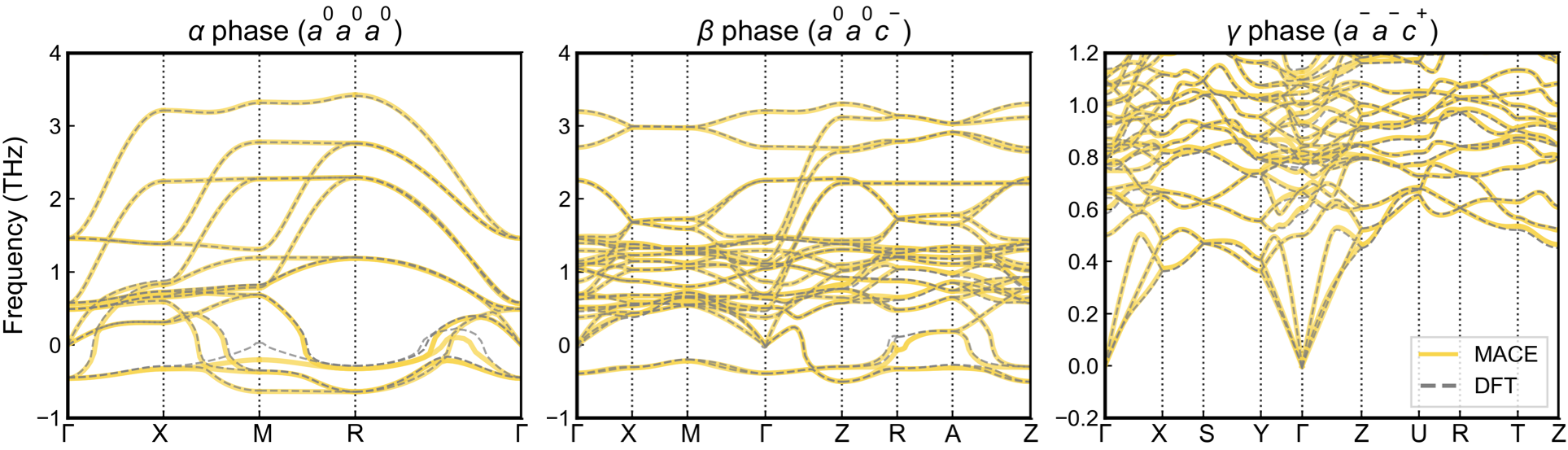}
    \caption{Comparison of MLFF and DFT phonon dispersions of \ce{CsPbI3} in the $\alpha$ ($a^{0}a^{0}a^{0}$), $\beta$ ($a^{0}a^{0}c^{-}$), and $\gamma$ ($a^{-}a^{-}c^{+}$) phases. The unified MLFF reproduces the DFT lattice-dynamical spectra well across all three structures.}
    \label{si:phonon_cspbi3}
\end{figure}

To assess the lattice-dynamical response, Fig.~\ref{si:phonon_cspbi3} compares MLFF and DFT phonon dispersions of \ce{CsPbI3} in its $\alpha$, $\beta$, and $\gamma$ phases. The agreement is consistently good across all three structures, showing that the unified MLFF reproduces the inorganic-framework lattice dynamics of this endpoint system with high fidelity. We do not use an analogous harmonic-phonon benchmark as a primary validation for \ce{FAPbI3}, because in the molecular iodide perovskite the orientational disorder of the FA sublattice and the resulting strong anharmonicity make a single-reference harmonic dispersion less uniquely defined than in the inorganic endpoint.

\begin{figure}[t]
    \centering
    
    \includegraphics[width=0.52\textwidth]{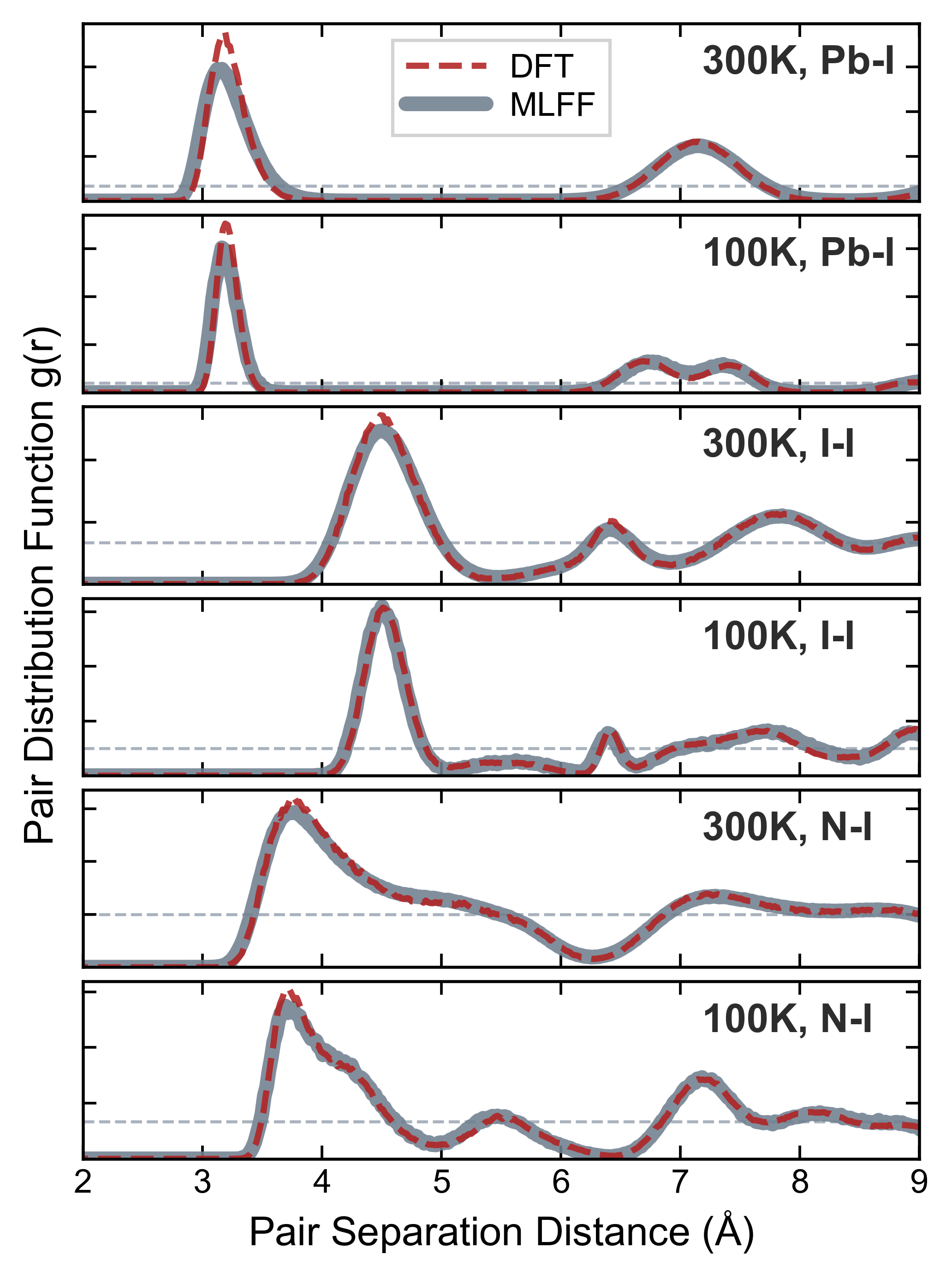}
    \caption{Radial distribution functions of \ce{FAPbI3} at 100~K (ordered $a^{+}a^{+}a^{+}$) and 300~K (cubic $a^{0}a^{0}a^{0}$) from DFT and MLFF trajectories. The Pb--I, I--I, and N--I pair correlations are shown.}
    \label{si:rdf}
\end{figure}

Finally, finite-temperature structural fidelity is tested by comparing DFT and MLFF radial distribution functions for \ce{FAPbI3} at 100 and 300~K (Fig.~\ref{si:rdf}). We focus on three representative pairs that capture the essential local physics: Pb--I for nearest-neighbour octahedral bonding, I--I for framework geometry and tilt-sensitive halide separations, and N--I for coupling between the FA cation and the inorganic cage. The MLFF reproduces all three distributions well at both temperatures, with only small differences in the first Pb--I peak. Taken together, these tests indicate that the unified potential reproduces the static, vibrational, mechanical, and finite-temperature structural observables most relevant to the large-scale MD analyses presented in the main text.

\clearpage

\subsection{Pulay Stress and Recomputed DFT Labels}

During variable-cell \emph{ab initio} MD, the simulation cell changes continuously, whereas the plane-wave basis is defined through a finite cutoff. In plane-wave DFT, this can introduce Pulay-stress errors because the basis is not fully complete with respect to changes in cell shape and volume, leading to systematic errors in energies and especially stresses if the instantaneous configurations are used directly as training labels~\cite{pulay_stress_bucko}. This issue is well known for cell relaxations and variable-cell simulations in VASP, where Pulay stress can bias the calculated stress tensor and distort the equilibrium volume unless the basis is sufficiently converged or the structures are recomputed with the updated basis~\cite{pulay_stress_francis}.

\begin{figure}[b]
    \centering
    
    \includegraphics[width=0.99\textwidth]{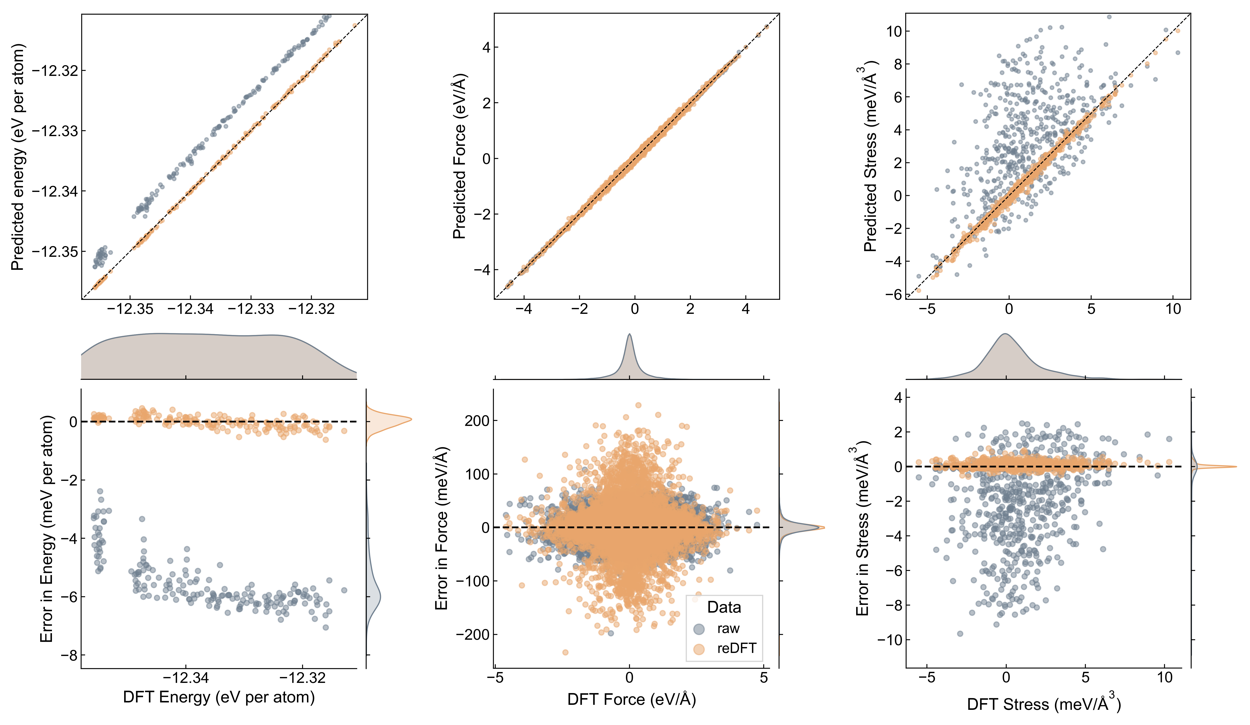}
    \caption{Comparison of \ce{MAPbI3} validation errors obtained from MLFFs trained on raw on-the-fly labels and on recomputed single-point DFT labels (\emph{reDFT}). Top row: predicted versus DFT values for energy, forces, and stress. Bottom row: corresponding prediction errors. The recomputed dataset removes the dominant systematic bias in the stress prediction and improves the consistency of the training labels.}
    \label{si:pulay_ma}
\end{figure}

To suppress this source of bias, all structures selected for final training were recomputed with static single-point DFT calculations using the converged labeling setup described in the Methods. This produced a recomputed dataset (\emph{reDFT}) with substantially reduced basis-set inconsistency relative to the raw variable-cell data obtained directly from on-the-fly trajectories. Fig.~\ref{si:pulay_ma} illustrates the effect for \ce{MAPbI3}: relative to a model trained on the raw labels, retraining on the recomputed labels removes the systematic shift in the energy and stress prediction and reduces the corresponding error distribution. In practice, this correction is particularly important for large-scale production MD, where even a modest systematic stress bias can accumulate into an artificial contraction or expansion of the simulation cell.

\clearpage

\subsection{Compositional Sampling Strategy for Surrogate Molecular Dynamics}
\label{sec:config}

\begin{figure}[htb]
    \centering
    
    \includegraphics[width=0.77\textwidth]{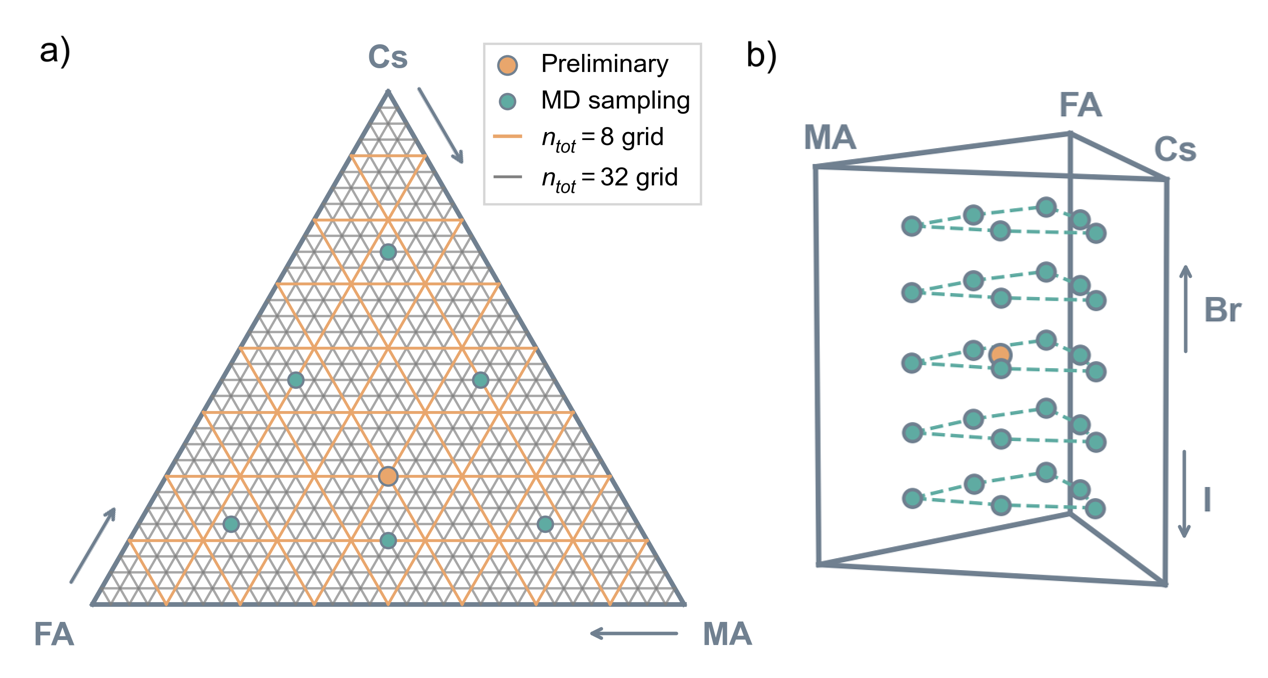}
    \caption{Sampling of the doubly mixed lead-halide perovskite composition space. (a) A-site ternary composition diagram showing the preliminary on-the-fly training composition and the set of A-site profiles used for surrogate MD sampling. The overlaid coarse and fine grids illustrate the $n_{\mathrm{tot}}=8$ and $n_{\mathrm{tot}}=32$ composition discretisations. (b) Schematic three-dimensional representation of the doubly mixed space spanned by the A-site and halide mixing axes, showing the preliminary composition and the mixed compositions sampled in the surrogate-MD stage.}
    \label{si:doubly_config}
\end{figure}

To sample the doubly mixed compositional space $(\mathrm{Cs}_{1-x-y}\mathrm{MA}_x\mathrm{FA}_y)\allowbreak \mathrm{Pb}\allowbreak (\mathrm{Br}_m\mathrm{I}_{1-m})_3$ with the surrogate MLFF, we introduce two independent mixing axes: the A-site sublattice and the halide sublattice. On the A-site sublattice, seven cation profiles $(n_{\mathrm{FA}},n_{\mathrm{MA}},n_{\mathrm{Cs}})$ are selected such that $n_{\mathrm{FA}}+n_{\mathrm{MA}}+n_{\mathrm{Cs}}=32$, namely $(11,11,10)$, $(5,5,22)$, $(14,14,4)$, $(4,14,14)$, $(5,22,5)$, $(22,5,5)$, and $(14,4,14)$. On the halide sublattice, 15, 31, 48, 65, or 81 bromine atoms are replaced by iodine using ICET, giving five mixed-halide compositions. Combining these two sets yields 35 distinct mixed structures, which are then paired with five idealised octahedral-tilt patterns to generate the surrogate-MD dataset described in the Methods. The temperatures are chosen such that the Br-rich compositions are sampled at lower temperatures than the corresponding I-rich compositions, consistent with the systematically lower phase transition temperatures of bromide-rich lead perovskites.

\begin{table}[htb]
\centering
\caption{Idealised initial tilt patterns and the corresponding temperature sets used for surrogate-MD sampling. Br-rich compositions are assigned lower temperatures than the corresponding I-rich compositions, consistent with the lower transition temperatures of bromide-rich lead perovskites.}
\begin{tabular}[t]{c@{\hspace{18pt}}c}
\toprule
Tilting mode & Temperatures (K) \\
\midrule
$a^{0}a^{0}a^{0}$ & 200, 300, 350, 400, 450 \\
$a^{0}a^{0}c^{-}$ and $a^{0}a^{0}c^{+}$ & 120, 150, 180, 210, 240 \\
$a^{-}a^{-}c^{+}$ and $a^{+}a^{+}a^{+}$ & 40, 55, 70, 85, 100 \\
\bottomrule
Temperatures (K) & Halide composition \\
\midrule
200/120/40 & \ce{I15Br81} \\
300/150/55 & \ce{I31Br65} \\
350/180/70 & \ce{I48Br48} \\
400/210/85 & \ce{I65Br31} \\
450/240/100 & \ce{I81Br15} \\
\bottomrule
\end{tabular}
\label{si:profiles}
\end{table}

\begin{figure}[htb]
    \centering
    
    \includegraphics[width=0.42\textwidth]{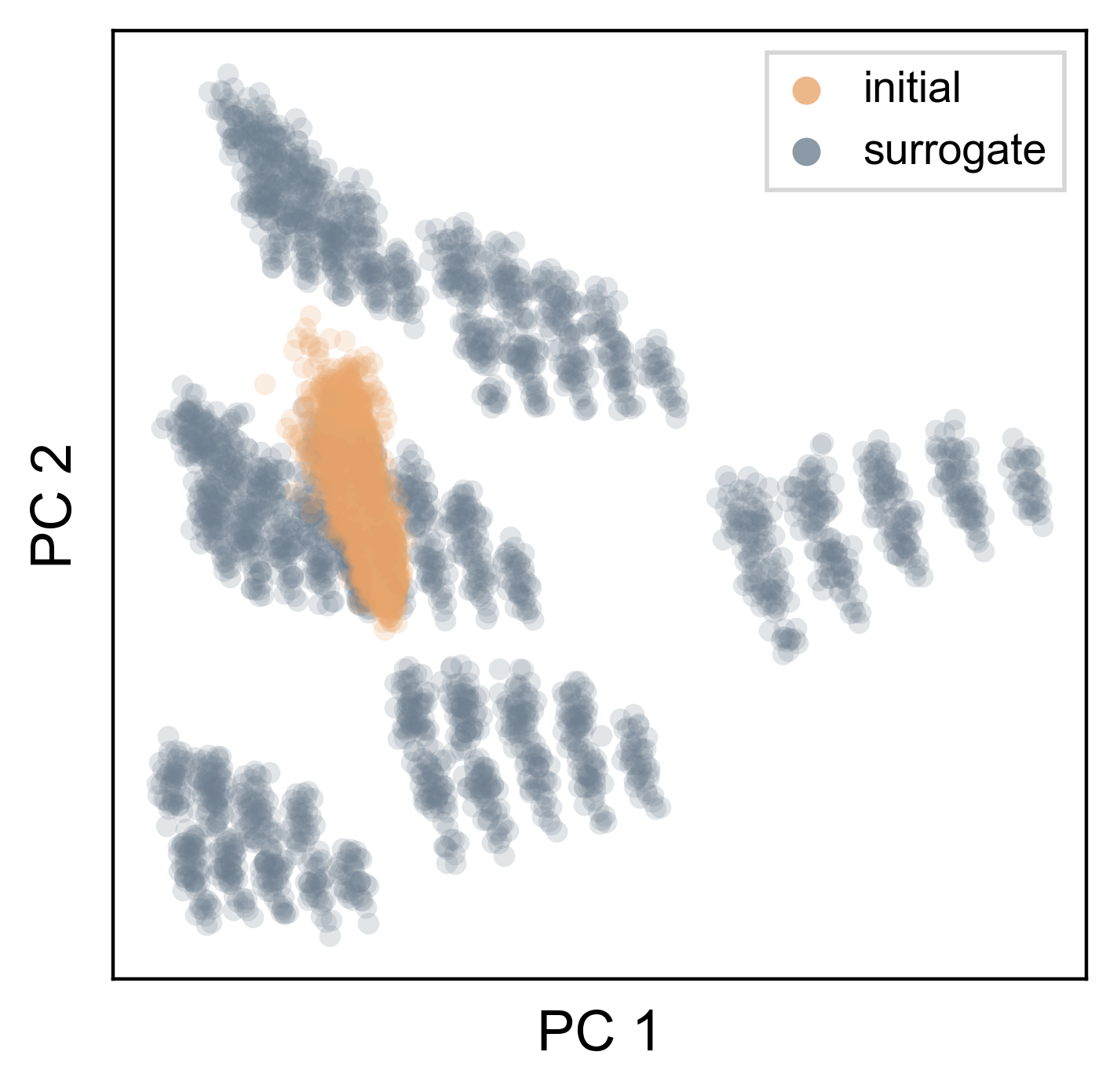}
    \caption{PCA projection of SOAP descriptors for structures obtained from the initial on-the-fly dataset and from the surrogate-MD sampling workflow. The surrogate-generated structures span a substantially broader region of descriptor space, showing that the initial dataset provides a chemically representative seed while the surrogate model expands the accessible local-environment diversity.}
    \label{si:soap_doubly1}
\end{figure}

The surrogate-sampling workflow generates 35{,}000 candidate structures from small-cell MD across the doubly mixed composition space. 3{,}000 structures were then selected by DIRECT for final DFT relabelling. Figure~\ref{si:soap_doubly1} shows that the surrogate-generated structures span a substantially broader region of SOAP descriptor space than the initial on-the-fly dataset obtained from the single reference composition \ce{Cs2MA3FA3Pb8Br12I12}. The initial on-the-fly structures therefore act as a chemically representative starting point near the centre of the targeted configuration space, while the surrogate model enables systematic exploration into a much wider range of local environments.

\begin{figure}[htb]
    \centering
    
    \includegraphics[width=0.98\textwidth]{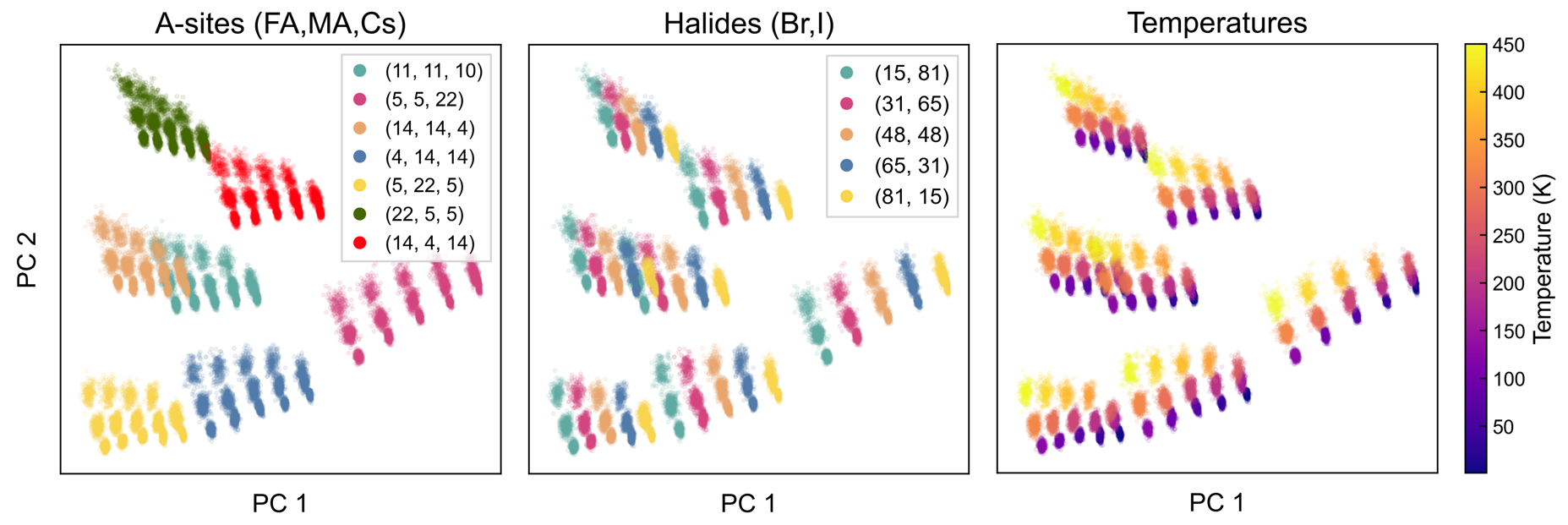}
    \caption{PCA projection of SOAP descriptors for the surrogate-generated doubly mixed perovskite structures, coloured by A-site composition, halide composition, and temperature. The descriptor space exhibits a hierarchical clustering in which A-site composition produces the strongest separation, followed by halide composition and then temperature.}
    \label{si:soap_doubly2}
\end{figure}

Within the surrogate-generated dataset, PCA of the SOAP descriptors reveals a clear hierarchical organisation of the local environments (Fig.~\ref{si:soap_doubly2}). The strongest separation is by A-site composition, indicating that changes in the relative fractions of FA, MA, and Cs dominate the first layer of structural diversity. Within each A-site cluster, halide composition produces further subdivision into narrower strands, consistent with systematic sampling along the Br--I mixing direction. Temperature then generates an additional progression within each compositional strand, reflecting the increasing diversity of local environments sampled at higher thermal energy. This separation supports the use of a structured surrogate-MD workflow rather than a purely random exploration of configuration space.

\begin{figure}[htb]
    \centering
    
    \includegraphics[width=0.98\textwidth]{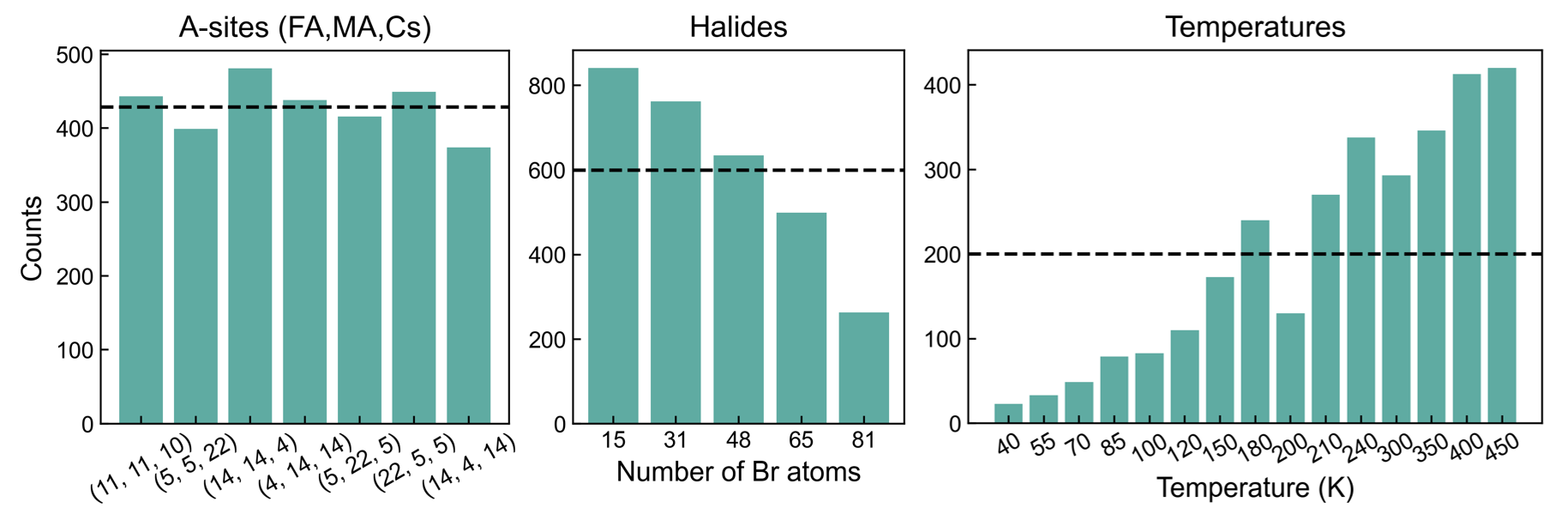}
    \caption{Number of DIRECT-selected structures across A-site compositions, halide compositions, and simulation temperatures. Dashed lines indicate the counts expected from a purely uniform selection. The selected dataset remains broadly balanced across A-site compositions but is enriched in I-rich and high-temperature structures, reflecting their greater structural diversity.}
    \label{si:doubly_selection}
\end{figure}

The selected 3{,}000-frame training subset is not distributed uniformly across all categories (Fig.~\ref{si:doubly_selection}). The A-site profiles remain comparatively balanced, whereas the selected structures are skewed towards I-rich compositions and towards higher temperatures. This bias is physically reasonable: I-rich systems and high-temperature trajectories exhibit stronger structural rearrangements and broader local environment diversity, and therefore contribute disproportionately to the set of distinct configurations retained by DIRECT.

\begin{figure}[htb]
    \centering
    
    \includegraphics[width=0.84\textwidth]{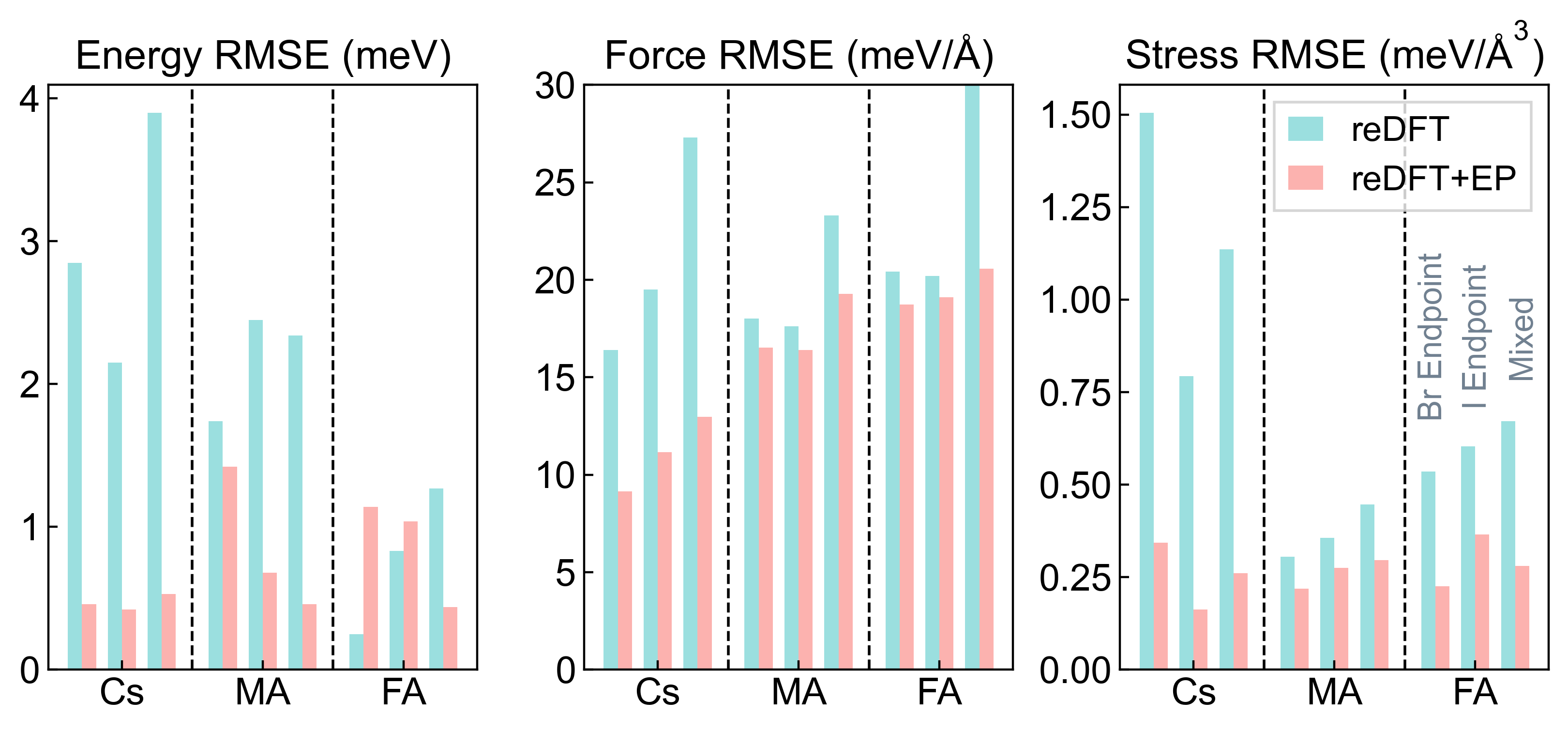}
    \caption{Validation RMSEs of models trained on the mixed-composition \emph{reDFT} dataset alone and on the \emph{reDFT+EP} dataset including endpoint-family structures. Results are grouped by A-site family and by Br endpoint, I endpoint, and mixed-halide validation sets. Including endpoint-family data improves predictive accuracy across most target regions of the doubly mixed chemical space.}
    \label{si:doubly_error}
\end{figure}

To further improve sampling near the boundaries of the chemical space, the final training set also includes endpoint-family data for \ce{CsPbX3}, \ce{MAPbX3}, and \ce{FAPbX3}, each contributing 1000 additional structures. The effect of this endpoint-family augmentation is shown in Fig.~\ref{si:doubly_error}. Relative to training on the mixed-composition \emph{reDFT} dataset alone, the \emph{reDFT+EP} model yields consistently improved errors across most validation sets. This improvement is especially clear for the energy and stress predictions, showing that explicit coverage of the edges of the A-site/halide composition space materially improves transferability within the interior of the doubly mixed space. This behaviour is consistent with the general aim of selection, namely to construct a training dataset that captures structural and chemical diversity across a large configuration manifold.

\clearpage

\subsection{Structural Dynamics Statistics}

\begin{figure}[htb]
    \centering
    
    \includegraphics[width=0.42\textwidth]{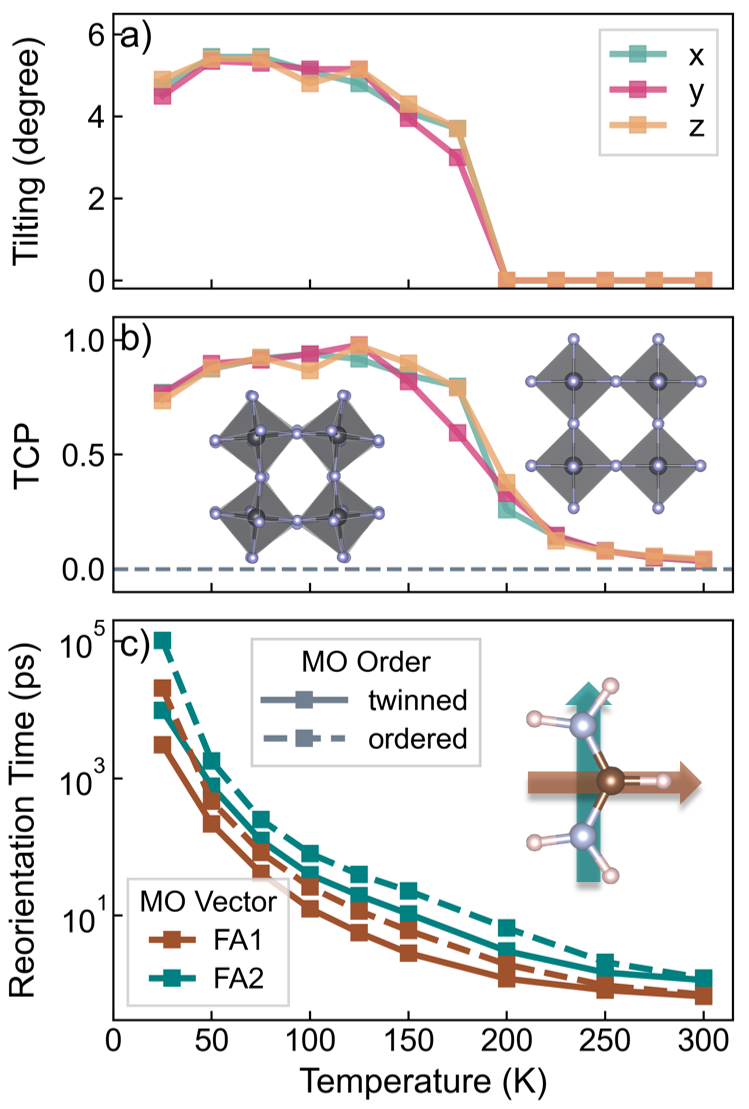}
    \caption{Temperature dependence of (a) the fitted mean tilt magnitude along the three crystallographic directions, extracted from the two-Gaussian fit of the dynamic tilt-angle distributions; (b) the TCP along the three crystallographic directions; and (c) characteristic FA molecular reorientation times extracted from the molecular-orientation autocorrelation functions.}
    \label{si:structure}
\end{figure}

Following the \textsc{PDynA} framework~\cite{pdyna2023liang}, each \ce{PbX6} octahedron is assigned a local tilt angle $\theta_{\alpha}(t;\mathbf{n})$ around axis $\alpha\in\{x,y,z\}$, where $\mathbf{n}=(n_x,n_y,n_z)$ indexes the octahedron in the supercell,
\begin{equation}
\theta_{\alpha}(t;\mathbf{n}).
\end{equation}
The three components together define the local instantaneous tilting state of an octahedron. In practice, the mean tilting magnitude shown in Fig.~\ref{si:structure}a is extracted separately for each crystallographic direction from the corresponding distribution of instantaneous tilt angles. For each $\alpha\in\{x,y,z\}$, the sampled distribution is fitted to a symmetric two-Gaussian form,
\begin{equation}
P_{\alpha}(\theta)=
A_{\alpha}\exp\!\left[-\frac{(\theta-\langle T_{\alpha}\rangle)^2}{2\sigma_{\alpha}^2}\right]
+
A_{\alpha}\exp\!\left[-\frac{(\theta+\langle T_{\alpha}\rangle)^2}{2\sigma_{\alpha}^2}\right],
\end{equation}
where $\langle T_{\alpha}\rangle$ is the fitted mean tilt magnitude about axis $\alpha$, and $\sigma_{\alpha}$ is the common width of the two peaks. This definition follows the use of Gaussian fitting to the dynamic tilt distributions in the original \textsc{PDynA} analysis. 

To quantify the local sign correlation of the tilt pattern, we use the nearest-neighbour correlation descriptor introduced in \textsc{PDynA}. For tilting around axis $\alpha$, correlated along axis $\beta$, the sign-sensitive local correlation with the $k$-th neighbour is
\begin{equation}
r_{\alpha,\beta}^{(k)}(t;\mathbf{n})=
\frac{
\theta_{\alpha}(t;\mathbf{n})\,\theta_{\alpha}(t;\mathbf{n}+k\hat{\beta})
}{
\sqrt{\left|\theta_{\alpha}(t;\mathbf{n})\,\theta_{\alpha}(t;\mathbf{n}+k\hat{\beta})\right|}
},
\end{equation}
which is effectively the sign of the tilt-product whenever both tilts are non-zero. Positive and negative values correspond to in-phase and out-of-phase local correlations, respectively. For the special case $\alpha=\beta$ and $k=1$, this quantity directly reflects the first-neighbour Glazer correlation along the tilt direction. 

The corresponding global spatial tilting correlation function is
\begin{equation}
R_{\alpha,\beta}(k)=
C\left\langle
\theta_{\alpha}(t;\mathbf{n})\,\theta_{\alpha}(t;\mathbf{n}+k\hat{\beta})
\right\rangle_{\mathbf{n},t,\pm k},
\end{equation}
where $C$ is chosen so that $R_{\alpha,\beta}(0)=1$, and $\langle \cdots \rangle_{\mathbf{n},t,\pm k}$ denotes averaging over all octahedra, all sampled times, and neighbours at $\pm k$ along $\beta$. The spatial extent of the tilt correlation is obtained by fitting the envelope of $R_{\alpha,\beta}(k)$ to
\begin{equation}
R_{\alpha,\beta}(k)=
\exp\!\left(-\frac{k}{\xi_{\alpha,\beta}}\right),
\end{equation}
where $\xi_{\alpha,\beta}$ is the fitted correlation length in the $\beta$ direction of the tilt around the $\alpha$ axis. In the main text, the real-space correlation lengths are reported as the relevant directional values. 

\begin{figure}[htb]
    \centering
    
    \includegraphics[width=0.82\textwidth]{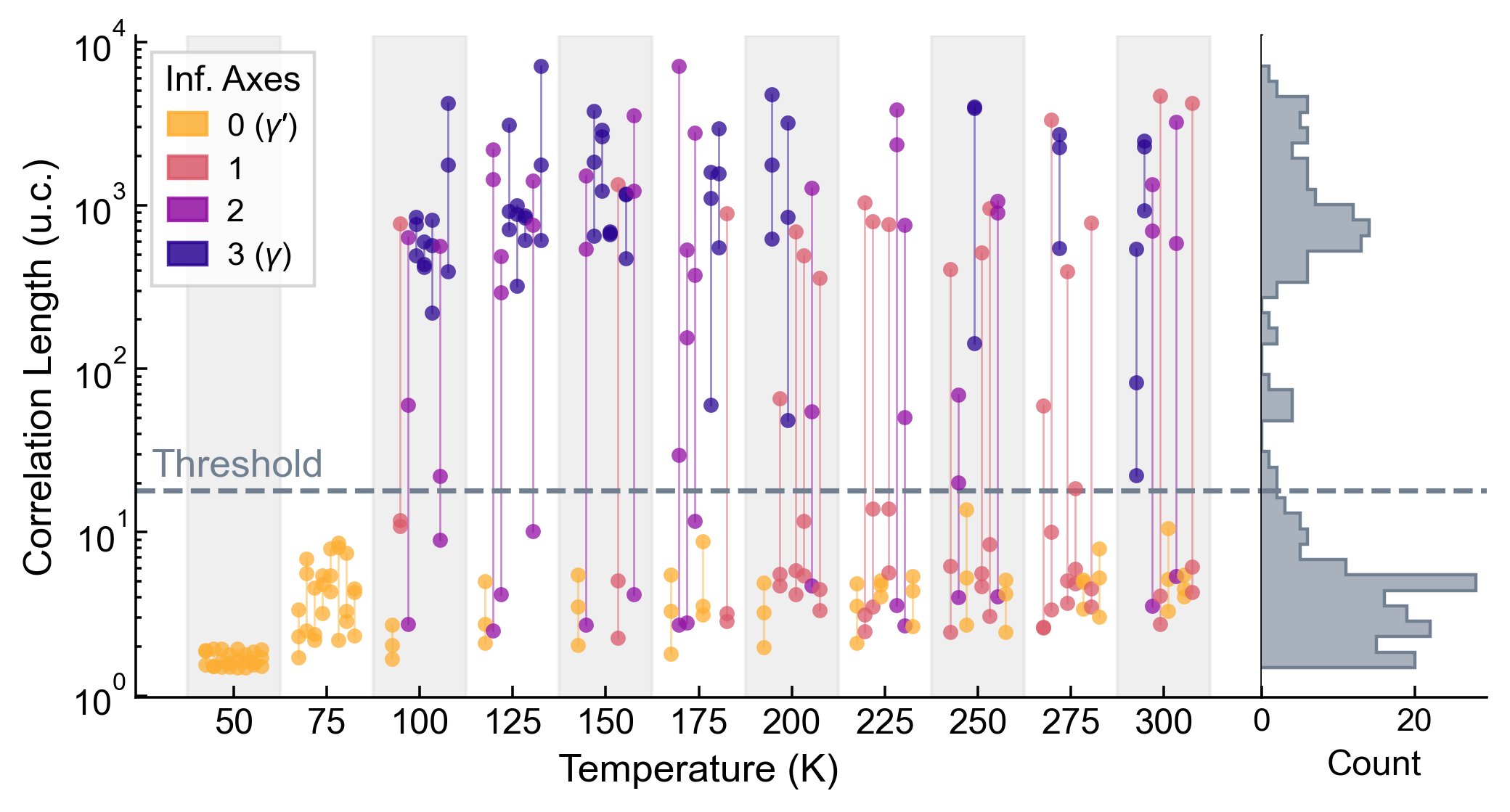}
    \caption{Temperature-dependent correlation length statistics for the controlled-orientation simulations, coloured by the number of fully correlated axes. Each vertical cluster shows the three directional correlation lengths ($x$, $y$, and $z$) from a single MD run at the indicated temperature, with repeat runs spread symmetrically around each temperature tick. Point colour encodes the number of effectively infinite axes identified for that run: 0 ($\gamma'$), 1, 2, or 3 ($\gamma$), thereby indicating the degree of long-range octahedral tilt coherence. The dashed horizontal line marks the threshold of 18 pseudocubic unit cells used to distinguish finite from effectively infinite correlation. The right panel shows the marginal distribution of all fitted correlation lengths on the same logarithmic scale.}
    \label{si:corrlen}
\end{figure}

For the controlled-orientation simulations discussed in Fig.~\ref{res:mo}c, the low-temperature structural outcomes were classified according to the number of crystallographic directions exhibiting system-spanning tilt correlation. In practice, a direction was labelled as fully correlated when the fitted real-space correlation length exceeded 18 pseudocubic unit cells, corresponding to the full linear size of the simulation box used in that analysis. Correlation lengths above this threshold were therefore treated as effectively infinite on the simulated length scale. This criterion distinguishes fully correlated single-domain $\gamma$-like configurations from heterogeneous low-temperature $\gamma'$ states with reduced long-range coherence. The corresponding fitted correlation lengths for all controlled-orientation simulations are shown explicitly in Fig.~\ref{si:corrlen}, which provides the underlying structural dataset used to classify the outcomes summarised in Fig.~\ref{res:mo}c.

The tilting correlation polarity (TCP) around axis $\alpha$ is defined from the populations of positive and negative first-neighbour correlations,
\begin{equation}
\delta_{\alpha}=
\frac{n_{\alpha}^{+}-n_{\alpha}^{-}}{n_{\alpha}^{+}+n_{\alpha}^{-}},
\end{equation}
where $n_{\alpha}^{+}$ and $n_{\alpha}^{-}$ count the instances of $r_{\alpha,\alpha}^{(1)}>0$ and $r_{\alpha,\alpha}^{(1)}<0$, respectively. As a result, $\delta_{\alpha}\in[-1,1]$, with values close to $1$, $0$, and $-1$ corresponding to Glazer superscripts $+$, $0$, and $-$, respectively. 

Time correlations are analysed for both octahedral tilting and molecular orientations. In both cases, the normalised autocorrelation functions are fitted to the same biexponential form used in the original \textsc{PDynA} formalism,
\begin{equation}
A(t)=
C\exp\!\left(-\frac{t}{\tau_1}\right)
+
(1-C)\exp\!\left(-\frac{t}{\tau_2}\right),
\label{eq:autocorr_biexp}
\end{equation}
where $\tau_1$ is the dominant correlation time and $\tau_2$ captures the short-time correction. In all fits considered here, $C>0.8$, so that $\tau_1$ is the dominant timescale and is therefore taken as the characteristic tilt or molecular reorientation time, while the second exponential term accounts for the fast initial decay correction. 

\begin{figure}[htb]
    \centering
    
    \includegraphics[width=0.98\textwidth]{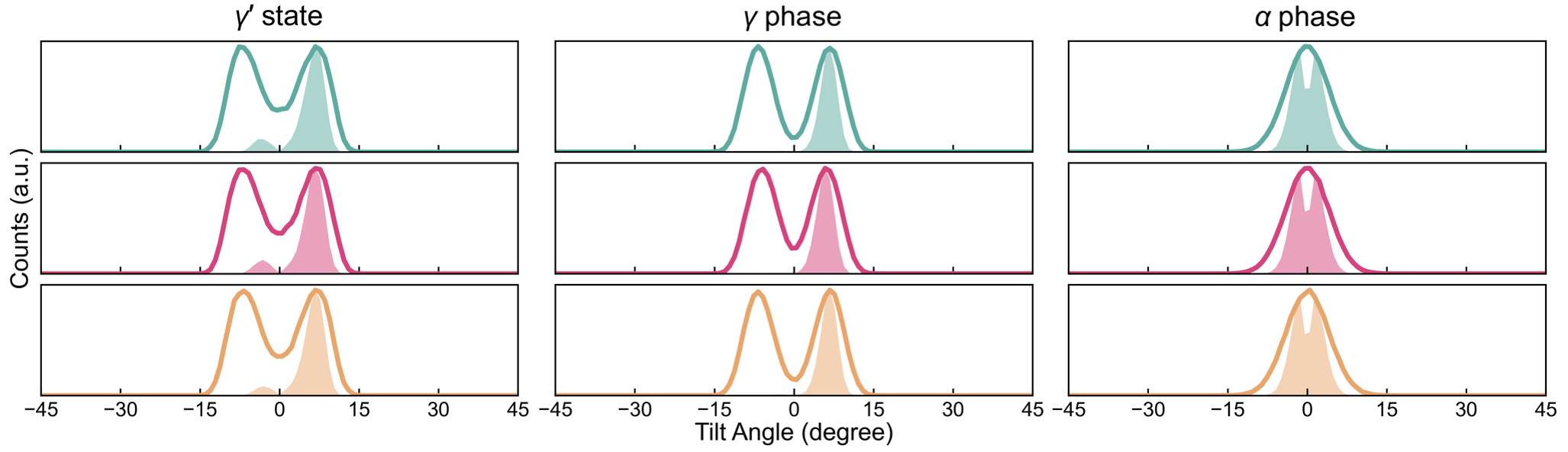}
    \caption{Explicit octahedral tilt angle distributions for the three key structural regimes of \ce{FAPbI3}: the high-temperature $\alpha$ phase, the ordered $\gamma$ phase, and the low-temperature $\gamma'$ state. Rows correspond to the three Cartesian tilt directions. }
    \label{si:3mode}
\end{figure}

\begin{figure}[htb]
    \centering
    
    \includegraphics[width=0.8\textwidth]{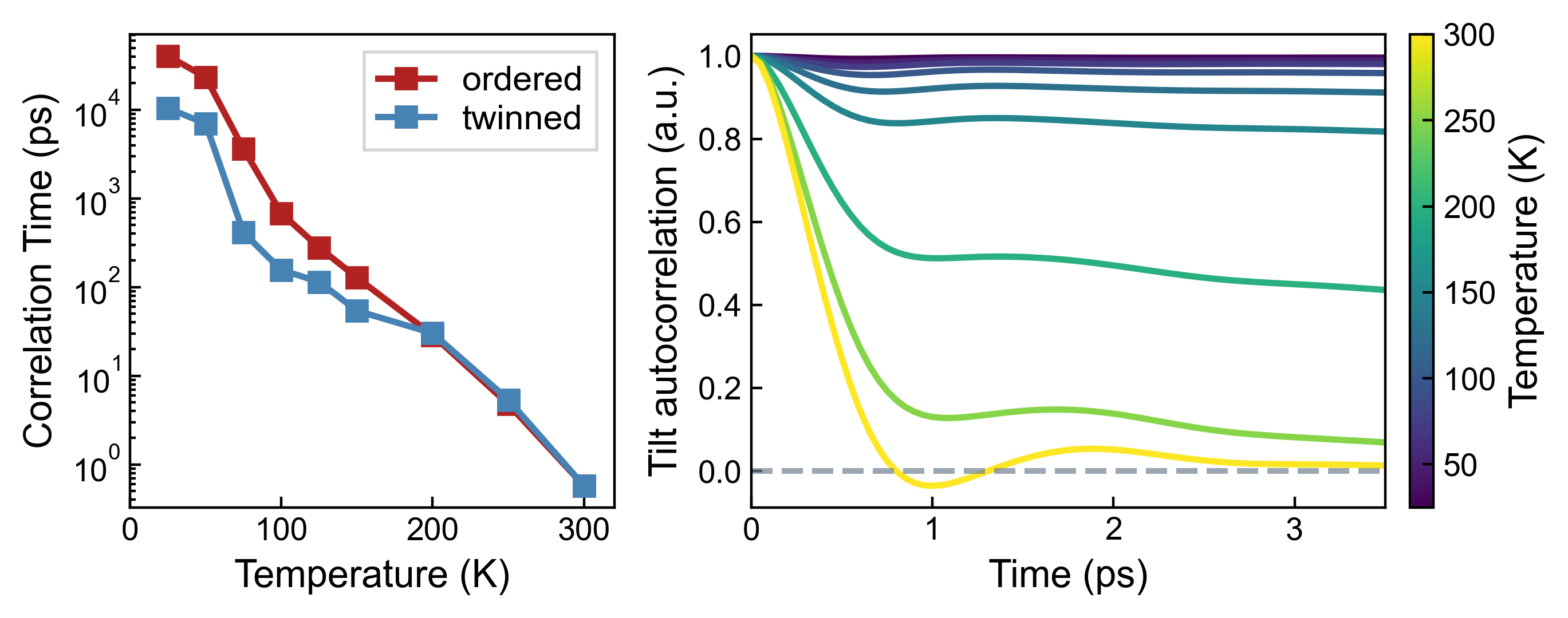}
    \caption{Tilt-autocorrelation analysis of \ce{FAPbI3}. Left panel: characteristic tilt correlation times extracted from the biexponential fit of eq.~\ref{eq:autocorr_biexp}. Right panel: normalised tilt autocorrelation functions, averaged over the three Cartesian directions, for representative temperatures. The low-temperature curves remain highly correlated over the accessible simulation window, whereas the 300~K response decays rapidly, indicating much shorter tilt memory.}
    \label{si:tilt_time}
\end{figure}

Figure~\ref{si:structure} summarises the temperature dependence of the fitted directional mean tilt angles, the corresponding TCP values, and the FA reorientation time. Panels a and b show not only the growth of $\gamma$-like in-phase tilt order on cooling from the high-temperature phase, but also its partial reduction at lower temperature as the system crosses into the twinned $\gamma'$ state, where both the fitted tilt amplitude and TCP decrease. Panel c shows the rapid increase in molecular reorientation time on cooling, providing the dynamical basis for the history-dependent crossover discussed in the main text. For completeness, Fig.~\ref{si:3mode} shows the explicit distributions of octahedral tilt angles for the three key structural regimes discussed in the main text, namely the high-temperature $\alpha$ phase, the ordered $\gamma$ phase, and the low-temperature $\gamma'$ state. In the $\alpha$ phase, the overall tilting is at zero degrees and with fluctuation up to 15 degrees, consistent with dynamic temporal local tilts. In the ordered $\gamma$ phase, the distributions sharpen into well-defined non-zero peaks, reflecting long-range in-phase tilt coherence. In the $\gamma'$ state, the distributions resemble those of the $\gamma$ phase but with finite out-of-phase correlation, which is attributed to the twin domain boundary. The corresponding tilt-autocorrelation analysis is shown separately in Fig.~\ref{si:tilt_time}, where the extracted correlation time rises sharply at low temperature, indicating long-lived tilt memory in the low-$T$ regime. The values from ordered structures are consistently higher than their twinned counterparts until they ultimately converge above 200~K. 

\clearpage

\subsection{Local Tilt Energetics and Bond-Length Coupling}
\label{sec:pes}

The tilt-resolved potential energy surfaces discussed in the main text should be distinguished from finite-temperature free-energy landscapes. In anharmonic halide perovskites, free energy includes both entropic contributions and thermal averaging and is therefore required to determine equilibrium phase stability~\cite{freeE_lanscape_fransson_2023}. Here, by contrast, the potential energy surfaces are used more narrowly as mechanistic probes of the local energetic hierarchy of idealised tilt distortions, without imposing long-range molecular order. They therefore clarify which aspects of the tilt landscape are weakly or strongly constrained locally, but do not by themselves determine the thermodynamically stable finite-temperature phase.

\begin{figure}[htb]
    \centering
    
    \includegraphics[width=0.8\textwidth]{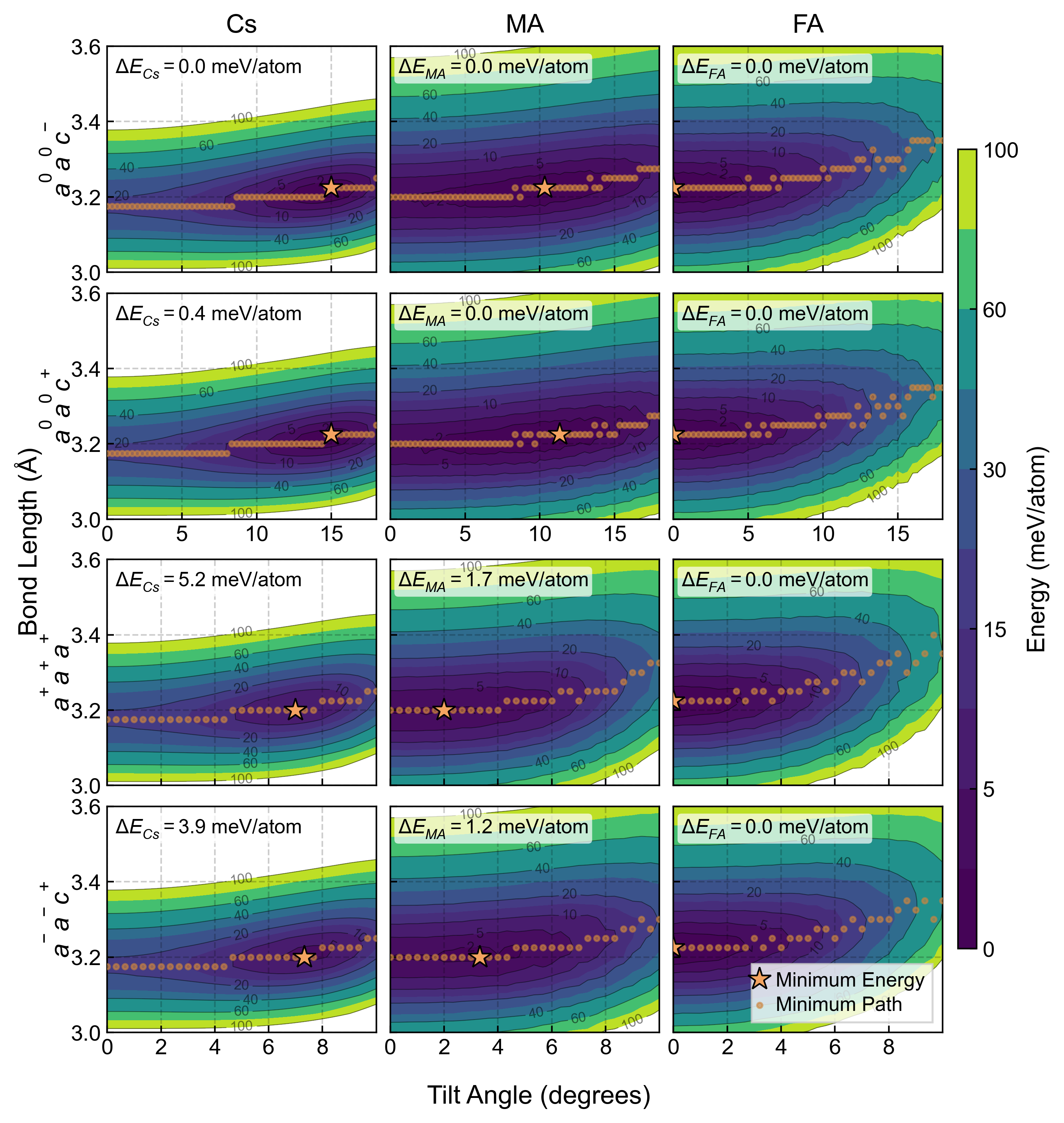}
    \caption{Two-dimensional potential energy surfaces of \ce{CsPbI3}, \ce{MAPbI3}, and \ce{FAPbI3} as functions of tilt angle and bond length for the idealised modes $a^{0}a^{0}c^{-}$, $a^{-}a^{-}c^{+}$, $a^{+}a^{+}a^{+}$, and $a^{0}a^{0}c^{+}$. The plotted surfaces are averaged over 50 independent samples. Stars mark the global minimum on each surface, and circles trace the minimum-energy path with respect to tilt angle, illustrating the coupling between octahedral tilting and bond length relaxation.}
    \label{si:pes}
\end{figure}

The bond-length-dependent potential energy surfaces in Fig.~\ref{si:pes} show that the tilt energetics cannot be interpreted independently of local lattice relaxation. For all three A-site compositions, the minimum-energy path shifts systematically in the two-dimensional space of tilt angle and bond length, indicating strong coupling between octahedral rotation and framework bond length relaxation. A fixed-bond-length cut therefore does not provide a physically complete description of the tilting energetics. For comparison between different tilt modes, the tilt coordinate is defined through the norm of the three octahedral tilt components. This is particularly important for comparing $a^{-}a^{-}c^{+}$ and $a^{+}a^{+}a^{+}$ distortions, because molecular-dynamics trajectories show that in the $a^{-}a^{-}c^{+}$ mode the typical relative amplitudes of the two equivalent in-plane tilts and the out-of-plane tilt are approximately $1:2$. Accordingly, for a variable angle parameter $\theta$, the $a^{-}a^{-}c^{+}$ mode is constructed as $\left[\theta/\sqrt{2},\theta/\sqrt{2},\sqrt{2}\theta\right]$, whereas the $a^{+}a^{+}a^{+}$ mode is written as $[\theta,\theta,\theta]$. With this choice, equal values of the variable angle correspond to equal norms of the three-component tilt vector, allowing the two modes to be compared consistently. From this perspective, \ce{CsPbI3} shows the clearest energetic preference for finite tilting, with minima that are lower than the cubic limit by up to $\sim 5$~meV/atom depending on the mode. It is noteworthy that the displacement of the Cs atom in the structure will favour the $a^{-}a^{-}c^{+}$ tilt mode over the single-tilt modes~\cite{inorg_tilt_prm_2018}. \ce{MAPbI3} is intermediate, with only a small energetic stabilisation of tilted structures, typically below 2~meV/atom. By contrast, \ce{FAPbI3} shows essentially no energetic preference for finite tilting at the local level: the lowest-energy configuration remains at, or extremely close to, the zero-tilt limit, and the $\Delta E<10$~meV/atom region spans a broad range of tilt angles, extending from approximately $0^\circ$ to $6^\circ$ for $a^{-}a^{-}c^{+}$ and $a^{+}a^{+}a^{+}$ modes and to nearly $10^\circ$ for $a^{0}a^{0}c^{-}$ and $a^{0}a^{0}c^{+}$. This broad, shallow basin reinforces the conclusion that in \ce{FAPbI3} the local potential energy surface alone only weakly constrains the tilt amplitude and does not uniquely select the final correlation pattern or tilting magnitude.

\clearpage

\subsection{Simulated Phase Transition Temperature}

The $\alpha$-to-$\gamma$ transition temperature in our simulations ($\sim$200\,K) falls approximately 80\,K below the experimental value ($\sim$280\,K). We attribute this to the following concurring contributions.

The MLFF is trained on r$^2$SCAN energies, forces, and stresses. While r$^2$SCAN represents a significant improvement over PBE for structural and thermodynamic properties~\cite{benchmark_r2scan_kingsbury_prm}, semi-local functionals generically underestimate the free energy barrier between competing structural phases in anharmonic systems because they underestimate the curvature of the potential energy surface near the high-symmetry reference structure. This bias is directly inherited by the MLFF through its training data and depresses the computed transition temperature. Indirect evidence for this in halide perovskites is provided by the observation that MLFF-based simulations trained on comparable DFT data consistently yield transition temperatures below experiment~\cite{pdyna2023liang,mixed_halide_liang_2025_chem-mater,mlff_perovskite_fransson_jpcc_2023}.

A further source of transition temperature depression is the systematic softening of the DFT potential energy surface that occurs during MLFF training, arising from the finite density of training configurations and the implicit regularisation of the model. This effect has been identified across multiple neural-network-potential architectures and leads to an underestimation of effective energy barriers and, consequently, of the temperature at which ordered phases stabilise~\cite{universal_softening_deng,exact_mlff_stefan_natcomm}. The present authors have verified this behaviour independently for Allegro~\cite{allegro_original} and GAP-based~\cite{vaspmlff_original} potentials applied to halide perovskite systems, consistent with the broader evidence that this is an architecture-independent limitation of current MLFF training strategies.

\clearpage

\subsection{Molecular Orientations}

To characterise the orientational distribution of FA, each molecular vector is projected onto spherical polar coordinates $(\phi,\theta)$ and accumulated into a two-dimensional hexagonal histogram. Denoting the normalised population of bin $k$ by

\begin{equation}
p_k=\frac{n_k}{\sum_j n_j},
\end{equation}
we quantify the degree of orientational order through the variance of the normalised bin counts,
\begin{equation}
\sigma_p^2=
\frac{1}{N_{\mathrm{bin}}}
\sum_{k=1}^{N_{\mathrm{bin}}}(p_k-\bar p)^2,
\qquad
\bar p=\frac{1}{N_{\mathrm{bin}}}\sum_{k=1}^{N_{\mathrm{bin}}}p_k.
\end{equation}

A nearly uniform orientational distribution gives a small variance, whereas localisation into preferred directions increases $\sigma_p^2$. In this way, the variance of the normalised hexbin counts provides a scalar order parameter complementary to the explicit pair-correlation analysis in the main text.

\begin{figure}[htb]
    \centering
    
    \includegraphics[width=0.94\textwidth]{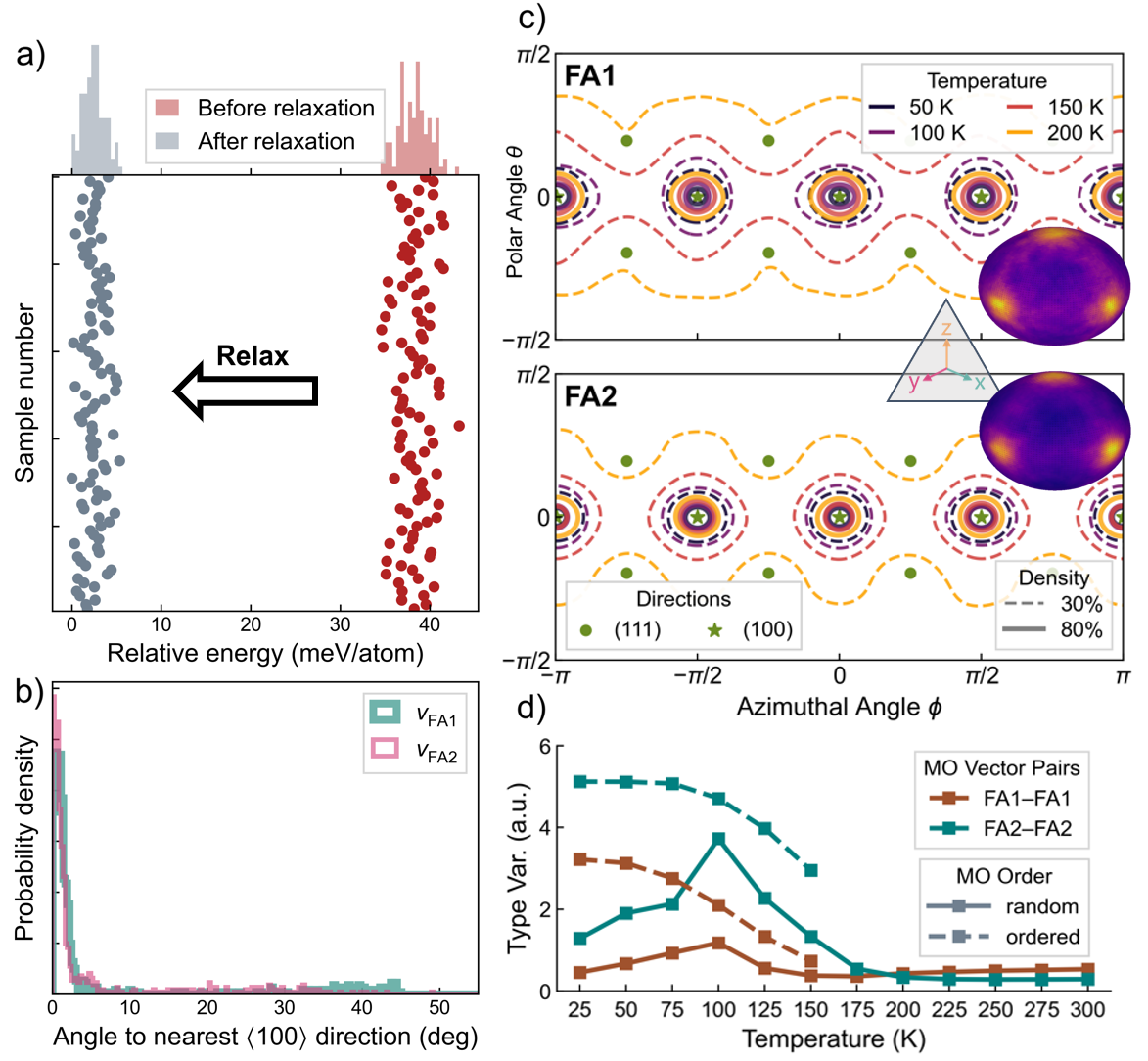}
    \caption{FA molecular orientations in \ce{FAPbI3}. (a) Relative energies of 100 structures with randomly initialised FA orientations before and after FA-only relaxation, referenced to the lowest relaxed configuration. The top histogram shows the corresponding energy distributions. (b) Distribution of the angle between each relaxed molecular vector and the nearest $\langle100\rangle$ direction, obtained by taking the maximum dot product with the six Cartesian $\langle100\rangle$ directions. (c) Polar-projection visualisations of the FA1 and FA2 molecular orientation distributions at different temperatures, with density contours highlighting the preferred $(100)$-type directions. The inset spheres illustrate the corresponding 3D distributions. (d), Temperature dependence of the orientational-order measure defined as the variance of the normalised hexbin counts; larger values indicate stronger localisation of the molecular orientations.}
    \label{si:molecule}
\end{figure}

To analyse orientational correlations beyond the polar-angle distributions, each FA cation is represented by two molecular vectors, $v_{\rm FA1}$ and $v_{\rm FA2}$, corresponding to the two orthogonal molecular axes used in the main-text pair-correlation analysis. Here $v_{\rm FA1}$ is defined as the vector from the midpoint of the two N atoms to the C atom, while $v_{\rm FA2}$ connects the two N atoms. To verify that the $\langle100\rangle$-based orientational analysis is not imposed by construction, we generated 100 independent \ce{FAPbI3} configurations by randomising all FA orientations in the equilibrium cell and then relaxing only the molecular degrees of freedom with the trained MLFF while keeping the inorganic framework fixed ($a^{0}a^{0}a^{0}$ mode). The energies before and after FA-only relaxation both separate into two clear groups (Fig.~\ref{si:molecule}a), indicating distinct molecular arrangements within the same fixed framework. More importantly, after relaxation both $v_{\rm FA1}$ and $v_{\rm FA2}$ are strongly concentrated near the symmetry-folded $\langle100\rangle$ direction (Fig.~\ref{si:molecule}b). This shows that even when initialised from fully random orientations, the FA molecules relax back into the same preferred orientational basin. The $(100)$-restricted initialisation used in the production MD therefore does not impose an artificial ordering pattern, but reflects the orientational manifold selected naturally by the dynamics.

Figure~\ref{si:molecule}c,d further shows the polar-projection distributions of the FA1 and FA2 molecular vectors as a function of temperature, together with the corresponding orientational-order measure. The highest-density regions remain centred on the $\langle100\rangle$ directions, while the distribution broadens substantially on heating. The temperature dependence of the variance confirms that orientational localisation becomes strongest near the low-temperature ordered regime and weakens again towards the high-temperature dynamically disordered phase.

For the nearest-neighbour pair-correlation analysis, a random benchmark was constructed by averaging over structures with fully randomised molecular orientations. Relative to this benchmark, the ordered $\gamma$ phase develops a symmetry-selected anisotropic molecular-correlation pattern: for a pair separated along one axis $\alpha$, FA1--FA1 favours transverse $\pm\beta_{1}$ relations, while FA2--FA2 favours transverse $\pm\beta_{2}$ relations, where $\alpha$, $\beta_{1}$, and $\beta_{2}$ denote the three principal directions as illustrated in Fig.~\ref{res:mo}a. This directional pattern is commensurate with long-range $a^{+}a^{+}a^{+}$ tilt coherence and is the molecular-ordering topology that fails to develop globally in the low-temperature $\gamma'$ state.

The molecular reorientation time is extracted from the temporal autocorrelation of the molecular orientation vectors, as in the \textsc{PDynA} formalism,
\begin{equation}
A_{\mathrm{MO}}(t)=
\left\langle
\mathbf{v}(t_0;\mathbf{n})\cdot \mathbf{v}(t_0+t;\mathbf{n})
\right\rangle_{t_0,\mathbf{n}},
\end{equation}
which is fitted to the same biexponential form of eq.~\ref{eq:autocorr_biexp}. Following the original implementation, $\tau_1$ is taken as the characteristic reorientation time because it dominates the decay, while $\tau_2$ accounts for the fast initial correction term.

\begin{figure}[htb]
    \centering
    
    \includegraphics[width=0.61\textwidth]{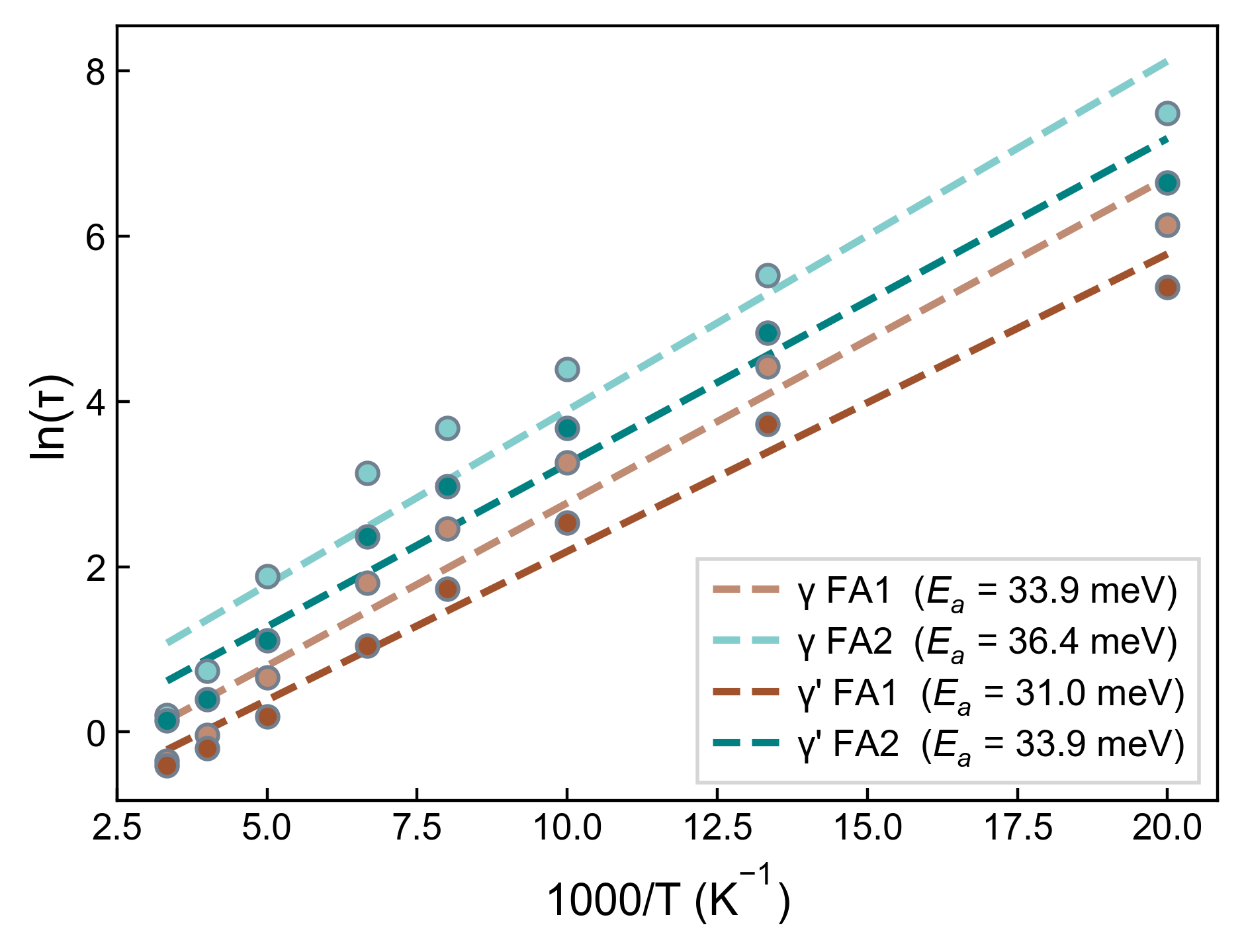}
    \caption{Arrhenius analysis of the FA reorientation times. Characteristic reorientation times extracted from the molecular-orientation autocorrelation functions are plotted as $\ln\tau_r$ versus $1000/T$, and the slope of the linear fit yields the effective activation energy $E_{\mathrm a}$. Separate fits are shown for FA1 and FA2 in the ordered $\gamma$ and twinned $\gamma'$ states.}
    \label{si:barrier}
\end{figure}

Assuming activated rotational dynamics, the temperature dependence of the fitted reorientation time is described by
\begin{equation}
\tau_r(T)=\tau_0 \exp\!\left(\frac{E_{\mathrm a}}{k_{\mathrm B}T}\right),
\end{equation}
or equivalently
\begin{equation}
\ln\tau_r=
\ln\tau_0+\frac{E_{\mathrm a}}{k_{\mathrm B}}\frac{1}{T}.
\end{equation}
A linear fit to $\ln\tau_r$ versus $1/T$ thus yields the effective activation energy $E_{\mathrm a}$ from the slope. We emphasise that the resulting $E_{\mathrm a}$ should be interpreted as a coarse-grained kinetic barrier associated with molecular reorientation in the fluctuating framework environment, rather than as a unique microscopic rotation barrier, because the motion is coupled to local lattice distortions and intermolecular correlations. In practice, the fit is restricted to the temperature window where approximately Arrhenius behaviour is observed. The resulting barriers obtained here are consistent with previously reported values for MD in hybrid lead-halide perovskites. For example, quasielastic neutron-scattering and spectroscopic studies have reported activation energies of order $\sim20$–30~meV for dipole reorientation of FA molecules around their principal axis in hybrid perovskites, while larger barriers of order $40$–100~meV have been associated with more complete molecular rotations involving stronger coupling to the inorganic framework~\cite{dynamics_molecule_iodide_fabini,rotational_dynamics_fapi_jpcl_2023}. The values extracted in the present work therefore fall within the range expected for thermally activated molecular reorientation in these materials. Additionally, the simulated values suggest that, at the same temperature, the rotation of molecules is favoured around the FA1 axis as well as in the twinned structure, which is consistent with the intuition from structural geometry.  

\clearpage

\subsection{Artificial Planar Disorder Domain and Energetic Competition}

Unless otherwise stated, the ordered $\gamma$ and heterogeneous $\gamma'$ structures used in the main text analyses were not imposed by hand, but selected from the large ensemble of cooling and equilibration MD trajectories described in the Methods. These trajectories generate many representative low-temperature configurations depending on temperature history. Structures labelled as ordered $\gamma$ were taken from trajectories that developed long-range $a^{+}a^{+}a^{+}$ coherence along all three crystallographic directions, whereas structures labelled as $\gamma'$ were taken from trajectories that remained locally $\gamma$-like but developed reduced long-range coherence and heterogeneous low-temperature textures. 

\begin{figure}[htb]
    \centering
    
    \includegraphics[width=0.80\textwidth]{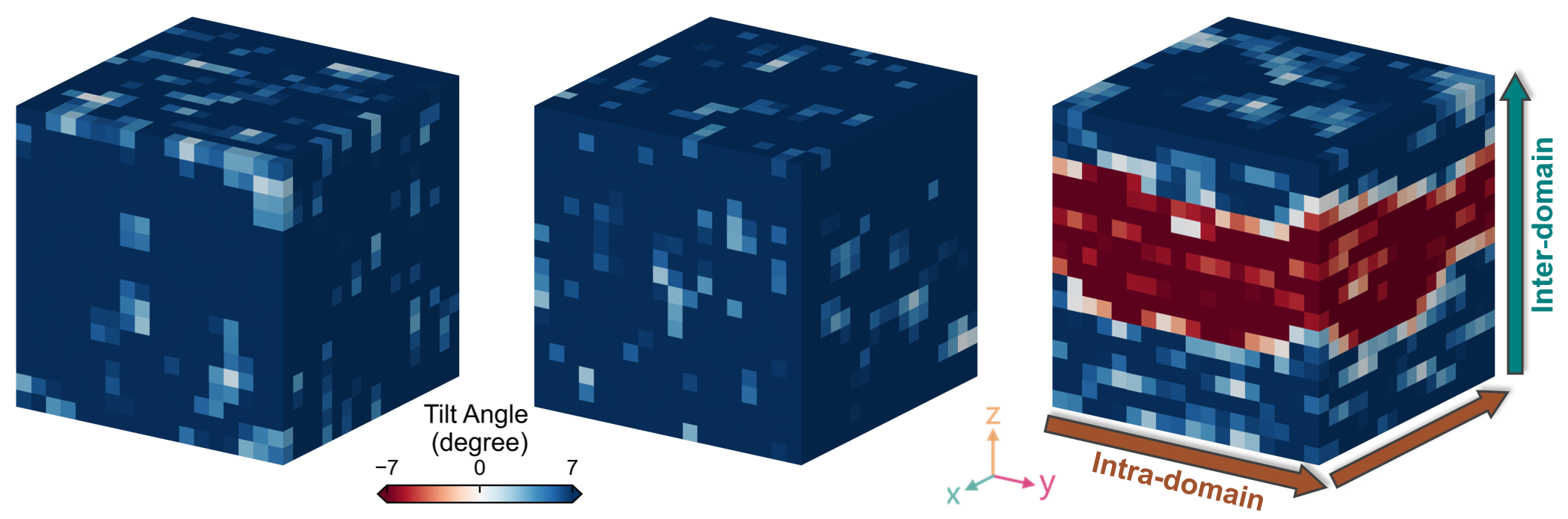}
    \caption{Three-dimensional tilt map of an instantaneous frame of \ce{FAPbI3} at 100~K containing an artificial planar twin boundary perpendicular to $z$. Tilt coherence is preserved within the domain planes but interrupted across the boundary, providing a control structure for identifying the reciprocal-space signature of a single twin interface.}
    \label{si:3Dvis}
\end{figure}

By contrast, an artificial structure was constructed to isolate the scattering signature of a single twin boundary. Starting from a perfect $a^{+}a^{+}a^{+}$ supercell, half of the octahedra in a central slab were shifted by one pseudocubic unit cell along the $y$ direction, thereby creating a planar discontinuity in the tilt pattern. This operation preserves coherent $x$- and $y$-directed tilts within the plane while introducing a twinned $z$-tilt across the slip planes. The structure was subsequently equilibrated at 100~K, where it remains metastable on the simulation timescale and can therefore be used as a clean control for the reciprocal-space signature of a single boundary. In Fig.~\ref{si:3Dvis}, the intra-domain directions (parallel to the boundary plane) and the inter-domain direction (perpendicular to the boundary) are explicitly illustrated, which provides a visual reference for the directional anisotropy of the resulting scattering features discussed in the main text (Fig.~\ref{res:dsf}).

Because the discontinuity is encountered only when crossing the boundary, the reciprocal-space response is strongly anisotropic. The $HK$ half-integer plane remains essentially identical to the ordered bulk $\gamma$ phase, whereas the $HL$ and $KL$ planes become sensitive to the boundary and develop the satellite-like diffuse features discussed in the main text.

\begin{figure}[bht]
    \centering
    
    \includegraphics[width=0.50\textwidth]{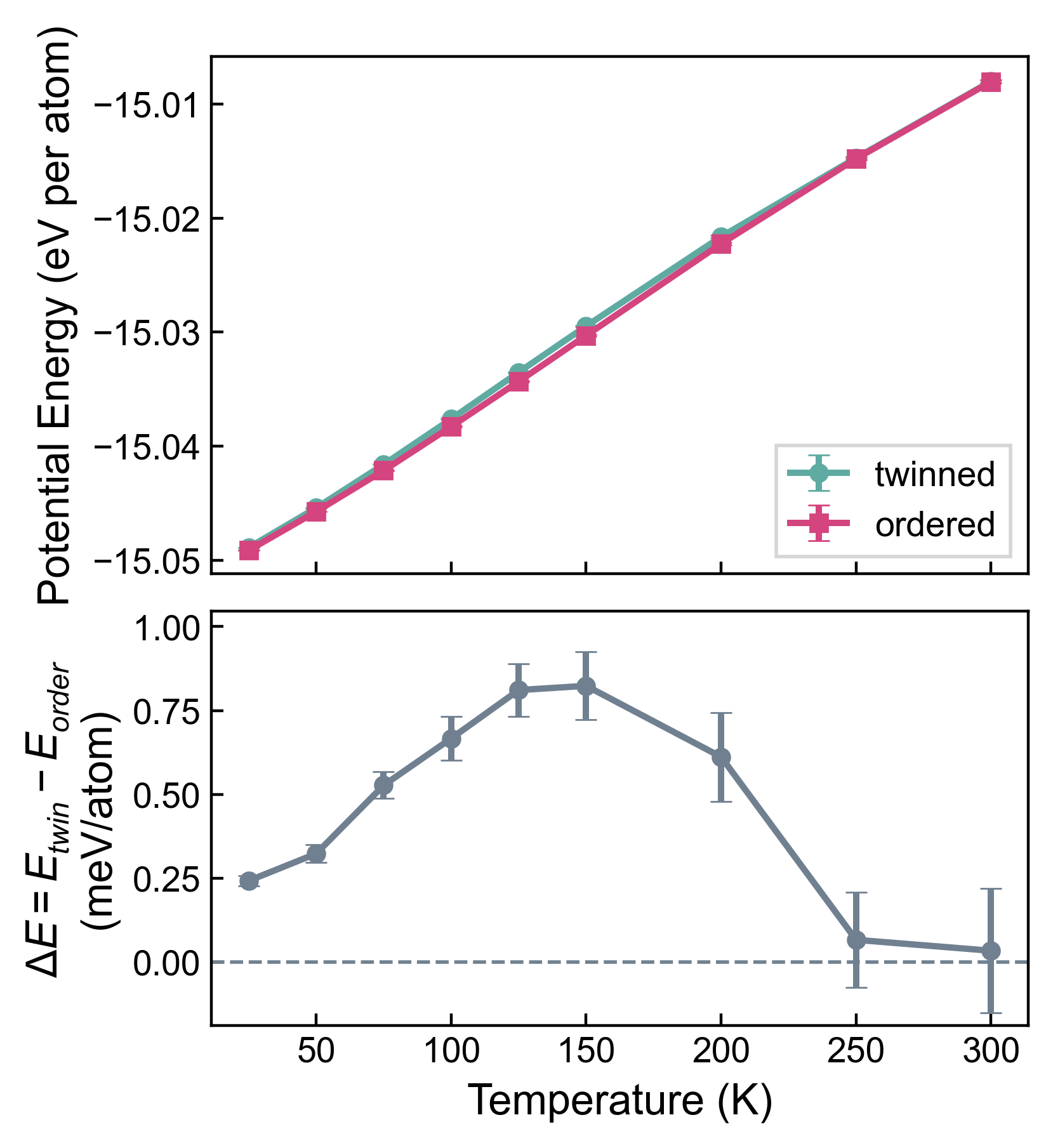}
    \caption{Potential energy per atom of the ordered $\gamma$ and twinned $\gamma'$ states as a function of temperature. Top panel: absolute potential energies. Bottom panel: energy difference $\Delta E=E_{\mathrm{twin}}-E_{\mathrm{order}}$. The ordered $\gamma$ state is always the energetic minimum, whereas the twinned $\gamma'$ state becomes nearly degenerate at high temperature, consistent with kinetic arrest rather than energetic stabilisation.}
    \label{si:epot}
\end{figure}

The energetic competition between the ordered $\gamma$ phase and the twinned $\gamma'$ state is summarised in Fig.~\ref{si:epot}. The ordered $\gamma$ structure is always lower in potential energy, confirming that it is the local energetic minimum. However, the energy difference is largest at $\gamma$ phase temperature and decreases strongly on heating, approaching zero within uncertainty at high temperature. This behaviour supports the interpretation advanced in the main text: the $\gamma'$ state is not the equilibrium minimum but a kinetically arrested metastable state, stabilised dynamically by the freezing of the molecular and tilt rearrangements rather than by a lower static energy~\cite{entropy_fapi_sciadv_2016}. The near-degeneracy at high temperature also explains why multiple low-temperature topologies can become accessible depending on the thermal history.

\clearpage

\subsection{Space Group Determination of the \texorpdfstring{$\gamma$}{gamma} Phase}

The low-temperature $\gamma$ phase of \ce{FAPbI3} has been assigned in the literature to multiple tilt systems, including the lower-symmetry tetragonal pattern $a^{0}a^{0}c^{+}$. Here we confirm experimentally that the $\gamma$ phase is best described by the $a^{+}a^{+}a^{+}$ tilt pattern (space group Im$\bar{3}$) by comparing the measured X-ray scattering with structure factors calculated for candidate tilt systems.

\begin{figure}[b]
    \centering
    \includegraphics[width=0.9\textwidth]{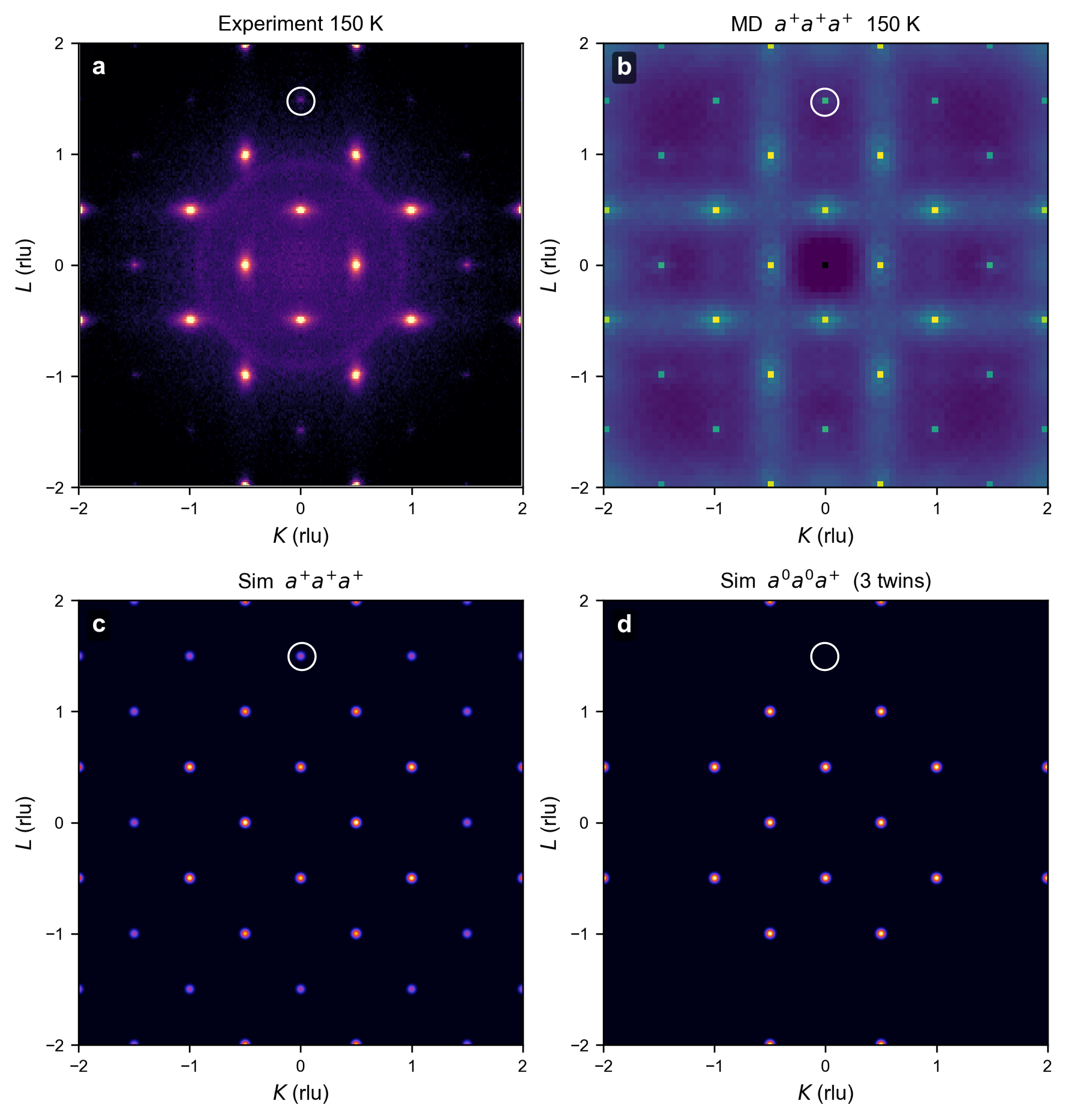}
    \caption{Space group determination of the $\gamma$ phase of \ce{FAPbI3}. (a) Experimental X-ray scattering pattern on the $HK1.5$ reciprocal-space plane at 150~K. White circles highlight superstructure reflections that are uniquely allowed by the $a^{+}a^{+}a^{+}$ tilt pattern. (b) Corresponding diffraction pattern computed from MD trajectories independently identified as $a^{+}a^{+}a^{+}$ in real space, reproducing the same superstructure reflections. (c) Structure factor simulation from CIF model with $a^{+}a^{+}a^{+}$ symmetry. (d) Structure factor simulation from an $a^{0}a^{0}c^{+}$ model with three symmetry-equivalent twin orientations superimposed. All diffraction patterns show $HK1.5$ reciprocal-space planes. The diffraction ring in (a) originates from ice formation on the crystal during the measurement.  }
    \label{si:sg_aaa}
\end{figure}

Figure~\ref{si:sg_aaa}a shows the experimental X-ray scattering pattern on the $HK1.5$ reciprocal-space plane measured at 150~K. This plane was chosen because it contains specific superstructure reflections, highlighted by white circles, that are uniquely allowed by the $a^{+}a^{+}a^{+}$ tilt pattern and forbidden in lower-symmetry tilt systems. The corresponding pattern obtained from our large-scale MD trajectories, which were independently identified as $a^{+}a^{+}a^{+}$ from the real-space tilt analysis, reproduces these characteristic superstructure reflections (Fig.~\ref{si:sg_aaa}b).

To further isolate the structural origin of these reflections, we computed structure factors directly from idealised CIF models for two candidate tilt patterns. For the $a^{+}a^{+}a^{+}$ model, Cs was placed on the A-site as a computationally convenient proxy, since the FA molecule contributes only weakly to the structure factor relative to Pb and I. The Cs substitution therefore introduces only a small perturbation to the diagnostic superstructure intensities. The resulting $a^{+}a^{+}a^{+}$ pattern (Fig.~\ref{si:sg_aaa}c) clearly reproduces the reflections highlighted in the experimental data.

By contrast, the $a^{0}a^{0}c^{+}$ model, which has previously been used to describe related FA lead-halide phases, does not generate these reflections even when three symmetry-equivalent twin orientations are superimposed to mimic the twinning scenario encountered in \ce{MAPbI3} (Fig.~\ref{si:sg_aaa}d). The diagnostic superstructure peaks are forbidden in each individual twin and therefore remain absent in their superposition.

\begin{figure}[b]
    \centering
    \includegraphics[width=0.9\textwidth]{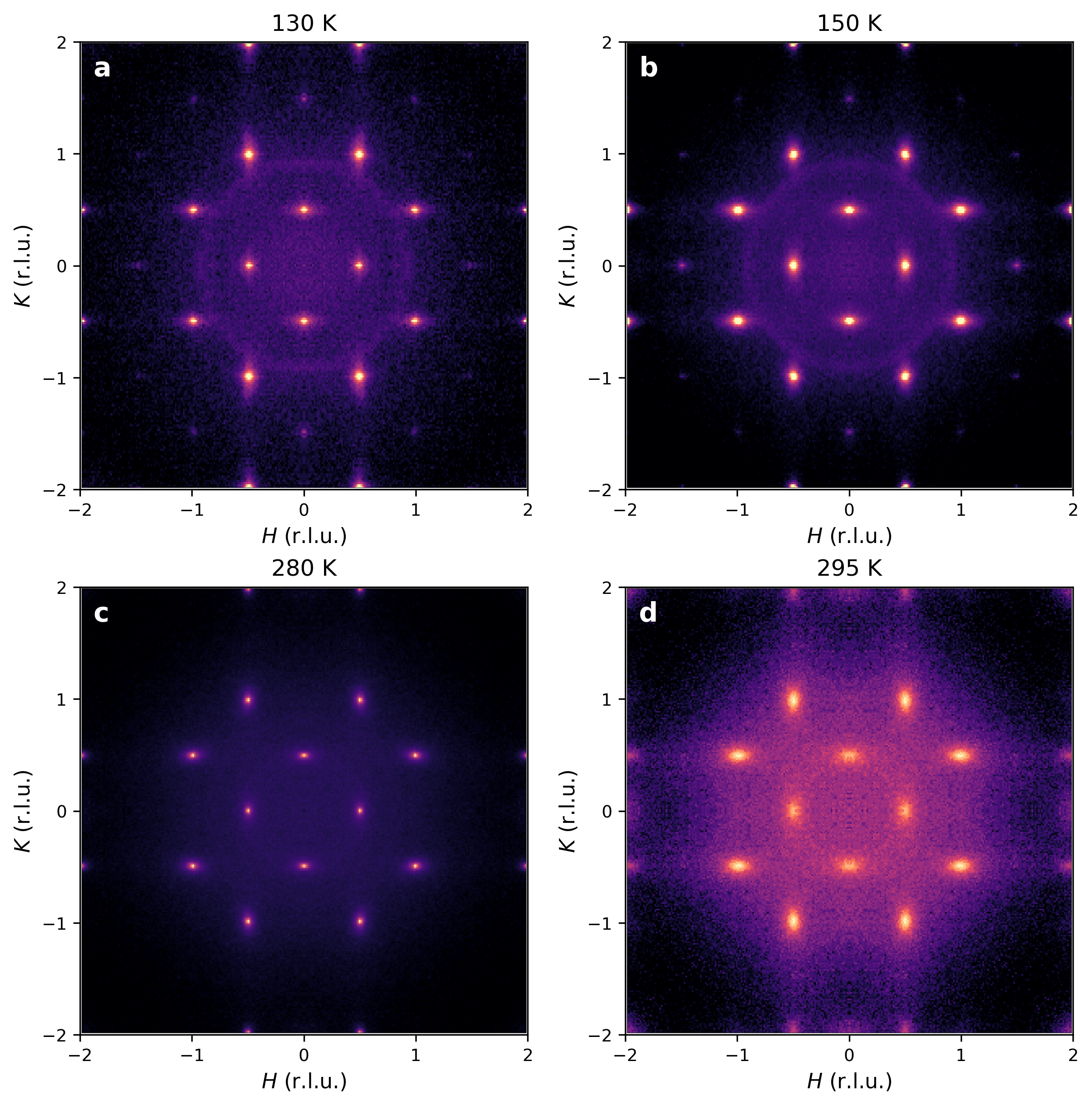}
    \caption{Temperature dependence of the experimental X-ray scattering of \ce{FAPbI3} on the $HK1.5$ reciprocal-space plane. (a)~130~K and (b)~150~K, in the $\gamma$ phase, where the characteristic $a^{+}a^{+}a^{+}$ superstructure reflections are present at the half-integer positions highlighted in Fig.~\ref{si:sg_aaa}a. (c)~280~K and (d)~295~K, above the ordering transition, where these superstructure reflections are no longer present and the pattern is instead dominated by diffuse scattering around the Bragg positions, characteristic of the dynamically disordered $\alpha$ phase.}
    \label{si:sg_Tseries}
\end{figure}

Taken together, the experimental observation of these reflections, their reproduction by MD-based diffraction simulations, and their absence in the twinned $a^{0}a^{0}c^{+}$ model provide a consistent and specific structural fingerprint that identifies the $\gamma$ phase of \ce{FAPbI3} as $a^{+}a^{+}a^{+}$.

The temperature dependence of the experimental scattering on this plane further supports the $a^{+}a^{+}a^{+}$ assignment (Fig.~\ref{si:sg_Tseries}). At 130~K and 150~K, the superstructure reflections are clearly visible at the half-integer positions highlighted in Fig.~\ref{si:sg_aaa}a, consistent with long-range $a^{+}a^{+}a^{+}$ tilt coherence within the measured volume. At 280~K and 295~K, these superstructure reflections are no longer observed, indicating that the static in-phase tilt order characteristic of the $\gamma$ phase is absent. Instead, the higher-temperature patterns retain pronounced diffuse scattering around the Bragg positions and near the zone boundary, reflecting the dynamic local tilt correlations of the $\alpha$-like regime rather than a frozen long-range tilt pattern. 

\clearpage

\subsection{Diffuse Scattering and Reciprocal Space Analysis}

The reciprocal-space observables in this work were computed from MD trajectories using the previously established real-space-to-scattering workflow for perovskites~\cite{MA_local_order_2023_toney}, which was subsequently tested successfully in related hybrid lead-halide systems~\cite{dynamic_domian_milos}. In this approach, the instantaneous atomic configurations from MD are converted into reciprocal-space intensities by evaluating the corresponding structure-factor expressions over the sampled trajectory, yielding both energy-integrated diffuse scattering and the dynamical structure factor $S(\mathbf{q},\omega)$. The full formalism is described in those earlier works; here we use the same workflow and focus on the specific reciprocal-space sampling, averaging, and correlation-length extraction procedures adopted for \ce{FAPbI3}.

For the energy-resolved reciprocal-space maps, the raw simulated $S(\mathbf{q},\omega)$ was first evaluated on a regular reciprocal-space grid spanning 0--5~r.l.u. along each reciprocal direction. Each sampled $\mathbf{q}$ point was then folded into the first Brillouin zone by reducing its coordinates modulo 1, and intensities contributing to the same reduced reciprocal-space bin were averaged to obtain a folded spectrum on a regular reduced-zone grid. For cubic structures, an additional averaging was performed over bins related by permutation of the $x$, $y$, and $z$ reciprocal axes. The folded spectrum was then sampled along the reduced-zone high-symmetry path $\Gamma\rightarrow M\rightarrow R\rightarrow \Gamma\rightarrow X\rightarrow M$, sorted along the energy axis, normalised by its maximum intensity, and plotted as the folded dynamical structure factor.

To compare equivalent diffuse features around the zone-boundary $M$ points, one-dimensional line profiles were extracted from the energy-integrated scattering intensity $S(\mathbf{q})$ on a cubic reciprocal-space grid. For the cubic case analysed here, we construct the 12 symmetry-equivalent line cuts
\[
(H,2.5,1.5), \qquad (2.5,K,1.5), \qquad (2.5,1.5,L),
\]
with $H,K,L \in \{0,1,2,3\}$, corresponding to scans through all symmetry-equivalent $M$-point environments in the chosen half-integer planes. For each cut, the intensity is averaged over a small transverse window in the orthogonal directions, cropped to a window around the nominal $M$-point region, and normalised to unit peak height. Because the reciprocal-space grid spacing of the MD data is limited to $1/18$ r.l.u., the individual line profiles are then linearly interpolated onto a common grid before averaging, which improves the numerical stability of the profile comparison and full width at half maximum (FWHM) extraction but does not remove the underlying resolution limit of the original data.

\begin{figure}[htb]
    \centering
    
    \includegraphics[width=0.7\textwidth]{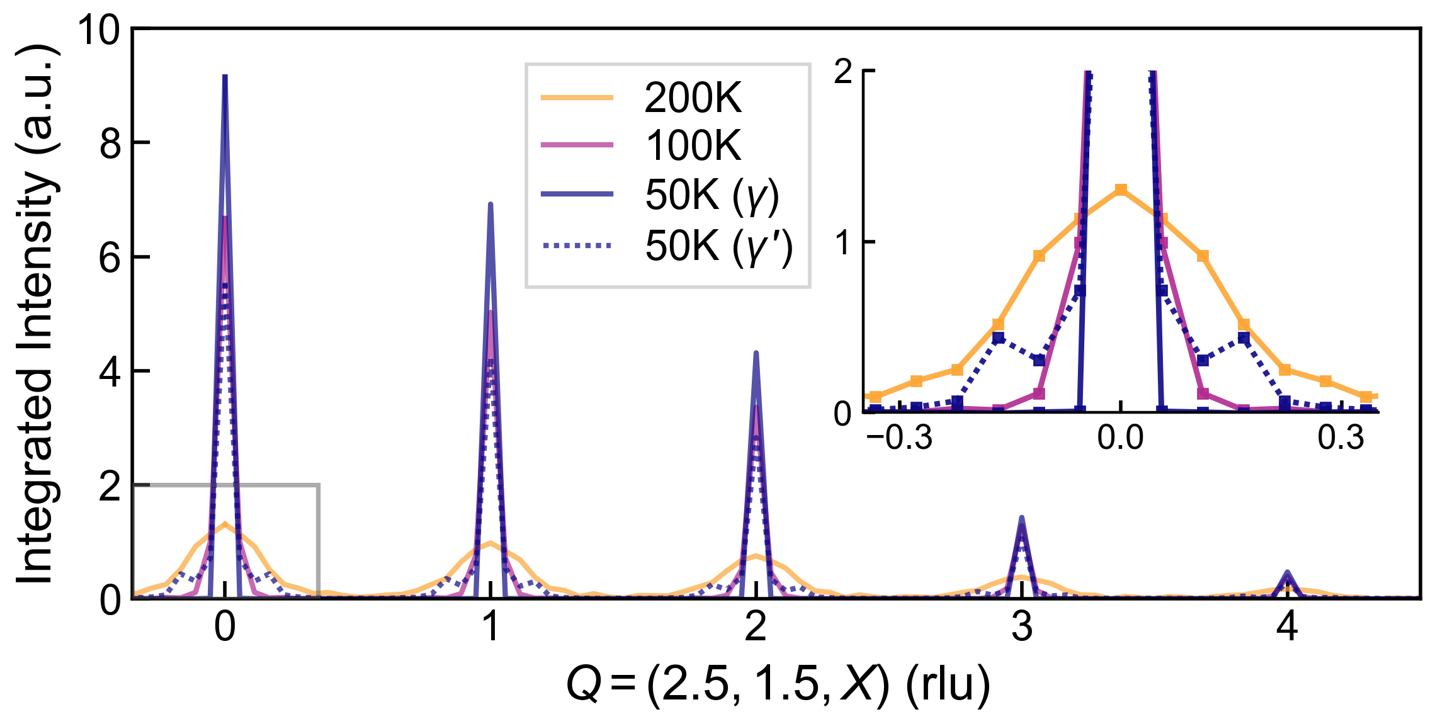}
    \caption{Energy-integrated X-ray scattering intensity $S(q)$ extracted along the reciprocal-space line profile $(2.5,1.5,X)$. Ordered and disordered structures are shown with solid and dashed lines, respectively. The inset highlights the near-$M$ line shape, including the satellite-like intensity characteristic of the twinned $\gamma'$ state.}
    \label{si:diffuse_profile}
\end{figure}

The reciprocal-space correlation length is extracted from the FWHM of this averaged profile, following the FWHM-based procedure used in the diffuse scattering analysis of related FA-based lead-halide perovskites~\cite{dynamic_domian_milos}. If $q_{\mathrm L}$ and $q_{\mathrm R}$ denote the left and right half-maximum crossing points of the peak, then
\begin{equation}
\mathrm{FWHM}=q_{\mathrm R}-q_{\mathrm L},
\end{equation}
and the associated reciprocal-space correlation length is estimated as
\begin{equation}
\xi_{\mathrm{rec}}=\frac{2a_{\mathrm{pc}}}{\pi\,\mathrm{FWHM}},
\end{equation}
where $a_{\mathrm{pc}}$ is the pseudocubic lattice constant. This expression corresponds to the standard Lorentzian/FWHM estimate used to connect reciprocal-space peak widths to a real-space correlation length.

\begin{figure}[htb]
    \centering
    
    \includegraphics[width=0.9\textwidth]{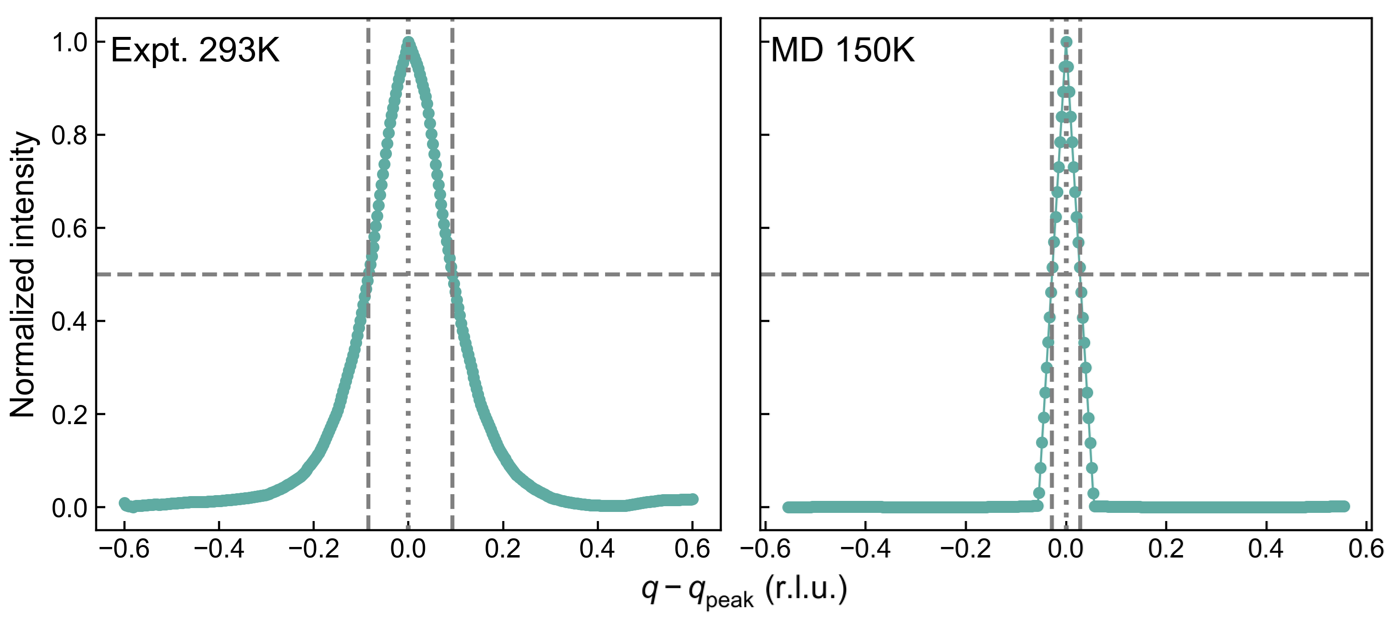}
    \caption{Examples of FWHM-based correlation length extraction from near-$M$ line profiles. Left: experimental 293~K data, for which the broadened single peak is well resolved. Right: MD 150~K data, where the peak width is close to the reciprocal-space resolution limit. Dashed vertical lines mark the half-maximum crossings.}
    \label{si:fwhm}
\end{figure}

Figure~\ref{si:fwhm} illustrates the fitting procedure. The experimental 293~K profile is well described by a single broadened peak, so the FWHM extraction is robust. In contrast, the MD 150~K ordered $\gamma$ profile is already so sharp that its width approaches the reciprocal-space grid spacing, making the fitted width effectively resolution-limited even after linear interpolation. For the low-temperature twinned $\gamma'$ state, the problem is more severe: the presence of satellite intensity means that the near-$M$ profile is no longer well described by a single peak at all. In such cases, the FWHM-based correlation length becomes intrinsically uncertain and should be interpreted only as a rough reciprocal-space metric rather than a rigorous single-domain correlation length. For this reason, the reciprocal-space correlation lengths compared with the real-space values in the main text should be interpreted as approximate metrics that capture the same overall trend, rather than as strictly equivalent quantities on a point-by-point basis.

\begin{figure}[htb]
    \centering
    
    \includegraphics[width=0.94\textwidth]{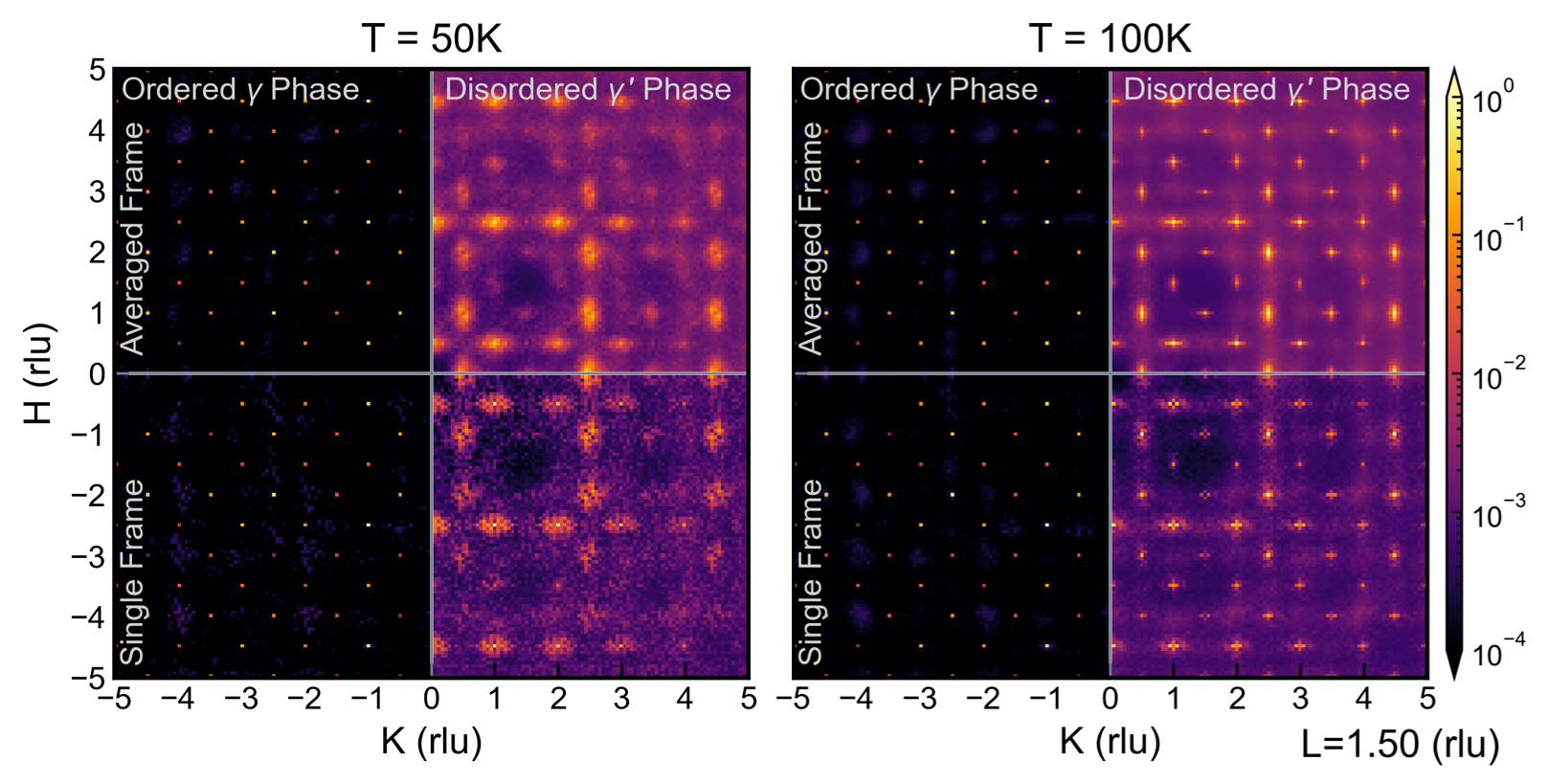}
    \caption{X-ray diffuse scattering of \ce{FAPbI3} on the half-integer plane $L=1.5$ r.l.u. at 50 and 100~K. The top row shows patterns averaged over four independent calculations and three symmetry-equivalent planes ($HK1.5$, $HL1.5$, and $KL1.5$); the bottom row shows one representative contribution from this 12-pattern average. The left and right columns correspond to the ordered $\gamma$ phase and the twinned $\gamma'$ state, respectively.}
    \label{si:diffuse}
\end{figure}

To account for the strong configurational heterogeneity of the low-temperature twinned $\gamma'$ state, the diffuse patterns shown in Fig.~\ref{si:diffuse} (and Fig.~\ref{res:dsf}a) were averaged over 12 independent MD calculations. In practice, this average combines four independent trajectories and the three symmetry-equivalent half-integer planes $HK1.5$, $HL1.5$, and $KL1.5$. The independence of these calculations was established through randomisation of both the A-site molecular orientations and the initial velocities. In each initial configuration, the FA1 vector of every molecule was assigned randomly to one of the six allowed $\langle100\rangle$ directions, and the corresponding FA2 vector was then assigned randomly to one of the two remaining perpendicular $\langle100\rangle$ directions, ensuring the geometrical constraint that FA1 and FA2 remain orthogonal. On top of this orientational randomisation, distinct initial velocity seeds were used in the MD runs to avoid generating similar initial octahedral configurations. This averaging is particularly important for the twinned $\gamma'$ state, whose reciprocal-space response varies substantially between independent realisations, whereas the bulk $\gamma$ and $\alpha$ phases show much less configurational variation and therefore do not require the same level of averaging.

In addition to the energy-integrated diffuse scattering, the low-energy neutron response was also compared between structures initialised from ordered $\gamma$ and disordered $\gamma'$ states. The twinned $\gamma'$ configurations consistently retain broader low-energy spectral features than the ordered $\gamma$ phase, even at the same nominal temperature, whereas on heating both structures towards the high-temperature regime their responses become progressively more similar. This supports the interpretation that the broadening observed in the low-temperature $\gamma'$ state arises from arrested structural heterogeneity rather than from thermal disorder alone.

\clearpage

\subsection{Electronic Disorder, Structural Featurisation, and Urbach Energy}
\label{sec:elec}

Additional details supporting the electronic-disorder analysis are provided in the figures below. Fig.~\ref{si:urbach_fit} shows an example of the Urbach-tail fitting procedure for the 50~K $\gamma'$ structure, illustrating the fitted low-energy tails near the valence and conduction band edges. In the main text, the Urbach energy is extracted from the statistical distribution of band-edge energies obtained from DFT calculations on subdivided $2\times2\times2$ subcells. For each subcell, the electronic band structure was computed, and the valence band maximum (VBM) and conduction band minimum (CBM) were identified. The sampled band-edge energies were then converted into normalised histogram-based density-of-states representations, and the low-energy tails were fitted to extract $E_U^{\mathrm{VBM}}$ and $E_U^{\mathrm{CBM}}$, from which the effective optical Urbach energy $E_U^{\mathrm{opt}}$ was obtained. Because these calculations were performed at the semilocal DFT level, the absolute gap values should be interpreted cautiously; the focus is on the relative variation of local band-edge energies across distinct structural environments rather than on quantitative prediction of the experimental bulk gap.

\begin{figure}[htb]
    \centering
    
    \includegraphics[width=0.5\textwidth]{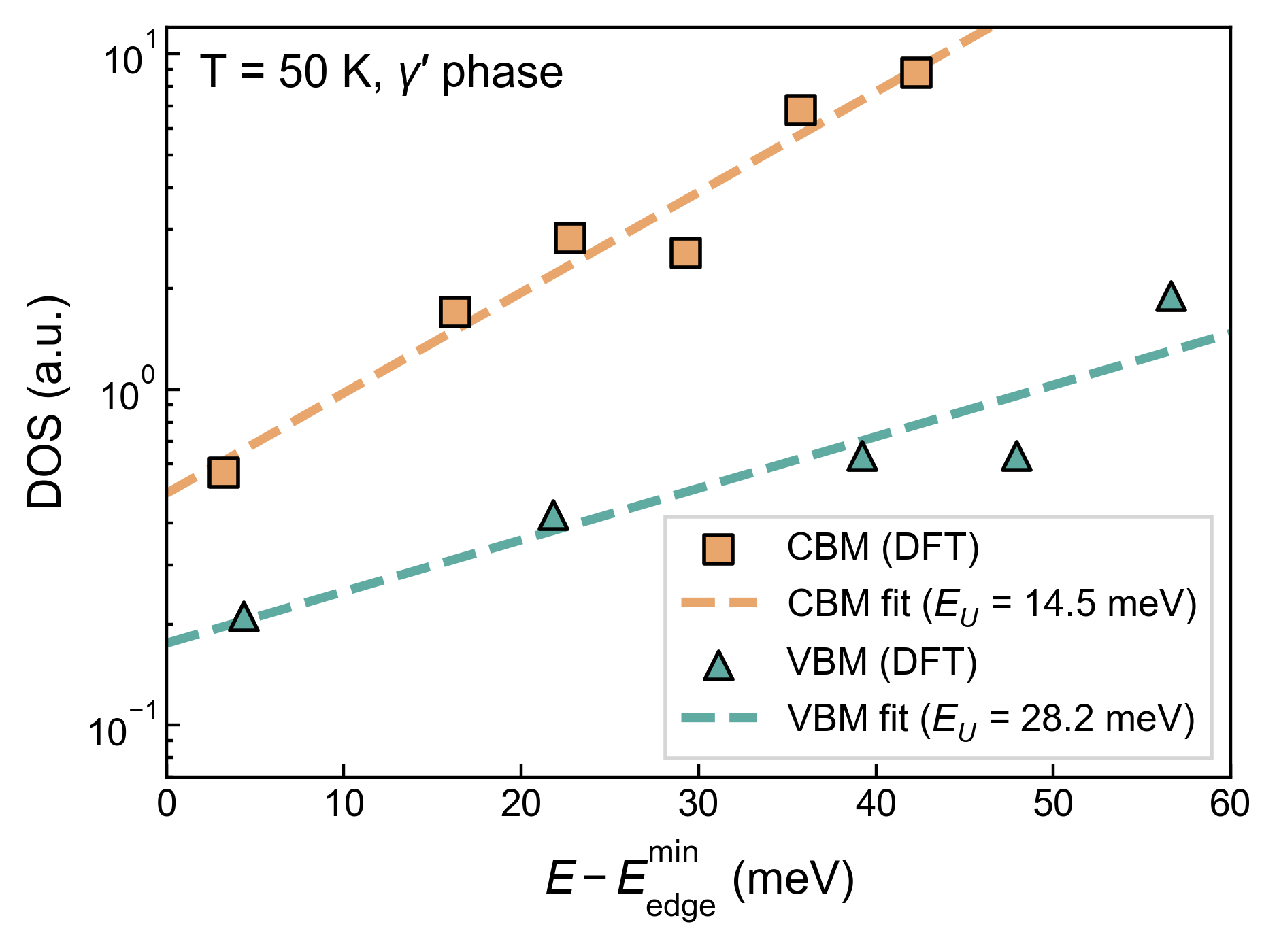}
    \caption{Example Urbach-tail fitting for the 50~K $\gamma'$ structure. Scatter points show the histogram-derived DOS near the valence and conduction band edges, and dashed lines show the exponential fits used to extract $E_U^{\mathrm{VBM}}$ and $E_U^{\mathrm{CBM}}$.}
    \label{si:urbach_fit}
\end{figure}

\begin{figure}[htb]
    \centering
    
    \includegraphics[width=0.98\textwidth]{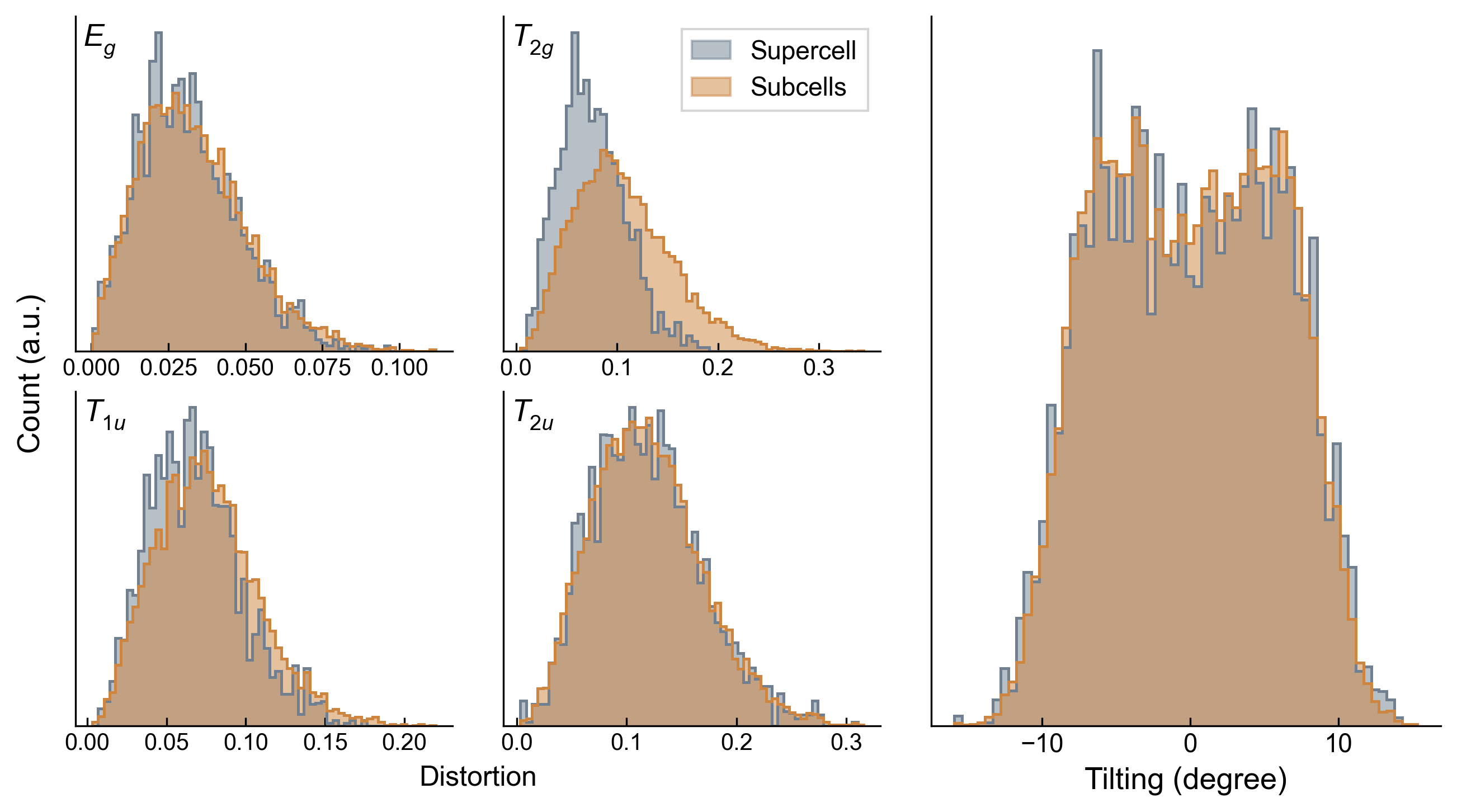}
    \caption{Comparison of structural descriptor distributions evaluated in the full MD supercell and in the extracted $2\times2\times2$ subcells used for the electronic analysis, from the 50~K $\gamma'$ structure. The right panel shows that the tilting distribution is largely unaffected by the subcell treatment. The left panels show that the octahedral distortion distributions are only modestly modified, with slight upward shifts most visible in the $T_{2g}$ and $T_{1u}$ modes, reflecting the periodic-boundary constraint imposed on the smaller cells.}
    \label{si:subcell_benchmark}
\end{figure}

Because the electronic analysis is carried out on periodically repeated $2\times2\times2$ subcells extracted from the large MD supercell, it is useful to assess how this subcell treatment modifies the local structural distributions. Figure~\ref{si:subcell_benchmark} compares the distributions of octahedral distortion and tilt descriptors evaluated over the full supercell and over the corresponding extracted subcells. The tilting distribution is largely preserved by the subcell construction, indicating that the local rotational landscape relevant to the electronic analysis is not perturbed. By contrast, the distortion distributions are shifted slightly towards larger values in the $T_{2g}$ and $T_{1u}$ modes after subcell treatment. We attribute this to the periodic boundary constraint imposed on the smaller cells, which introduces a modest additional incompatibility in the local distortion field while leaving the tilt statistics largely unchanged.

Across the full dataset, the local band gap spans approximately 1.65--1.90~eV. The lower end of this distribution overlaps the experimentally reported band gap range~\cite{dewolf_absorption_jpcl_2014,fapi_science_seok_2019,compositional_engineering_Nature_jeon} of black-phase \ce{FAPbI3}, whereas the computed values are systematically elevated overall. This offset is expected from the approximate electronic-structure treatment and should not be interpreted as a quantitative prediction of the bulk optical gap. This offset is expected from the approximate electronic-structure treatment. In lead iodide perovskites, spin-orbit coupling lowers the Pb-derived conduction-band minimum substantially, typically by several tenths of an electronvolt, and this effect is not included here~\cite{hhp_soc_effect_even}. In addition, the use of periodically repeated $2\times2\times2$ subcells can modestly amplify local distortions relative to the parent MD supercell, as discussed below.

This analysis is based on one representative equilibrated MD snapshot at each temperature, subdivided into several hundred $2\times2\times2$ subcells. This choice is justified by the fact that, at fixed temperature, the octahedral-tilt distributions are statistically stationary across equilibrated snapshots of the same trajectory. Although the instantaneous tilt of any given octahedron fluctuates in time, the ensemble distribution of local tilts is unchanged within statistical uncertainty from one representative frame to another. Because the electronic analysis is performed over the full set of subcells extracted from a large supercell, the resulting distributions of local band gaps and band-edge energies are therefore expected to be robust with respect to the particular frame selected. In this sense, the dominant source of variation is the equilibrium structural distribution at a given temperature rather than the choice of one specific equilibrated snapshot.

The resulting electronic bands were unfolded onto the primitive Brillouin zone to enable direct comparison across the distinct local environments generated by the MD trajectories. Each extracted $2\times2\times2$ subcell contains eight \ce{PbI6} octahedra and therefore 24 local tilt components, so a reduced structural representation is required to relate the electronic variation to the local geometry in a physically interpretable way. We therefore tested multiple structural featurisations based on both octahedral tilts and distortions, and compared their correlations with the computed local band gap. As discussed below, the strongest correlation is obtained for the octahedron-averaged tilt norm, defined by first taking the norm of the three tilt components for each octahedron and then averaging this scalar quantity over all octahedra in the subcell.

\begin{figure}[htb]
    \centering
    
    \includegraphics[width=0.98\textwidth]{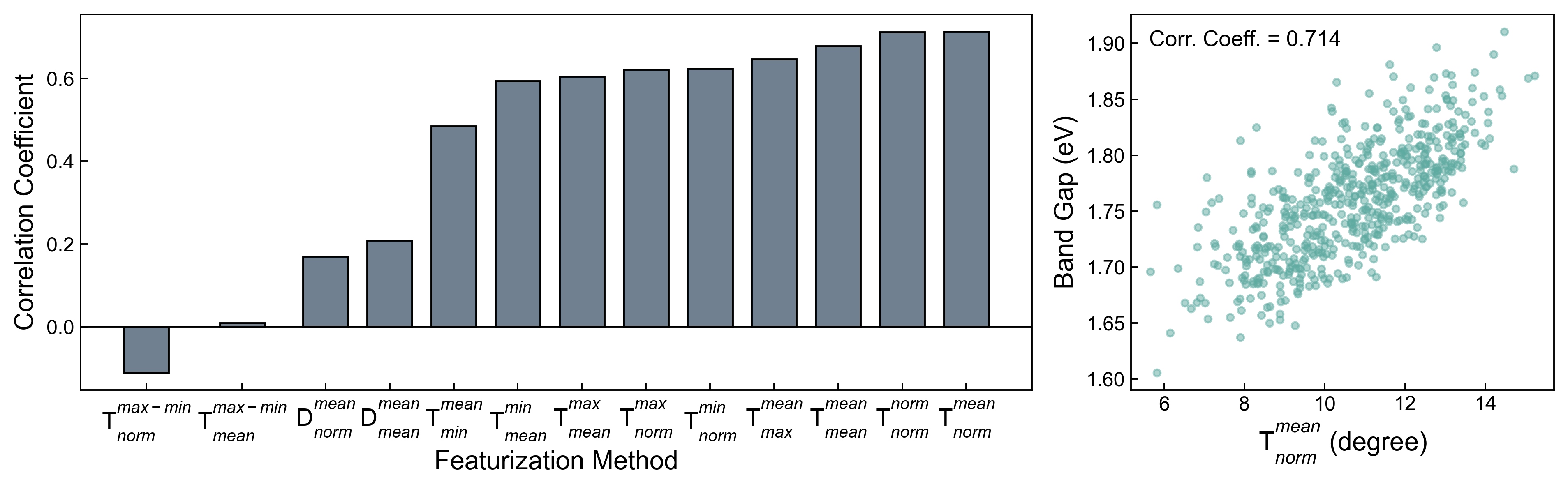}
    \caption{Correlation coefficients obtained for different structural featurisation strategies that map the local $2\times2\times2$ subcell geometry to the DFT band gap. In the notation $X_{\mathrm{sub}}^{\mathrm{sup}}$, $X$ denotes the underlying quantity ($T$: octahedral tilt, $D$: octahedral distortion), the subscript denotes the reduction at the single-octahedron level, and the superscript the reduction across all octahedra in the subcell. The strongest correlation is obtained for the octahedron-averaged tilt norm $T_{\mathrm{norm}}^{\mathrm{mean}}$, whose corresponding band gap distribution is shown in the right panel.}
    \label{si:feature}
\end{figure}

The comparison of candidate structural descriptors is summarised in Fig.~\ref{si:feature}. To relate the local band gap to the local structure, several reduced structural descriptors were tested. Each $2\times2\times2$ subcell contains eight octahedra, each described either by tilt components or by octahedral distortion measures. In the notation used in Fig.~\ref{si:feature}, the capital letter specifies the underlying quantity ($T$ for tilt-based features, $D$ for distortion-based features), the subscript denotes the reduction applied at the level of a single octahedron, and the superscript denotes the reduction applied across all octahedra in the subcell. For example, $T_{\mathrm{norm}}^{\mathrm{mean}}$ means first computing the norm of the three tilt components for each octahedron and then averaging this scalar quantity over all octahedra in the subcell. Among the tested featurisation strategies, the octahedron-averaged tilt norm $T_{\mathrm{norm}}^{\mathrm{mean}}$ gave the strongest correlation with the computed local band gap and was therefore adopted in the main text.

In the unfolded band structures, the smaller-gap subcells also tend to retain stronger spectral weight at the band edges, whereas more weakly weighted edge states are associated with stronger local symmetry breaking and spectral broadening. This trend is secondary to the main gap--tilt relation discussed in the main text, but is consistent with the view that the most weakly perturbed local environments remain more Bloch-like when projected onto the primitive-cell basis.

\begin{figure}[htb]
    \centering
    
    \includegraphics[width=0.75\textwidth]{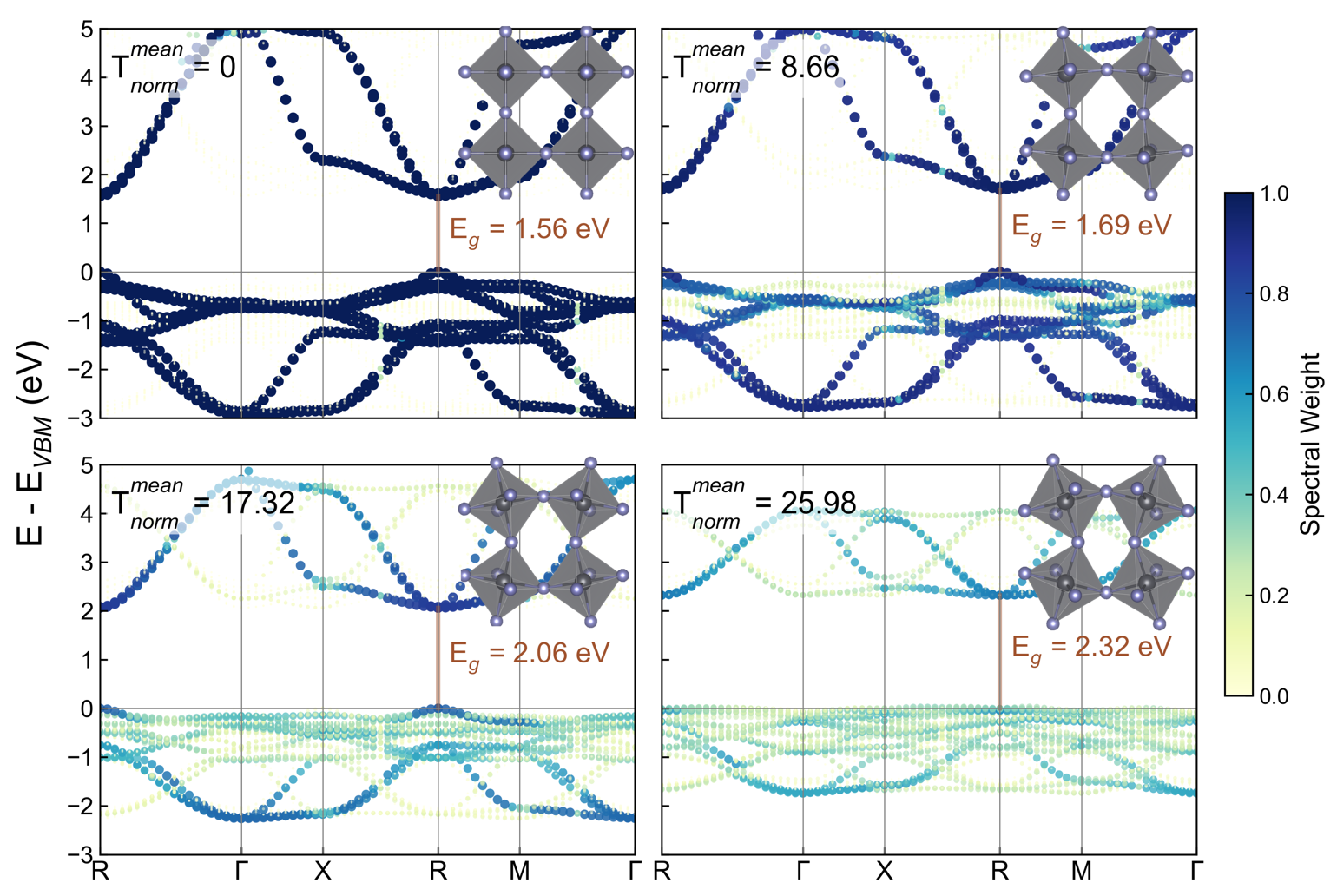}
    \caption{Idealised inorganic $a^{+}a^{+}a^{+}$ \ce{FAPbI3} band structures used to define the reference band gap--tilt relation. The four structures shown have identical \ce{Pb-I} bond lengths (3.25 \AA) and single-axis tilt angles of 0, 5, 10, and 15 degrees, respectively.}
    \label{si:ideal_band}
\end{figure}

To further define the structural baseline used for the band gap--tilt relation, we constructed an idealised $a^{+}a^{+}a^{+}$ series in which the tilt angle was varied systematically from zero to finite physical values. For this reference model, the A-site cations were removed so that the electronic structure is determined solely by the inorganic \ce{PbI3} framework. This does not qualitatively alter the band gap trend of interest here, because the band-edge states in lead halide perovskites are derived predominantly from Pb and I orbitals rather than from direct A-site contributions. Within this idealised framework, the band structure is therefore governed only by the octahedral tilt amplitude and the Pb--I bond length. Fig.~\ref{si:ideal_band} shows four representative structures with the same bond length and $a^{+}a^{+}a^{+}$ tilt angles of 0, 5, 10, and 15 degrees, corresponding to $T_{\mathrm{norm}}^{\mathrm{mean}}$ values of 0, 8.66, 17.32, and 25.98 degrees. The band gaps obtained from this series define one of the ideal reference lines plotted in Fig.~\ref{res:electro}c.

\end{document}